# Concealing-Restoring System for Physical Layer Data Based on Stochastic Filtering Theory

Tomohiro Fujii[a], Masao Hirokawa[b,*]

[a] *Internet Initiative Japan, Kyushu Branch*
[b] *Graduate School of Information Science and Electrical Engineering, Kyushu University*

**Abstract**

We propose a concealing-restoring system (CRS) for data on physical layer of the OSI reference model. CRS conceals those data by disturbing them with some random noises, and restores the data from the concealed ones to the original ones by using the noise elimination based on a proper stochastic filtering theory. Although we introduced the outline of the almost linear version of CRS in our previous work [1], we explain its details, and study its nonlinearization to improve the security of CRS in this paper.

*Keywords:* data concealing, data on physical layer, data restoring, noise elimination, random noise disturbance, stochastic filtering, stochastic process estimation
*2010 MSC:* 60H40, 68P25, 94A60

## 1. Introduction

The practical use of microdevices for the Internet of Things (IoT) interfaces has made remarkable advance in recent years. Due to its current cutting-edge technologies, IoT including the brain-machine interface (BMI)/ brain-computer interface (BCI) has been turned into the reality. For instance, Benabid et al succeed in controlling an exoskeleton by brain signals of a tetraplegic patient through an epidural wireless BMI [2]. Neuralink reports the news that they have been developing the N1 Link, a fully-implanted, wireless, high-channel count BMI chip [3, 4]. Flesher et al experimentally show that tactile percepts of signals from a robotic, prosthetic arm can be evoked by using a BCI by establishing an afferent channel to the BCI to mimic sensory input from the skin of a hand [5]. It is naturally feared that someone hacks into some embedding type medical devices and hijacks them. The serious apprehension may be beginning to become a reality. It is reported by the US Government Accountability Office (GAO) that a cardiac pacemaker device can be tampered from remote place by radiocommunication [6]. A demonstration of hacking a live jellyfish and the controlling its neural signals is performed by Xu and Dabiri [7]. These are becoming increasingly alarming problems, and we must establish the security in the microdevices [8]. In addition to the security problems above, there are some other problems for the drone (i.e., the flying IoT system in our real life): the hijack of the drone operation, and the exploitation of data on it. We should mind that someone can tap and steal signals between a drone and its remote controller.

*Corresponding author
*Email address:* hirokawa@inf.kyushu-u.ac.jp (Masao Hirokawa)

We are interested in the security for data in the space which has too small arithmetic capacity to install an encryption technology. The scenes we envision also include countermeasures for the firmware attack and side-channel attack in a low layer of the computer architecture. The firmware attack bypasses some softwares for antivirus and encryption on the higher-layer, and infects the lower-layer data in a device [9, 10, 11, 12]. The side-channel attack bypasses the cryptographic technique based on mathematical complexity and taps the cryptographic key [13, 14, 15, 16, 17]. Several sorts of side-channel attacks have been proposed, and many new side-channel attacks have been presented [18, 19, 20, 21, 22, 23, 24, 25, 26, 27]. In particular, CacheBleed [28] and TLBleed [29] have come under the industrial spotlight.

In this paper, we propose a concealing-restoring system (CRS) with some secret common keys for the data on the physical layer of the OSI reference model. Here, OSI is the abbreviation of the open systems interconnection, and the OSI reference model consists of 7 layers: the physical layer, the data link layer, the network layer, the transport layer, the session layer, the presentation layer, and the application layer from the lowest layer to the highest one [30]. We restrict our idea to scenes such as the instances described above, and we do not expect general scenes in wireless communication. Thus, the specification and construction of CRS should be for exclusive use among the device users, not be opened to the public. The secret common keys of CRS must be shared by the device users in advance with another method prior to the use of CRS.

Many endeavors have studied the security on physical layers [23, 31, 32,

33, 34, 35, 36, 37, 38, 39, 40, 41, 42] using individual, physical properties. In the light of noise, Lai, Gamal, and Poor use the random noise to hide the information of the secret key [43, 44]. Tomaru uses the random noise to generate secret keys in the common-key cryptography (symmetric key cryptography) [45, 46]. We also use random noises in our CRS, however, that is because we make the random bit flips (i.e., bit errors) directly in bit words on the physical layer. To the best of the authors' knowledge, the concealing-restoring method with using the random noises directly for the data (i.e., message in terms of cryptography) on physical layer is not established yet. We use some mathematical notions as our secret keys for the noise disturbance and noise elimination in CRS. The secret key generation by random noise is among ours. See §4.1. At first glance, the formation of the equations for CRS looks like a generalization of that for the man-in-the-middle (MiM) attacks. However, how to use of the equations in CRS is different from those in the MiM attacks. Also, see §4.1.

We suppose that we maintain the security of the data on physical layer over a proper period of time by installing CRS on the data link layer. Because the data link layer is situated between the physical layer and the network layer, and administers and controls the relation between those two layers. We handle the data on physical layer from the data link layer; CRS should be simple but effective as much as possible. The data concealing is performed by using the random noise disturbance introduced from the data link layer. The data restoring is achieved by the noise elimination in the data link layer. The introduction of the noise disturbance and the noise elimination are based on a proper stochastic filtering theory [47]. We showed its prototype which is the easiest and simplest CRS based on the linear Kalman filtering theory [1]. In this paper, we report how we can change the sort of the noise disturbance and the noise elimination in our CRS to improve its security ability by introducing and using nonlinearity.

The construction of this paper is in the following. In Section 2, we introduce some mathematical set-ups for CRS. In Section 3, we explain the general framework and the methods of CRS. In Section 4, we give a concrete example of CRS to show how CRS works. In Section 5, we optimize some parameters appearing in that example, and make its accuracy estimates based on statistical analysis. In Section 6, we apply the example of CRS to some digital pictorial images.

## 2. Mathematical Set-Ups

CRS is primarily for signal data $X_t$, $t \in \mathbb{R}$. Setting the initial data $X_t^1$ as $X_t^1 = X_t$, CRS is mathematically described by a simultaneous equation system (SES):

$$F_i(X_t^i, \dot{X}_t^i, U_t^i, W_t^{1,i}) = 0, \qquad i = 1, 2, \cdots, N, \tag{1}$$

$$X_t^{i+1} = f_i(X_t^i, W_t^{2,i}), \qquad i = 1, 2, \cdots, N, \tag{2}$$

$$U_t^{N+1} = f_{N+1}\left(X_t^{N+1}\right), \tag{3}$$

where $\dot{X}_t^i$ stands for the differential of $X_t^i$, i.e., $\dot{X}_t^i = dX_t^i/dt$. In this SES, each functional $F_i$ makes a stochastic differential equation (SDE), and each map $f_i$ determines a form of the linear or nonlinear equation. For these SDEs, we prepare $2N$ random noises $W_t^{j,i}$, $j = 1, 2$; $i = 1, 2, \cdots, N$. The stochastic processes $U_t^i$, $i = 1, 2, \cdots, N, N+1$, are $N+1$ *concealed data* from the original data $X_t$; that is, $X_t$ is concealed and shared in the form of $U_t^i$, $i = 1, 2, \cdots, N, N+1$ by CRS.

In order to reduce computing power for CRS, we may assume the map $f_i(\cdot, w) : x \mapsto f_i(x, w)$ is *bijective*. More specifically, regarding $w$ as a parameter and fixing it, the map $f_i(\cdot, w) : x \mapsto y = f_i(x, w)$ has the inverse map $f_i^{-1}(\cdot, w)$ such that $x = f_i^{-1}(y, w)$. The *nonlinearity* in this paper means that the linearity in the following sense is broken. We say that CRS is *linear* if each $F_i$ is a linear functional (i.e., $F_i(c_1\vec{x}_1 + c_2\vec{x}_2) = c_1F_i(\vec{x}_1) + c_2F_i(\vec{x}_2)$ for arbitrary vectors $\vec{x}_j = (x_j, y_j, u_j, w_j)$ and constants $c_j$, $j = 1, 2$), and each $f_i$ is a linear map (i.e., $f_i(c_1\vec{x}_1 + c_2\vec{x}_2) = c_1f_i(\vec{x}_1) + c_2f_i(\vec{x}_2)$ for arbitrary vectors $\vec{x}_j = (x_j, w_j)$ and constants $c_j$, $j = 1, 2$). CRS is said to be *almost linear* if $F_i$ and $f_i$, $i = 1, \cdots, N$, are linear but $f_{N+1}$, and *nonlinear* if some of $F_i$ and $f_i$, $i = 1, \cdots, N$, are nonlinear. The prototype [1] is an almost linear CRS.

In the stage of concealing data, we use the noise disturbance introduced by Eq.(1). We amplify the disturbance through Eqs.(2) and (3). In the stage of restoring data, meanwhile, the data restoration is achieved by a noise elimination. Thus, based on a proper stochastic filtering theory, we remove the random noises from each concealed data $U_t^i$, and we estimate the data $X_t^i$. In this paper, we denote the estimate by $\widehat{X}_t^i$, and call it the *estimated data* or *estimate* for the data $X_t^i$. The last estimate $\widehat{X}_t^1$ is our desired *restoration* of the original data $X_t$ in CRS. We denote it by $\widehat{X}_t$.

Throughout this paper, the data which should be concealed consist of bit words. We suppose that Alice conceals the bit words and gives them to Bob, and then, he restores the bit words from the concealed ones to the original ones. We envision the scene, for instance, Alice saves the concealed bit words and Bob restores and uses them, or she sends the concealed bit words and he receives and restores them. Here, Alice and Bob may be names of not only persons but also devices. We suppose that Eve tries to eavesdrop on or wiretap the communication consisting of bit words between Alice and Bob. Alice and Bob keep the forms of $F_i$ and $f_i$ secret in order to use them as secret common keys known only by themselves. They also use the distributions of the random noises as secret common keys. Therefore, it is the randomness and nonlinearity that they use as the secret common keys of CRS.

### *2.1. D/A & A/D Encodings*

In the OSI reference model, the transformation between analog data called signals and binary data called bit words is made on the physical layer. The data on the physical layer are conducted from the data link layer. Thus, we install our CRS on the data link layer to conceal and restore the data on the physical layer. We then need two transformations: the digital-to-analog encoding and the analog-to-digital encoding. Following the so-called Active-High (i.e., positive logic), we denote 'low' by '0' and 'high' by '1' to obtain the bit words. Of course, we may adopt the Active-Low (i.e., negative logic) instead of the Active-High. Concatenating $K+1$ bits, $a_0, a_1, \cdots, a_K \in \{0, 1\}$, we have the bit word, $a_0 a_1 \cdots a_K$. Thus, it is important how we determine 'low' and 'high' for the signals.

**Digital-to-Analog Encoding** (DAE): Obtaining the analog signal $X_t$ from the bit word, $a_0 a_1 \cdots a_K$, we use the linear interpolation in the following. First, we define $X_i$ by

$$X_i = \begin{cases} 1 & \text{if } a_i = 1, \\ 0 & \text{if } a_i = 0, \end{cases} \qquad i = 0, 1, \cdots, K.$$

Next, we connect the adjacent the data, $X_i$ and $X_{i+1}$, with a straight line for each $i = 0, 1, \cdots, K-1$,

$$X_t = (X_{i+1} - X_i)(t - i) + X_i, \qquad i \leq t \leq i + 1.$$

Then, we obtain the polygonal line $X_t$ on the interval, $0 \leq t \leq K$. We call this continuation to obtain $X_t$ from $X_i$ the *DA continuation.* Since this signal $X_t$ is obtained from the bit word $a_0 a_1 \cdots a_K$, we call $X_t$ a *binary pulse* of the $(K+1)$-bit word $a_0 a_1 \cdots a_K$. For example, we have the binary pulse $X_t$ of the 11-bit word, 11001001110, as in Figure 1.

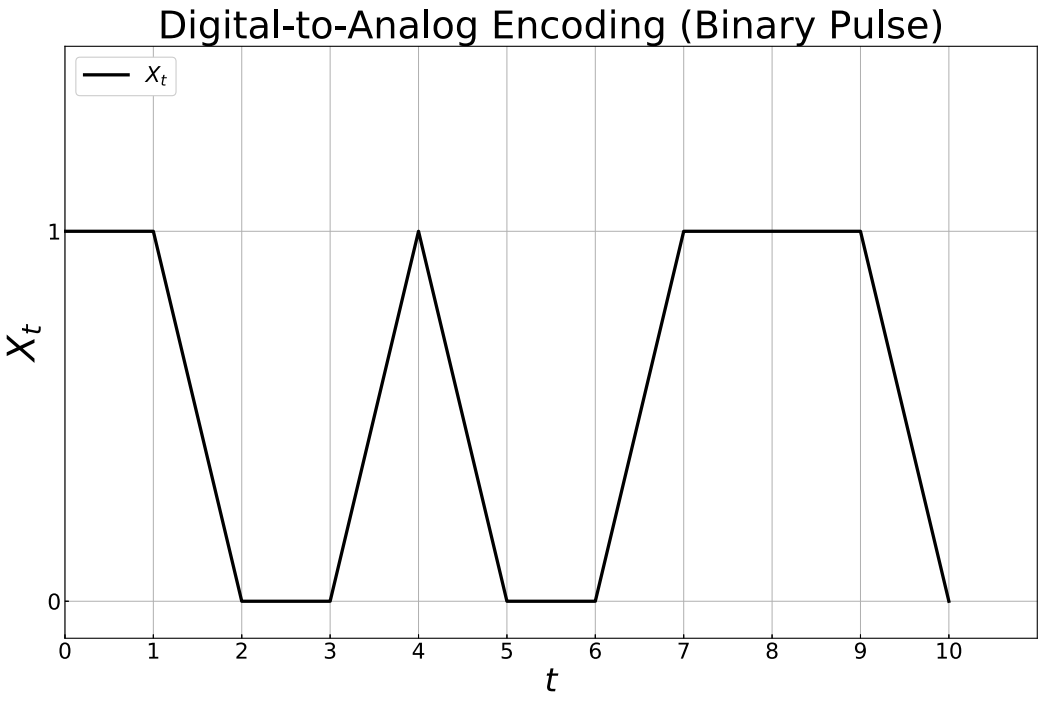

Figure 1: Binary Pulse by DAE. The binary pulse $X_t$ transformed from the 11-bit word, 11001001110.

**Analog-to-Digital Encoding** (ADE): We give a method for making the bit word, $\widehat{a}_0 \widehat{a}_1 \cdots \widehat{a}_K$, from a signal $\widehat{X}_t$. We now follow the digital abstraction with a logic high range (LHR) and a logic low range (LLR) [48]. We denote by $V_{\mathrm{LHR}}$ and $V_{\mathrm{LLR}}$ the minimum of LHR and the maximum of LLR, respectively. We will discuss how to determine them in §5.4. Our signal $\widehat{X}_t$ is deeply affected by the random noises, and each $\widehat{X}_i$ deviates from the correct bit, 0 or 1. Thus, we need the *noise margin*, the amount of noises that should be added to a worst-case value of $\widehat{X}_i$ such that $\widehat{X}_i$ can be interpreted as the correct bit. Thus, the values of $V_{\mathrm{LHR}}$ and $V_{\mathrm{LLR}}$ are determined such that each $\widehat{X}_i$ is always in either LHR or LLR by taking a proper noise margin into consideration as in Figure 2. The range between $V_{\mathrm{LLR}}$ and $V_{\mathrm{LHR}}$ is the so-called *forbidden zone* (FZ).

A bit $\widehat{a}_i$ is defined by

$$\widehat{a}_i = \begin{cases} 1 & \text{if } \widehat{X}_i \geq V_{\mathrm{LHR}}, \\ 0 & \text{if } \widehat{X}_i \leq V_{\mathrm{LLR}}. \end{cases}$$

For simplicity, in this paper, we assume that the forbidden zone is empty. Namely, we assume $V_{\mathrm{LLR}} = V_{\mathrm{LHR}}$. Under this assumption, we define the threshold $V_{\mathrm{thd}}$ by

$$V_{\mathrm{thd}} = V_{\mathrm{LLR}} = V_{\mathrm{LHR}}. \tag{4}$$

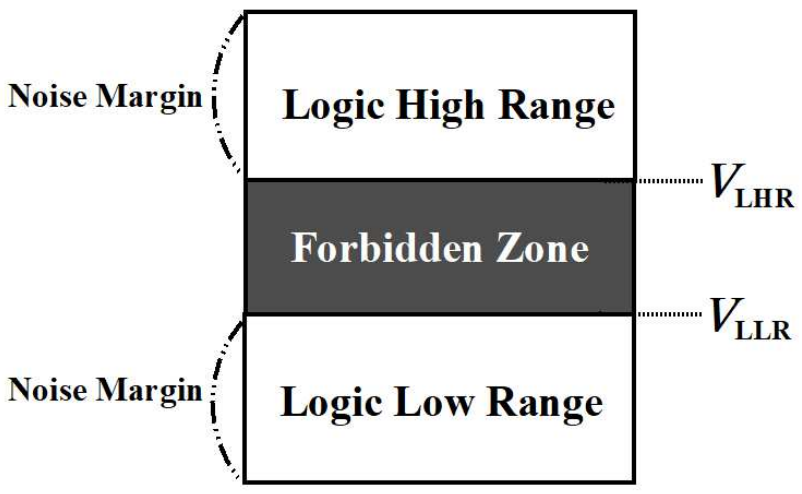

Figure 2: Logic Level & Noise Margin.

Then, for each $i = 0, 1, \cdots, K$, we define the bit $\widehat{a}_i$ by

$$\widehat{a}_i = \begin{cases} 1 & \text{if } \widehat{X}_i \geq V_{\text{thd}}, \\ 0 & \text{if } \widehat{X}_i < V_{\text{thd}}. \end{cases}$$

We call $\widehat{a}_k$ the *restored bit.* Concatenating the bits, $\widehat{a}_0, \widehat{a}_1, \cdots, \widehat{a}_K$, we obtain a bit word, $\widehat{a}_0\widehat{a}_1 \cdots \widehat{a}_K$. We call $\widehat{a}_0\widehat{a}_1 \cdots \widehat{a}_K$ the *restored bit word* from $\widehat{X}_t$.

For example, setting our threshold $V_{\text{thd}}$ as 0.5, we obtain the bit word, 01010100110, for the analog signal $\widehat{X}_t$ given in Figure 3.

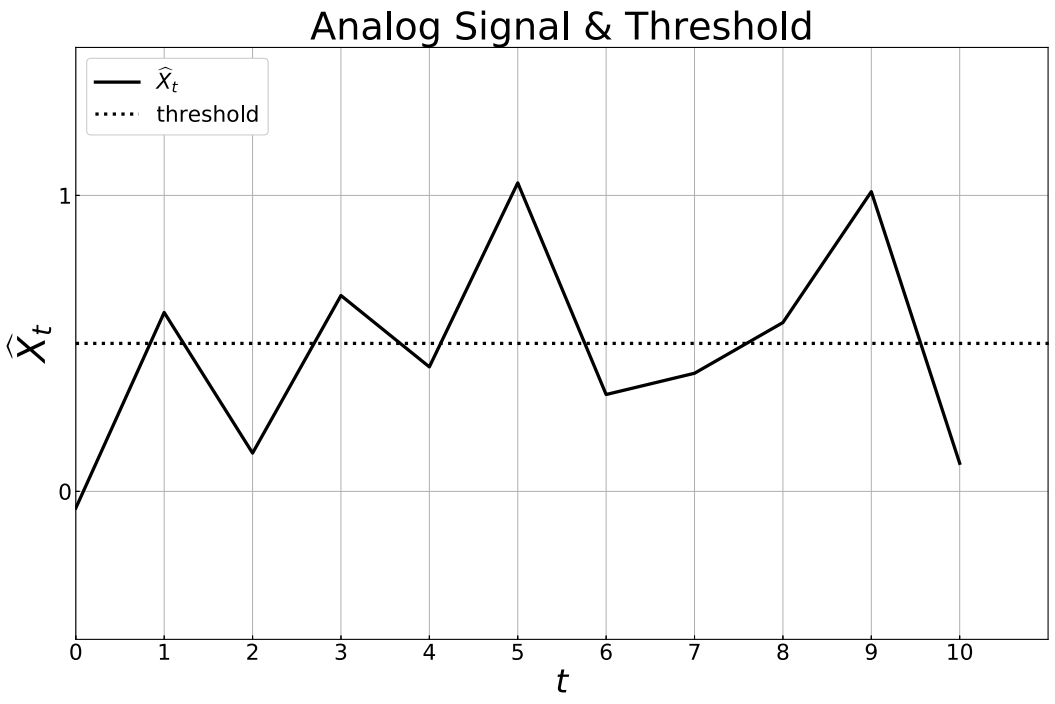

Figure 3: Analog Signal & Threshold for ADE. The analog signal $\widehat{X}_t$ takes the values, $\widehat{X}_0 = -0.05647363$, $\widehat{X}_1 = 0.60455548$, $\widehat{X}_2 = 0.12933681$, $\widehat{X}_3 = 0.66184304$, $\widehat{X}_4 = 0.42057086$, $\widehat{X}_5 = 1.04245415$, $\widehat{X}_6 = 0.32729871$, $\widehat{X}_7 = 0.3991026$, $\widehat{X}_8 = 0.57042864$, $\widehat{X}_9 = 1.01275446$, $\widehat{X}_{10} = 0.09490226$. The threshold is set as $V_{\text{thd}} = 0.5$.

**Remark on Forbidden Zone**: We can use $V_{\text{LHR}}$ and $V_{\text{LLR}}$ as secret common keys. In the case of our simple example, the secret common key is the threshold $V_{\text{thd}}$. We can introduce another tip into CRS by using $V_{\text{LHR}}$ and $V_{\text{LLR}}$. We here assume $0 < V_{\text{LLR}} < V_{\text{LHR}} < 1$. We now give characters $a_i$ a permission to take the value of $b = (V_{\text{LHR}} + V_{\text{LLR}})/2$ for instance. We define the component $X_i$ by

$$X_i = \begin{cases} 1 & \text{if } a_i = 1, \\ 0.5 & \text{if } a_i = b, \\ 0 & \text{if } a_i = 0, \end{cases} \qquad i = 0, 1, \cdots, K,$$

instead of that of the binary pulse. If we can determine the value of $V_{\text{LHR}}$ and $V_{\text{LLR}}$ such that all $\widehat{X}_i$ always fall into LHR for $i$ with $a_i = 1$; into LLR for $i$ with $a_i = 0$; into FZ for $i$ with $a_i = b$, then we can use $b$ for concealing bit words. We make $b$ lurk in the original bit word randomly. We define the character $\widehat{a}_i$ by

$$\widehat{a}_i = \begin{cases} 1 & \text{if } \widehat{X}_i \geq V_{\text{LHR}}, \\ \text{␣} & \text{if } V_{\text{LLR}} < \widehat{X}_i < V_{\text{LHR}}, \\ 0 & \text{if } \widehat{X}_i \leq V_{\text{LLR}}. \end{cases}$$

Ignoring '␣' in the output, $\widehat{a}_0\widehat{a}_1 \cdots \widehat{a}_K$, then we can reach the original bit word thinking each '␣' a space and narrowing the space between bits. For instance, if our output is 1␣0␣01␣1␣01␣01␣1␣0, then it is equal to 10011010110. This method with $V_{\text{LHR}}$ and $V_{\text{LLR}}$ introduces the effect of the Schmitt triggers of electric circuit into RS, while the method with $V_{\text{thd}}$ has the same effect as that of comparator of electric circuit.

**Remark on Analog Decoding & Encoding in CRS**: In our CRS, we always set the initial restored bit (i.e., initial restored value) as $\widehat{a}_0 (= \widehat{X}_0) = 0$. Thus, for the original bit word $a_0 a_1 \cdots a_K$, the part which we conceal should be given by $a_1 \cdots a_K$ so that the CRS works. Then, the part of the bit word which we need to restore by CRS is $\widehat{a}_1 \cdots \widehat{a}_K$. In other words, the concealed data of the original bit word $a_0 a_1 \cdots a_K$ input into CRS have an ancilla bit $a_0$, and then, the whole data are always restored as $0\,\widehat{a}_1 \cdots \widehat{a}_K$ by CRS.

The flow of the processes of our CRS is described in Figure 4. Firstly, Alice transforms a bit word to its binary pulse with DAE. Secondly, she conceals the binary pulse with the concealing system (CS) of CRS, and gives the concealed data to Bob. Bob first restores the concealed data to its restoration with the restoring system (RS) of CRS. After that, he transforms the restoration to the restored bit word with ADE.

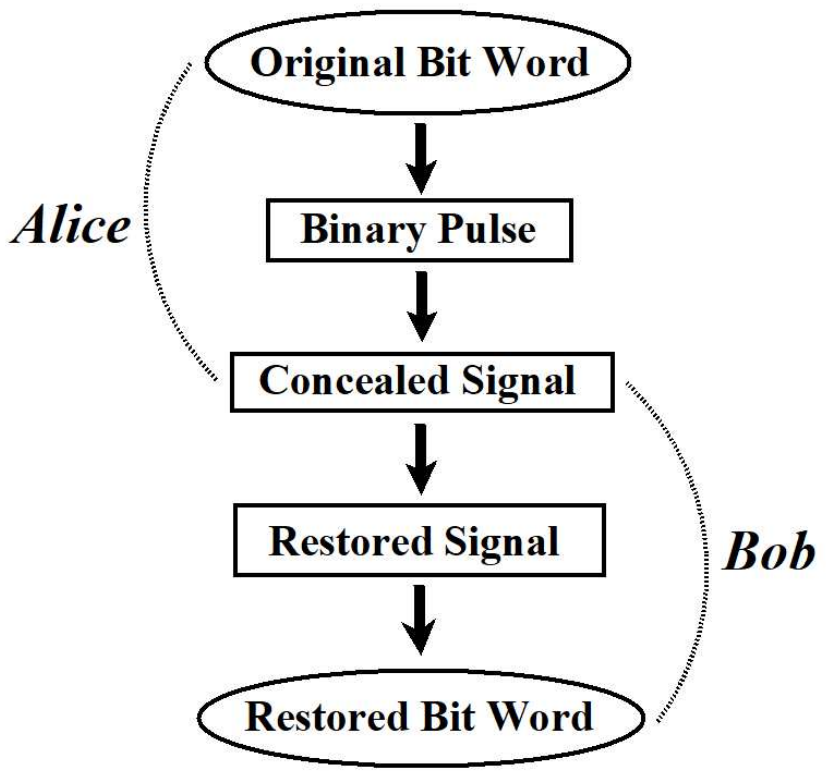

Figure 4: Flow Chart of CRS.

### *2.2. Schemes for Nonlinearization*

We here introduce nonlinearity into our CRS with using nonlinear functions. We will show this introduction of the nonlinearity brings an improvement of concealing in Section 6. What is the most important in the schemes is that any nonlinear function will do, so long as it is bijective. We prepare a bijective, nonlinear function $g_\nu$ of one variable. We assume that Alice and Bob know the concrete forms of the functions and the individual inverse functions. They can use $g_\nu$ as a secret common key.

Even in the case where the functionals, $F_i$, $i = 1, 2, \cdots, N$, are linear, we can nonlinearize them with some easy ways such as the nonlinearization schemes below. In the following nonlinearization schemes, we suppose that Bob can compute the inverse function $g_\nu^{-1}$ easily.

**Nonlinearization Scheme A**: We suppose that Alice sends the concealed data,

$$U_t^{\mathrm{nl},i} = g_\nu(U_t^i), \quad i = 1, \cdots, N, \tag{5}$$

to Bob instead of $U_t^i$. We also suppose that Eve happens to know the concrete form of each $F_i$, and moreover, manages to obtain all $U_t^{\mathrm{nl},i}$. As long as Alice and Bob keep the common key $g_\nu$ secret from somebody else, Eve is bound to restore the original bit word by using

$$F_i(X_t^i, \dot{X}_t^i, U_t^{\mathrm{nl},i}, W_t^{1,i}) = 0, \quad i = 1, 2, \cdots, N, \tag{6}$$

instead of Eq.(1). Then, her functional is $F_i^{\mathrm{Eve}}(x, y, u, w) = F_i(x, y, g_\nu(u), w)$, and it is nonlinear with respect to $\vec{x} = (x, y, u, w)$. On the other hand, Bob knows the concrete form of $g_\nu$ as the secret common key. He can immediately compute the data $U_t^i = g_i^{-1}(U_t^{\mathrm{nl},i})$; he has only to use the original, linear functional $F_i$ for restoring the concealed bit word.

We think that this method is almost simplest to introduce into CRS and to reduce computing power for RS.

**Nonlinearization Scheme B**: We suppose that Alice employs $g_\nu$ to define each map $f_i$ by

$$f_i(x, w) = g_\nu(x + w), \quad i = 1, 2, \cdots, N, N + 1. \tag{7}$$

Then, she has Eq.(2) as

$$X_t^{i+1} = f_i(X_t^i, W_t^{2,i}) = g_\nu(X_t^i + W_t^{2,i}),$$

and Eq.(3) as

$$U_t^{N+1} = f_{N+1}(X_t^N) = g_\nu(X_t^{N+1}).$$

We now define a signal $X_t^{\mathrm{nl},i}$ by

$$X_t^{\mathrm{nl},i} = g_\nu^{-1}(X_t^i), \quad i = 1, 2, \cdots, N, \tag{8}$$

namely, we define $X_t^{\mathrm{nl},i}$ so that we have $X_t^i = g_\nu(X_t^{\mathrm{nl},i})$. Then, we can rewrite Eqs.(1) and (2) as

$$F_i(g_\nu(X_t^{\mathrm{nl},i}), dg_\nu(X_t^{\mathrm{nl},i})/dt, U_t^i, W_t^{1,i}) = 0, \tag{9}$$

$$X_t^{\mathrm{nl},i+1} = g_\nu(X_t^{\mathrm{nl},i}) + W_t^{2,i}. \tag{10}$$

We again suppose that Eve happens to know each functional $F_i$ but the secret common key $g_\nu$. She is sure to think that Eqs.(9) and (10) make a nonlinear system for the signal $X_t^{\mathrm{nl},i}$. Bob knows the secret common key $g_\nu$; he can transform Eq.(9) to Eq.(1) which is linear for the signal $X_t^i$. Moreover, he has the equation, $X_t^{i+1} = g_\nu(X_t^i + W_t^{2,i})$, from Eq.(10). He is not puzzled by this nonlinearity because he can obtain $g_\nu^{-1}$ easily.

We note that Nonlinearization Scheme B needs much more computing power than Nonlinearization Scheme A for $i = 1, \cdots, N$. In this paper, therefore, we will employ Nonlinearization Scheme A for $i = 1, \cdots, N$ and Nonlinearization Scheme B for $i = N + 1$ to introduce nonlinearity into CRS.

We give some examples of such nonlinear functions below, and use them in this paper to show our concrete results. Although they are simple, they are efficiently effective and tractable.

We define a simple function $g_{\mathrm{s}}$ on the interval $[0, 1]$ by

$$g_{\mathrm{s}}(x) = \begin{cases} x + 0.75 & \text{if } 0 \leq x < 0.25, \\ x + 0.25 & \text{if } 0.25 \leq x < 0.5, \\ x - 0.25 & \text{if } 0.5 \leq x < 0.75, \\ x - 0.75 & \text{if } 0.75 \leq x \leq 1. \end{cases} \tag{11}$$

We can take more complicated one as

$$g_{\mathrm{c}}(x) = \begin{cases} x + 0.5 & \text{if } 0 \leq x < 0.125, \\ x + 0.75 & \text{if } 0.125 \leq x < 0.25, \\ x - 0.125 & \text{if } 0.25 \leq x < 0.375, \\ x & \text{if } 0.375 \leq x < 0.5, \\ x - 0.5 & \text{if } 0.5 \leq x < 0.625, \\ x + 0.125 & \text{if } 0.625 \leq x < 0.75, \\ x - 0.5 & \text{if } 0.75 \leq x < 0.875, \\ x - 0.25 & \text{if } 0.875 \leq x \leq 1. \end{cases} \tag{12}$$

We can make *infinitely many* these type of nonlinear functions with using some straight lines. The great advantage of them is that they are easy to handle by computer. In this way, we can take several kinds of nonlinear bijections $g_\nu$ in countless numbers, $\nu = 1, 2, \cdots$. Because we use the nonlinear function $g_\nu$ as our secret common keys, we obtain many candidates of the secret common keys by altering the definition of the function $g_\nu$.

## 3. Methods of Concealing & Restoring

In this section, we briefly explain how to conceal the signal $X_t$ and how to restore $X_t$ from the concealed data $U_t^i$ to the restoration $\widehat{X}_t$.

### *3.1. How to Conceal Data*

We here explain the method of our CS. We say CS is *linear* if all the functionals, $F_i$, $i = 1, \cdots, N$, and functions, $f_i$, $i = 1, \cdots, N, N+1$, are linear, and it is *nonlinear* if $F_i$ and $f_i$, $i = 1, \cdots, N$, are nonlinear.

We make the concealing procedures in the ascending order of the parameter $i$ by using Eqs.(1) and (2). To enter the concealing procedures, we prepare the initial data $X_t^1$ of SES, and use the original signal $X_t$ for it,

$$X_t^1 = X_t.$$

At the first step of the concealing procedures, we generate a random noise $W_t^{1,1}$. We input the initial data $X_t^1$, together with $W_t^{1,1}$, into Eq.(1), and have SDE,

$$F_1(X_t^1, \dot{X}_t^1, U_t^1, W_t^{1,1}) = 0.$$

We seek $U_t^1$ in this SDE, and we obtain a concealed data $U_t^1$. At the second step of the concealing procedures, inputting $X_t^1$ and $W_t^{2,1}$ into Eq.(2), we have the data,

$$X_t^2 = f_1(X_t^1, W_t^{2,1}),$$

for the next step. Preparing a random noise $W_t^{1,2}$, we input the data $X_t^2$, together with $W_t^{1,2}$, into Eq.(1), and have next SDE,

$$F_2(X_t^2, \dot{X}_t^2, U_t^2, W_t^{1,2}) = 0.$$

We then obtain the concealed data $U_t^2$. We repeat these similar steps, and obtain the concealed data, $U_t^1, U_t^2, \cdots, U_t^N$, by turns, and hide the data, $X_t^1, X_t^2, \cdots, X_t^N$, with the random noise disturbance. At the last step of the concealing procedures, we input the concealed data $X_t^N$ into Eq.(2) and obtain the data $X_t^{N+1}$. To exit the concealing procedures, we input the data $X_t^{N+1}$ into Eq.(3). We finally obtain the last concealed data $U_t^{N+1}$.

This is the concealing procedures of our CS. Following them, we can create the sequence of the concealed data, $U_t^1, U_t^2, \cdots, U_t^N, U_t^{N+1}$.

### *3.2. How to Restore Data*

We here explain the method of our RS. We say RS is *linear* if all the functionals, $F_i$, $i = 1, \cdots, N$, and functions, $f_i$, $i = 1, \cdots, N, N+1$, are linear, and it is *nonlinear* if $F_i$ and $f_i$, $i = 1, \cdots, N$, are nonlinear.

The restoring procedures are made in the descending order of the parameter $i$. To begin with, we estimate the data $X_t^{N+1}$ from the concealed data $U_t^{N+1}$ by using Eq.(3), and have the estimate $\widehat{X}_t^{N+1}$. If we assume that $f_{N+1}$ is bijective, we can obtain $\widehat{X}_t^{N+1}$ exactly as $X_t^{N+1}$. To enter the restoring procedures, we must find or invent a proper stochastic filtering theory to accomplish the noise elimination in RS. In that stochastic filtering theory,

Eqs.(1) and (2) should play roles of the state equation and the observation equation, respectively.

At the first step of the restoring procedures, we input the $\widehat{X}_t^{N+1}$ into the slot of $X_t^{N+1}$ of Eq.(2), and the concealed data $U_t^N$ into Eq.(1). We then have simultaneous equations to seek the data $X_t^N$,

$$\begin{aligned} &F_N(X_t^N, \dot{X}_t^N, U_t^N, W_t^{1,N}) = 0, \\ &\widehat{X}_t^{N+1} = f_N(X_t^N, W_t^{2,N}). \end{aligned}$$

Although we cannot completely reproduce the original random noises, $W_t^{1,N}$, $W_t^{2,N}$, we know their distributions as secret common keys instead. Thus, we can estimate the stochastic process $X_t^N$ with the help of the proper stochastic filtering theory. We then obtain the estimated data $\widehat{X}_t^N$ for $X_t^N$. At the second step of the restoring procedures, inputting the estimated data $\widehat{X}_t^N$ into the slot of $X_t^N$ of Eq.(2), and the concealed data $U_t^{N-1}$ into Eq.(1), we reach simultaneous equations to seek the data $X_t^{N-1}$,

$$\begin{aligned} &F_{N-1}(X_t^{N-1}, \dot{X}_t^{N-1}, U_t^{N-1}, W_t^{1,N-1}) = 0, \\ &\widehat{X}_t^N = f_{N-1}(X_t^{N-1}, W_t^{2,N-1}). \end{aligned} \tag{13}$$

In the same way as in the first step, the stochastic filtering theory gives us the next estimated data $\widehat{X}_t^{N-1}$. We repeat the similar steps, and obtain the estimates, $\widehat{X}_t^N, \widehat{X}_t^{N-1}, \cdots, \widehat{X}_t^2, \widehat{X}_t^1$, by turns. We pick up the last estimate $\widehat{X}_t^1$, and this is the restoration $\widehat{X}_t$ of CRS for the original signal $X_t$.

**Reminder on Initial Bit & Value of CRS**: The initial value of the restoration $\widehat{X}_t$ is always $\widehat{X}_0 = 0$ in our RS. Thus, we must omit the first restored bit $\widehat{a}_0$ from the output $\widehat{a}_0\widehat{a}_1 \cdots \widehat{a}_K$ of RS, and then, the part of our restored bit word which we need is always $\widehat{a}_1 \cdots \widehat{a}_K$. Therefore, when we conceal a bit word, the bit word that we conceal should be set as $a_1 \cdots a_K$, and the bit $a_0$ is used an ancilla bit for the whole bit word $a_0 a_1 \cdots a_K$ in CS.

*3.3. Mathematical Parameters for Secret Common Keys*

In CRS, some mathematical parameters play roles of secret common keys as in Table 1. Even if Eve obtains the concealed data, $U_t^i$, $i = 1, \cdots, N$, she must correctly grasp SES consisting of Eqs.(1), (2), (3). In particular, since the concealing is made by some random noises, she is forced to make the noise elimination based on a proper stochastic filtering to restore the bit word communicated to Bob by Alice. In addition, she needs to know the individual distributions of the random noises that Alice and Bob use. Because the binary pulse that Alice conceals and sends is signal, Eve is required to guess right about the method of ADE, especially, their noise margin.

| Mathematical Tools | Secret Common Keys |
|---|---|
| SDE | $F_i, \quad i = 1, \cdots, N$ |
| Nonlinearity | $F_i, \quad i = 1, \cdots, N$<br>$f_i, \quad i = 1, \cdots, N, N+1$ |
| Probabilistic Property | distribution |
| Noise Margin | $V_{\text{LLR}}, \quad V_{\text{LHR}}$ |

Table 1: Parameters' Roles as Secret Common Keys

## 4. CRS with A Simple SES

Regarding how to determine each of functionals, $F_i$, $i = 1, 2, \cdots, N$, its any definition is fine so long as a noise elimination (and thus, stochastic filtering theory) is established for RS. To eliminate the random noise, generally speaking, we must know the concrete forms of the functionals, $F_i$, $i = 1, \cdots, N$. Although we did not show the algorithms for the implementation of almost linear CRS in Ref.[1], we will show them in this section.

We suppose that Alice and Bob's nonlinearization strategy for introduction of nonlinearity into CRS is the following. We denote $U_t^{\text{nl},i}$ just by $U_t^i$ as the abbreviation in their strategy.

**Nonlinearization Strategy**: Alice and Bob employ Nonlinearization Scheme A for $i = 1, \cdots, N$, and Nonlinearization Scheme B for $i = N + 1$. Alice uses a bijective, nonlinear function $g_\nu$ for CS. Then, Bob uses the inverse function $g_\nu^{-1}$ for RS, and has only to use linear RS based on the 'linear' Kalman filtering theory.

*4.1. Example of Linear Functionals*

We here consider the linear CRS. We give an example of concrete definition of $F_i$ though there are many definitions. In the actual field, the forms of the functionals, $F_i$, $i = 1, \cdots, N$, should be confidential only for Alice and Bob's secret common keys. The example in this paper is just to explain how CRS works. We employ the *Langevin equation* for Eq.(1). That is, Eq.(1) can read: $\dot{X}_t^i$ = 'viscous resistance' + 'random force'. Determining the 'viscous resistance' and 'random force' in concrete, we prepare a constant $A^i$, non-zero constants $b_u^i, b^i$, and a function $v^i(t)$ for each $i$. We give them as 'viscous resistance' $= -(1 - A^i)X_t^i$ and 'random force' $= b_u^i U_t^i + b^i v^i(t) - b_u^i W_t^{1,i}$. Here, $1 - A^i$ plays the role of the resistance coefficient in the Langevin equation, and thus, it should usually be positive, i.e., $1 \geq A^i$, in physical sense. We can employ a random noise as $v^i(t)$, and then, we suppose that its distribution is given by $N(0, (\sigma_v)^2)$, the normal distribution whose mean and standard deviation are respectively 0 and $\sigma_v$. We define each functional $F_i$ such that it makes a SDE,

$$dX_t^i = -(1 - A^i)X_t^i dt + b_u^i U_t^i dt + b^i v^i(t)dt - b_u^i dB_t^i. \tag{14}$$

That is,

$$\dot{X}_t^i = -(1 - A^i)X_t^i + b_u^i U_t^i + b^i v^i(t) - b_u^i W_t^{1,i}. \tag{15}$$

We suppose that each $W_t^{1,i}$ is Gaussian white noise whose mean $m^{1,i}$ and variance $V^{1,i}$ are respectively 0 and $(\sigma_1^i)^2$. $B_t^i$ is the Brownian motion given by $W_t^{1,i} = dB_t^i/dt$, $i = 1, 2, \cdots, N$. In this example, for simplicity, we employ linear function of $\vec{x} = (x, w)$ for $f_1, \cdots, f_N$: $f_i(\vec{x}) = x + w$, $i = 1, 2, \cdots, N$. Then, Eq.(2) becomes

$$X_t^{i+1} = f_i(X_t^i, W_t^{2,i}) = X_t^i + W_t^{2,i}. \tag{16}$$

We note that RHS of Eq.(16) is the simplest version of observations between the Alice's transmitter and Bob's receiver in the standard model of the secret key generation in the wireless communication (see §V of Ref.[39]). We moreover hide $X_t^i$, $i = 1, \cdots, N$, with Eq.(15). Eq.(15) reminds us of a generalization of the MiM attacks in which $b_u^i U_t^i + b^i v^i(t)$ plays the role of Eve's signal to steal some information about the secret key. We, however, use $U_t^i$ and $v^i(t)$ as the concealed data and one of common secret keys of CRS, respectively, between Alice and Bob. We also suppose that each $W_t^{2,i}$ is Gaussian white noise whose mean $m^{2,i}$ and variance $V^{2,i}$ are respectively 0 and $(\sigma_2^i)^2$. We assume that the noises $W_t^{1,i}$ and $W_t^{2,i}$ are independent for each $i = 1, 2, \cdots, N$, but the noises $W_t^{2,i}$, $i = 1, 2, \cdots, N$, may not always be independent.

Since Eqs.(14) and (16) make a linear system and respectively play the roles of the state equation and observation equation in the stochastic filtering theory, we can employ the 'linear' Kalman filtering theory [47, 49, 50, 51] for the noise elimination.

Due to Eq.(14), the secret common key $F_i$ is reduced to the parameters $A^i, b^i, b_u^i$, and the function $v^i(t)$; they become the secret common keys instead of $F_i$. In particular, $v^i(t)$ becomes a secret common key generated by random noise. Another secret common key is the distribution of the white noises $W_t^{j,i}$. Following Nonlinearization Strategy, Alice and Bob employ a bijective, nonlinear function $g_\nu$. Since Bob knows the inverse function of $g_\nu$, he can make the restoration merely by using 'linear' Kalman filtering. The nonlinear function also becomes Alice and Bob's secrete common key then. Thus, the candidates of Alice and Bob's secret common keys in this example are as in Table 2.

| Secret Common Keys | |
|---|---|
| Source | Our Choice |
| $F_i$ | $A^i, b^i, b_u^i$, and $v^i(t)$; $g_\nu$ |
| $f_i$ | $g_\nu$ |
| $W_t^{j,i}$ | $N(m^{j,i}, V^{j,i})$ |
| $V_{\text{LLR}}$, $V_{\text{LHR}}$ | $V_{\text{thd}}$ |

Table 2: Choice of Secret Common Keys in Concrete.

The residence coefficient $(1 - A^i)$ in the leading term of RHS of Eq.(15) is controlled by the parameter $A^i$. The amplitude of the concealed data $U_t^i$ and the white noise $W_t^{1,i}$ are governed by the parameter $b_u^i$. The amplitude of the secret common key $v^i(t)$ can be enhanced by the parameter $b^i$. The details of the optimization for determining these parameters are in §5.1.

### *4.2. Discrete Version of Linear CRS*

We use numerical analysis to obtain the restoration $\widehat{X}_t$ by linear CRS. We discretize Eq.(1); we approximate the differential by the forward difference,

$$\frac{dX_t^i}{dt} \approx \frac{X_{t+\Delta t}^i - X_t^i}{\Delta t},$$

for $t = k\Delta t$ with $k = 0, 1, 2, \cdots, n$. Using $\Delta t$ as the time unit, the differential is approximated as $dX_t^i/dt \approx X_{k+1}^i - X_k^i$, and therefore, Eq.(14) is discretized as

$$X_{k+1}^i = A^i X_k^i + b_u^i U_k^i + b^i v_k^i - b_u^i W_k^{1,i}. \tag{17}$$

Here, we denote $v^i(k)$ by $v_k^i$. With this discretization, we consider the simplest Eqs.(16) and (3) with the forms,

$$\begin{aligned} X_k^{i+1} &= X_k^i + W_k^{2,i}, \quad i = 1, 2, \cdots, N, \\ U_k^{N+1} &= X_k^{N+1}, \end{aligned} \tag{18}$$

by setting $f_i(x, w) = x + w$, $i = 1, \cdots, N$, and $f_{N+1}(x) = x$. In addition to the concealed data $U_k^{N+1}$, we can obtain the other concealed data $U_k^i$ as

$$U_k^i = \frac{1}{b_u^i} \left\{ X_{k+1}^i - A^i X_k^i - b^i v_k^i \right\} + W_k^{1,i} \tag{19}$$

by Eq.(17). Though we did not describe the algorithm to create the concealed data in our previous paper [1], it is in Algorithm 1.

---

**Algorithm 1** Linear CS

---

Determine secret common keys, $A^i, b^i, b_u^i; m^{j,i}, \sigma_j^i$.
Determine a common key, $v_k^i$.
Define white noises, $W_k^{j,i}$, $j = 1, 2$; $i = 1, 2, \cdots, N$, with the individual mean $m^{j,i}$ and variance $(\sigma_j^i)^2$.
Determine $N$, how many SDEs you want.
Determine $n$, how many data you handle.
**for** $k = 0, 1, \cdots, n$ **do**
  Set $X_k^1 := X_k$
**end for**
**for** $i = 1, 2, \cdots, N$ **do**
  **for** $k = 0, 1, \cdots, n$ **do**
    Set $U_k^i := (b_u^i)^{-1} \left\{ X_{k+1}^i - A^i X_k^i - b^i v_k^i \right\} + W_k^{1,i}$
    Set $X_k^{i+1} := X_k^i + W_k^{2,i}$
  **end for**
**end for**
**for** $k = 0, 1, \cdots, n$ **do**
  Set $U_k^{N+1} := X_k^{N+1}$
**end for**

---

Conversely, we can estimate the data, $X_k^N$, $X_k^{N-1}$, $\cdots$, $X_k^1$, from the concealed data, $U_k^N$, $U_k^{N-1}$, $\cdots$, $U_k^1$, by using the linear Kalman filtering theory [47, 49, 50, 51]. We can make an algorithm to obtain the estimated data, $\widehat{X}_k^N$, $\widehat{X}_k^{N-1}$, $\cdots$, $\widehat{X}_k^1$, in the following. We denote the priori estimate by $\widehat{X}_k^{-,i}$, the variance by $P_k^i$, the priori variance by $P_k^{-,i}$, and the Kalman gain by $g_k^i$ for each $i = N, \cdots, 1$. We then repeat the procedures consisting of Prediction Stage and Filtering Stage of the linear Kalman filtering from $i = N$ to $i = 1$ as described below.

Before entering the procedures of the Kalman filtering, we prepare the data $\widehat{X}_k^{N+1}$ simply by

$$\widehat{X}_k^{N+1} = U_k^{N+1}, \quad k = 0, 1, \cdots, n.$$

Prediction Stage and Filtering Stage of the linear Kalman filtering are as follows:

**Prediction Stage**:

$$\begin{aligned} \widehat{X}_k^{-,i} &= A^i \widehat{X}_{k-1}^i + b_u^i U_{k-1}^i + b^i v_{k-1}^i, \\ P_k^{-,i} &= (A^i)^2 P_{k-1}^i + (\sigma_1^i)^2 (b_u^i)^2. \end{aligned}$$

**Filtering Stage**:

$$\begin{aligned} g_k^i &= \frac{P_k^{-,i}}{P_k^{-,i} + (\sigma_2^i)^2}, \\ \widehat{X}_k^i &= \widehat{X}_k^{-,i} + g_k^i \left( \widehat{X}_k^{i+1} - \widehat{X}_k^{-,i} \right), \\ P_k^i &= \left(1 - g_k^i\right) P_k^{-,i}. \end{aligned}$$

Although the algorithm to restore the concealed data was not described in our previous paper [1], it is given in Algorithm 2.

When we introduce nonlinearity into linear CRS, we adopt Nonlinearization Strategy in Algorithms 1 and 2.

### *4.3. Explanation of CRS with Examples*

We explain how CRS works with using a simple example. We apply both linear and nonlinear CRSs to a bit word. We, for instance, prepare the bit word,

$$a_0 a_1 \cdots a_{20} = 101111001000011110111. \tag{20}$$

**Algorithm 2** Restoration from Concealed Data $U_t^i$, $i = 1, 2, \cdots, N, N+1$

Get secret common keys, $A^i, b^i, b_u^i; m^{j,i}, \sigma_j^i$.
Get the common key, $v_k^i$.
Obtain the concealed data, $U_k^i$.
**for** $k = 0, 1, \cdots, n$ **do**
  Set $\widehat{X}_k^{N+1} := U_k^{N+1}$
**end for**
**for** $i = N, N-1, \cdots, 0$ **do**
  Determine initial values, $\widehat{X}_0^{-,i}$, $P_0^i$, $g_0^i$, $\widehat{X}_0^i$.
  **for** $k = 1, 2, \cdots, n$ **do**
    Set $\widehat{X}_k^{-,i} := A^i \widehat{X}_{k-1}^i + b_u^i U_{k-1}^i + b^i v_{k-1}^i$
    Set $P_k^{-,i} := (A^i)^2 P_{k-1}^i + (\sigma_1^i)^2 (b_u^i)^2$
    Set $g_k^i := P_k^{-,i} \{P_k^{-,i} + (\sigma_2^i)^2\}^{-1}$
    Set $\widehat{X}_k^i := \widehat{X}_k^{-,i} + g_k^i (\widehat{X}_k^{i+1} - \widehat{X}_k^{-,i})$
    Set $P_k^i := (1 - g_k^i) P_k^{-,i}$
  **end for**
**end for**
**for** $k = 0, 1, \cdots, n$ **do**
  Set $\widehat{X}_k := \widehat{X}_k^1$
**end for**

We make Eqs.(14), (16), and (3) for $N = 2$ with $A^i = 0.1$ and $b^i = b_u^i = 1$ for each $i = 1, 2$. We will explain the reason why we chose the parameters, $A^i$, $b^i$, $b_u^i$, like the above in §5.1. For the introduction of nonlinearity, we use the nonlinear function $g_c$ given by Eq.(12) as $g_\nu$ in Eq.(5) for $i = 1, 2$ and in Eq.(7) for $i = 3$. We define a random sequence, $v_k^i$, $k = 1, \cdots, n$, of bits for each $i$ by applying ADE to normal random numbers with $N(0, (\sigma_v)^2)$. We assume that the individual means of white noises, $W_t^{j,i}$, $i = 1, \cdots, N$, $j = 1, 2$, are all 0, the standard deviation of the white noise $W_t^{j,1}$ is $\sigma_1^i = \sigma_1$, and that of the white noise $W_t^{j,2}$ is $\sigma_2^i = \sigma_2$.

Following DAE, we can make the binary pulse $X_t$ from the bit word $a_0 a_1 \cdots a_{20}$ in Eq.(20) as in Figure 5. From this binary pulse $X_t$, linear CS makes the concealed data, $U_t^1$, $U_t^2$, $U_t^3$, as in Figure 6. We here recall Reminder on Initial Bit & Value of CRS in §3.2. We must omit the first restored ancilla bit $\widehat{a}_0$ from the output $\widehat{a}_0 \widehat{a}_1 \cdots \widehat{a}_{20}$, and then, our restored bit word is always $\widehat{a}_1 \cdots \widehat{a}_{20}$. Thus, 0 is the initial bit $\widehat{a}_0$ of CRS though the original bit $a_0$ is 1.

We show the restoration $\widehat{X}_t$ by linear RS and its restored bit word in Table 3.

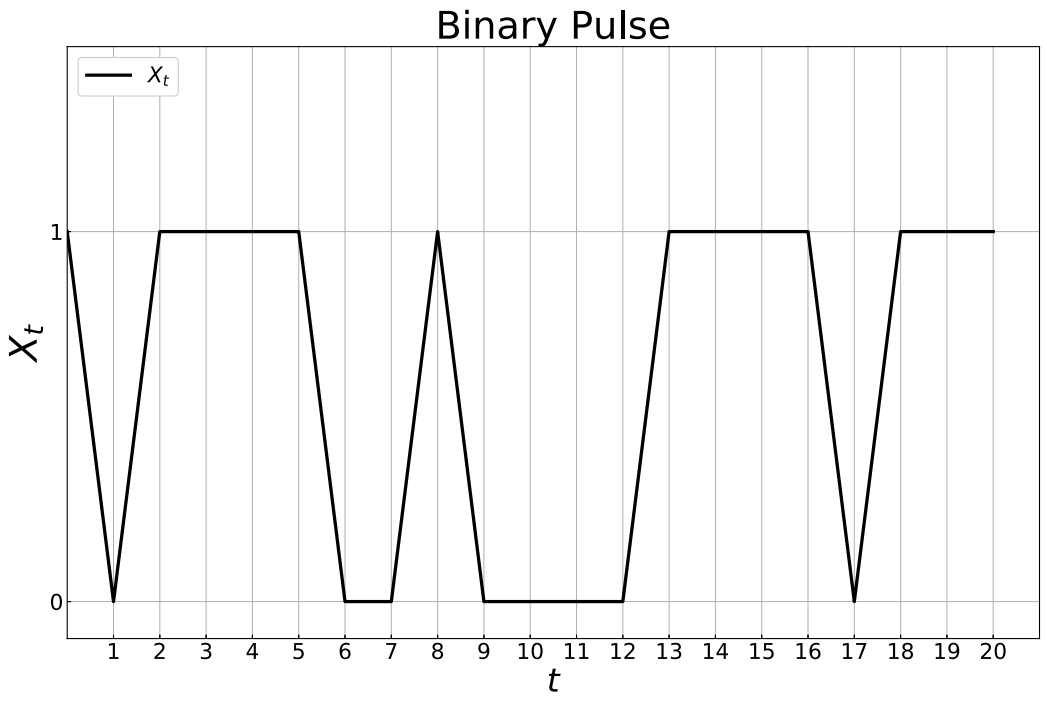

Figure 5: Binary Pulse. The binary pulse is made from the 21-bit word given by Eq.(20) by DAE with $V_{\mathrm{thd}} = 0.5$.

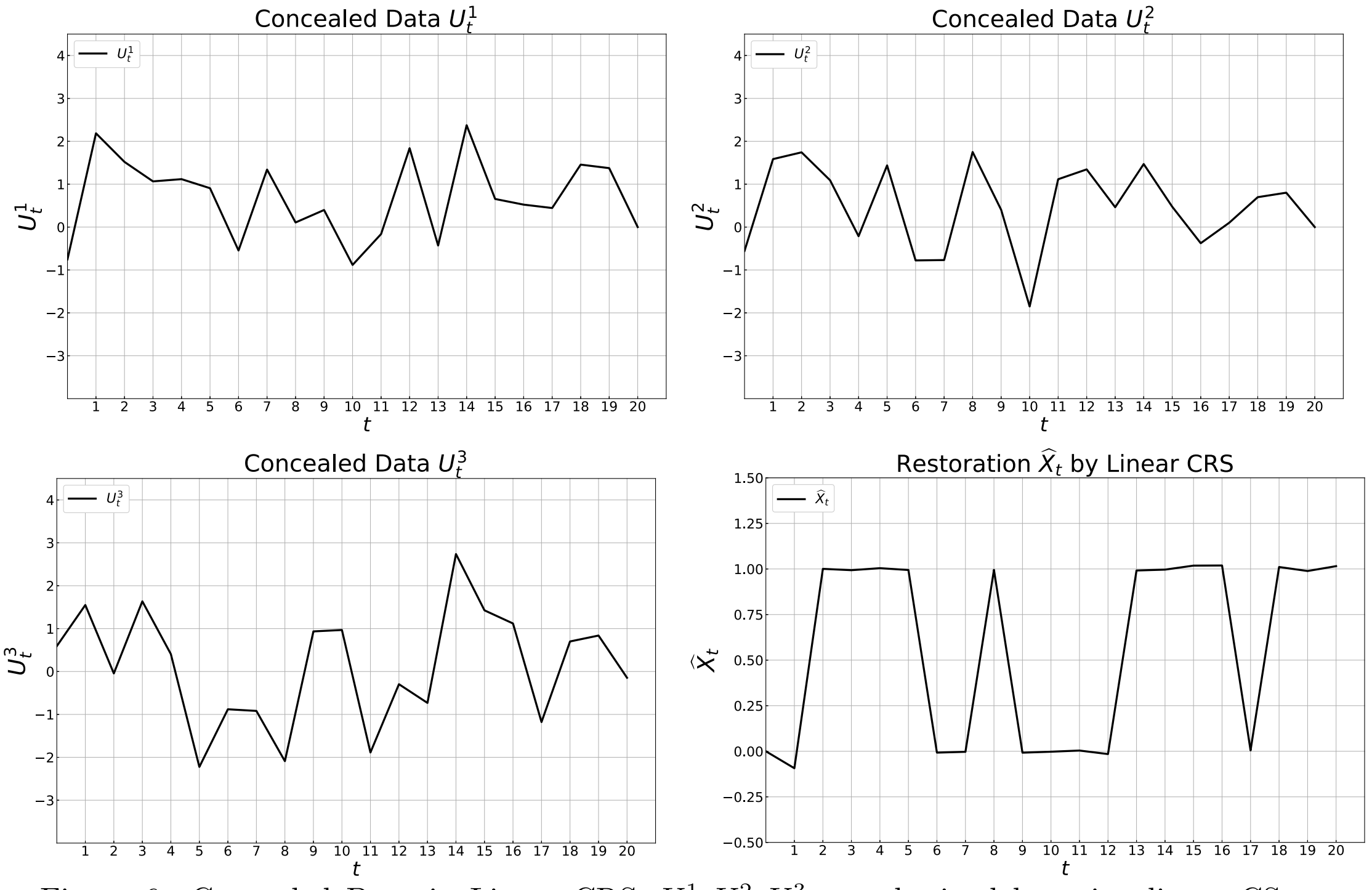

Figure 6: Concealed Data in Linear CRS. $U_t^1, U_t^2, U_t^3$ are obtained by using linear CS for the bit word given in Figure 5. DA continuation is used to obtain each graph of $U_t^i$. Following Reminder on Initial Bit & Value of CRS in §3.2, the value of $\widehat{X}_0$ is always 0 as the initial value of the restoration $\widehat{X}_t$, which implies that the value of $\widehat{a}_0$ is different from that of the ancilla bit $a_0$ in this case. Thus, we discard $\widehat{a}_0$ from the output of our RS. We set parameters as $A = 0.1$, $b = 1$, $b_u = 1$, $\sigma_1 = 0.01$, and $\sigma_2 = \sigma_v = 1$.

We now assume that Eve succeeds in wiretapping and obtaining one of $U_t^1$, $U_t^2$, $U_t^3$. Moreover, we assume that she realizes that Active-High is used, and knows the ADE method with $V_{\rm thd} = 0.5$. She tries to get a bit word from the obtained data $U_t^i$ without the noise elimination. Then, she obtains bit words as in Table 3. The table says that Bob succeeds in obtaining the correct restoration, and in recovering the original bit word $a_1 \cdots a_{20}$, but the ancilla bit $a_0$, from the restoration. On the other hand, Eve fails to restore the bit words communicated by Alice to Bob.

| $t$ | 0 | 1 | 2 | 3 | 4 | 5 | 6 | 7 | 8 | 9 | 10 | 11 | 12 | 13 | 14 | 15 | 16 | 17 | 18 | 19 | 20 |
|---|---|---|---|---|---|---|---|---|---|---|---|---|---|---|---|---|---|---|---|---|---|
| $X_t$ | 1 | 0 | 1 | 1 | 1 | 1 | 0 | 0 | 1 | 0 | 0 | 0 | 0 | 1 | 1 | 1 | 1 | 0 | 1 | 1 | 1 |
| $U_t^1$ | -0.753 | 2.186 | 1.519 | 1.065 | 1.117 | 0.906 | -0.546 | 1.341 | 0.107 | 0.398 | -0.882 | -0.162 | 1.838 | -0.43 | 2.374 | 0.655 | 0.523 | 0.444 | 1.456 | 1.374 | 0 |
| ADE | 0 | 1 | 1 | 1 | 1 | 1 | 0 | 1 | 0 | 0 | 0 | 0 | 1 | 0 | 1 | 1 | 1 | 0 | 1 | 1 | 0 |
| $U_t^2$ | -0.556 | 1.585 | 1.743 | 1.092 | -0.213 | 1.438 | -0.777 | -0.769 | 1.751 | 0.403 | -1.85 | 1.115 | 1.345 | 0.463 | 1.47 | 0.475 | -0.376 | 0.101 | 0.697 | 0.801 | 0 |
| ADE | 0 | 1 | 1 | 1 | 0 | 1 | 0 | 0 | 1 | 0 | 0 | 1 | 1 | 0 | 1 | 0 | 0 | 0 | 1 | 1 | 0 |
| $U_t^3$ | 0.592 | 1.55 | -0.045 | 1.637 | 0.407 | -2.224 | -0.881 | -0.919 | -2.09 | 0.935 | 0.968 | -1.888 | -0.296 | -0.732 | 2.739 | 1.426 | 1.12 | -1.178 | 0.701 | 0.838 | -0.146 |
| ADE | 1 | 1 | 0 | 1 | 0 | 0 | 0 | 0 | 0 | 1 | 1 | 0 | 0 | 0 | 1 | 1 | 1 | 0 | 1 | 1 | 0 |
| $\widehat{X}_t$ | 0 | -0.093 | 1.001 | 0.993 | 1.004 | 0.994 | -0.007 | -0.003 | 0.995 | -0.008 | -0.002 | 0.004 | -0.015 | 0.992 | 0.996 | 1.018 | 1.019 | 0.005 | 1.011 | 0.989 | 1.015 |
| ADE | 0 | 0 | 1 | 1 | 1 | 1 | 0 | 0 | 1 | 0 | 0 | 0 | 0 | 1 | 1 | 1 | 1 | 0 | 1 | 1 | 1 |

Table 3: Values at $t = 0, 1, \cdots, 20$ for Figure 6. Eve fails to restore the original bit words by ADE from $U_t^i$, while Bob succeeds in doing it by ADE from $\widehat{X}_t$. Following Reminder on Initial Bit & Value of CRS in §3.2, we use $a_0$ as an ancilla bit, and discard $\widehat{a}_0$ from the output of RS.

Meanwhile, nonlinear CS makes the concealed data, $U_t^1$, $U_t^2$, $U_t^3$, as in Figure 7 from the binary pulse $X_t$ in Figure 5.

Bob can obtain the restoration $\widehat{X}_t$ by nonlinear RS and the restored bit word from it. Following Reminder on Initial Bit & Value of CRS in §3.2, the restored ancilla bit $\widehat{a}_0$ of RS is 0, while the ancilla bit $a_0$ of CS is 1. We assume that Eve again succeeds in obtaining one of $U_t^1$, $U_t^2$, $U_t^3$ and ADE method with $V_{\rm thd} = 0.5$, and she tries to get a bit word from the obtained data $U_t^i$. As before, however, she does not know how to eliminate noises. Then, she obtains the bit word as in Table 4. Bob succeeds in making the correct restoration and recovering the original bit word, while Eve fails to eavesdrop on the bit words between Alice and Bob.

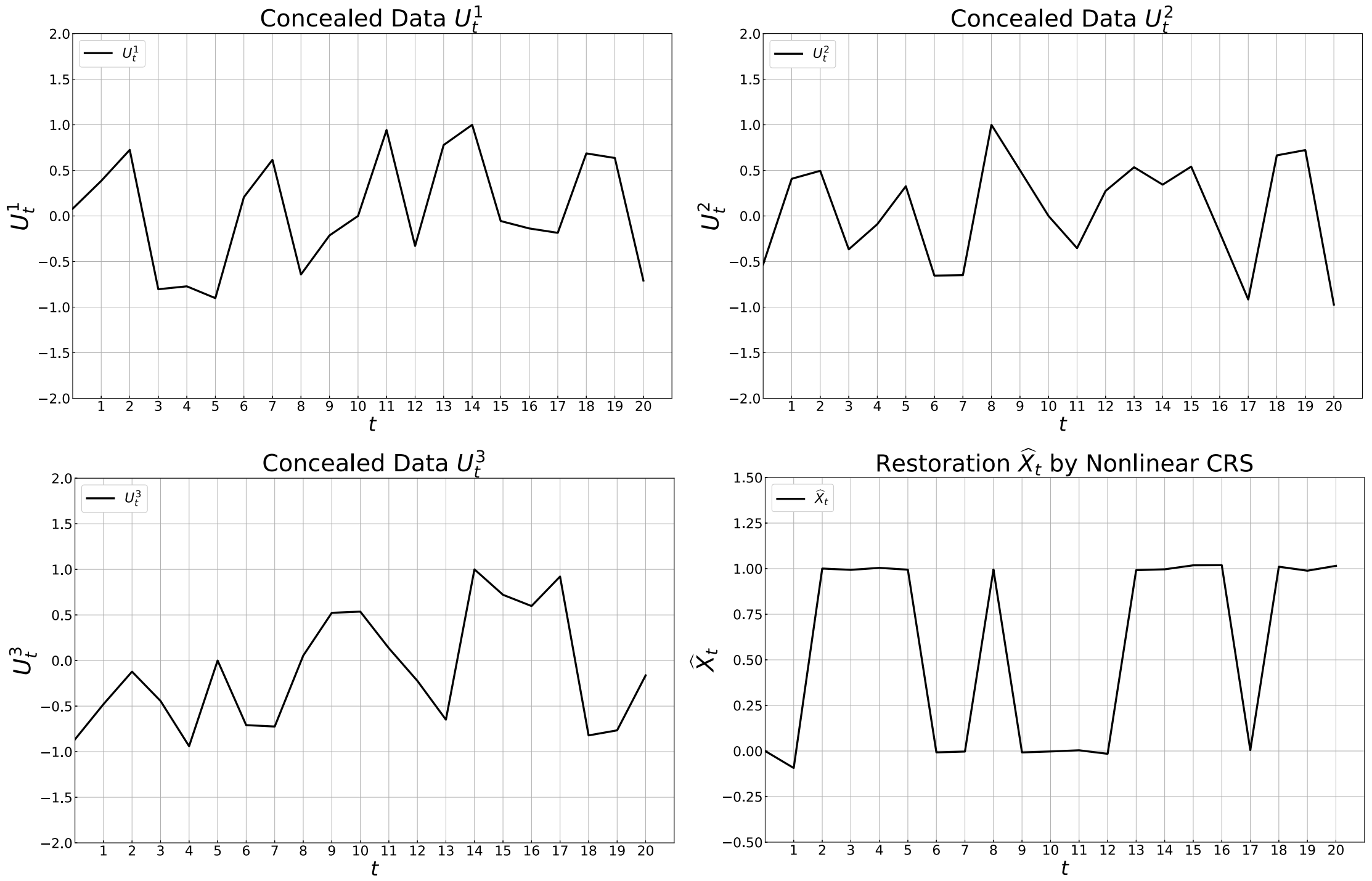

Figure 7: Concealed Data in Nonlinear CRS. $U_t^1, U_t^2, U_t^3$ are for the bit word given in Figure 5 by nonlinear CS with $g_{\mathrm{c}}$ defined by Eq.(12), where we denote $U_t^{\mathrm{nl},i}$ just by $U_t^i$ according to the abbreviation in Nonlinearization Strategy in Section 4. Following Reminder on Initial Bit & Value of CRS in §3.2, the value of $\widehat{X}_0$ is always 0 as the initial value of the restoration $\widehat{X}_t$, which implies that the value of $\widehat{a}_0$ is different from that of the ancilla bit $a_0$ in this case. Thus, we discard $\widehat{a}_0$ from the output of our RS. DA continuation is used to obtain each graph of $U_t^i$. We set parameters as $A = 0.1$, $b = 1$, $b_u = 1$, $\sigma_1 = 0.01$, and $\sigma_2 = \sigma_v = 1$.

| $t$ | 0 | 1 | 2 | 3 | 4 | 5 | 6 | 7 | 8 | 9 |
|---|---|---|---|---|---|---|---|---|---|---|
| $X_t$ | 1 | 0 | 1 | 1 | 1 | 1 | 0 | 0 | 1 | 0 |
| $U_t^1$ | 0.08 | 0.385 | 0.725 | -0.804 | -0.772 | -0.901 | 0.206 | 0.615 | -0.642 | -0.214 |
| ADE | 0 | 0 | 1 | 0 | 0 | 0 | 0 | 1 | 0 | 0 |
| $U_t^2$ | -0.531 | 0.408 | 0.495 | -0.366 | -0.091 | 0.326 | -0.654 | -0.649 | 1 | 0.501 |
| ADE | 0 | 0 | 0 | 0 | 0 | 0 | 0 | 0 | 1 | 1 |
| $U_t^3$ | -0.865 | -0.479 | -0.122 | -0.444 | -0.94 | 0 | -0.709 | -0.724 | 0.054 | 0.523 |
| ADE | 0 | 0 | 0 | 0 | 0 | 0 | 0 | 0 | 0 | 1 |
| $\widehat{X}_t$ | 0 | -0.093 | 1.001 | 0.993 | 1.004 | 0.994 | -0.007 | -0.003 | 0.995 | -0.008 |
| ADE | 0 | 0 | 1 | 1 | 1 | 1 | 0 | 0 | 1 | 0 |

| $t$ | 10 | 11 | 12 | 13 | 14 | 15 | 16 | 17 | 18 | 19 | 20 |
|---|---|---|---|---|---|---|---|---|---|---|---|
| $X_t$ | 0 | 0 | 0 | 1 | 1 | 1 | 1 | 0 | 1 | 1 | 1 |
| $U_t^1$ | 0 | 0.942 | -0.329 | 0.778 | 1 | -0.056 | -0.137 | -0.185 | 0.686 | 0.636 | -0.708 |
| ADE | 0 | 1 | 0 | 1 | 1 | 0 | 0 | 0 | 1 | 1 | 0 |
| $U_t^2$ | 0 | -0.353 | 0.274 | 0.534 | 0.344 | 0.541 | -0.181 | -0.916 | 0.665 | 0.723 | -0.972 |
| ADE | 0 | 0 | 0 | 1 | 0 | 1 | 0 | 0 | 1 | 1 | 0 |
| $U_t^3$ | 0.536 | 0.135 | -0.223 | -0.649 | 1 | 0.721 | 0.598 | 0.921 | -0.821 | -0.766 | -0.163 |
| ADE | 1 | 0 | 0 | 0 | 1 | 1 | 1 | 1 | 0 | 0 | 0 |
| $\widehat{X}_t$ | -0.002 | 0.004 | -0.015 | 0.992 | 0.996 | 1.018 | 1.019 | 0.005 | 1.011 | 0.989 | 1.015 |
| ADE | 0 | 0 | 0 | 1 | 1 | 1 | 1 | 0 | 1 | 1 | 1 |

Table 4: Values at $t = 0, 1, \cdots, 20$ for Figure 7. Eve fails to restore the original bit words by ADE from $U_t^i$, while Bob succeeds in doing it by ADE from $\widehat{X}_t$. Following Reminder on Initial Bit & Value of CRS in §3.2, we use $a_0$ as an ancilla bit, and discard $\widehat{a}_0$ from the output of RS.

Judging from Tables 3 and 4, Bob succeeds in restoring the bit word sent by Alice. We can actually make perfect restoration in our numerical experiments. To show this, we will make some statistical error estimates of CRS in §5.2. The statistical estimation reveals a symptoms of the perfectness of CRS.

## 5. Parameter Optimization & Accuracy Estimates of CRS

Since Bob employs Nonlinearization Strategy in Section 4 to introduce the nonlinearity into CRS, he has only to use linear RS to obtain the restoration. In §5.1 and §5.2, therefore, we show some parameter optimization and accuracy estimates for linear CRS only. In our numerical estimates in this section, we take common parameters, $A$, $b$, $b_u$, and $\sigma_j$ over all $i = 1, \cdots, N$, and set our parameters as $A^i = A$, $b^i = b$, $b_u^i = b_u$, and $\sigma_j^i = \sigma_j$.

### *5.1. Parameter Optimization for CRS*

Let the original bit word be given by $a_1 \cdots a_K$. We here remember that the ancilla bit is $a_0$ in CS, and that $\widehat{a}_0$ and $\widehat{X}_0$ are both 0 in RS following Reminder on Initial Bit & Value of CRS in §3.2. Thus, the bit word which we should restore by CRS is $\widehat{a}_1 \cdots \widehat{a}_K$ for the original bit word $a_1 \cdots a_K$.

We introduce two kinds of norms to estimate and optimize the parameters, $A$, $b$, and $b_u$. We define the restored bit word $\widehat{\boldsymbol{a}}$ by $\widehat{\boldsymbol{a}} = \widehat{a}_1 \cdots \widehat{a}_K$. We note that $\widehat{\boldsymbol{a}}$ does not include the ancilla restored bit $\widehat{a}_0$.

**Maximum Norm**: We define a norm $\| \ \|_{\max}$ for the restored bit word $\widehat{\boldsymbol{a}} = \widehat{a}_1 \cdots \widehat{a}_K$ by

$$\|\widehat{\boldsymbol{a}}\|_{\max} = \max_{k=1,2,\cdots,K} |a_k - \widehat{a}_k|. \tag{21}$$

We call $\| \ \|_{\max}$ the *maximum norm* for the restored bit word $\widehat{\boldsymbol{a}}$. Since each of $a_k$ and $\widehat{a}_k$, $k = 1, \cdots, K$, is a bit (i.e., $a_k, \widehat{a}_k \in \{0, 1\}$), we have

$$|a_k - \widehat{a}_k| = \begin{cases} 1 & \text{if } a_k \neq \widehat{a}_k, \\ 0 & \text{if } a_k = \widehat{a}_k. \end{cases} \tag{22}$$

Thus, we know that $\|\widehat{\boldsymbol{a}}\|_{\max} = 1$ if and only if there is at least one $k \in \{1, \cdots, K\}$ satisfying $a_k \neq \widehat{a}_k$. In other words, if $\|\widehat{\boldsymbol{a}}\|_{\max} = 0$, then $a_k = \widehat{a}_k$ for all $k \in \{1, \cdots, K\}$. Therefore, we can judge the bit flip as

$$\|\widehat{\boldsymbol{a}}\|_{\max} = \begin{cases} 1 & \text{if there is a } k \text{ satisfying } a_k \neq \widehat{a}_k, \\ 0 & \text{if } a_k = \widehat{a}_k \text{ for all } k. \end{cases} \tag{23}$$

**Mean Norm**: We define a norm $\| \ \|_{\text{mean}}$ for the restored bit word $\widehat{\boldsymbol{a}} = \widehat{a}_1 \cdots \widehat{a}_K$ by

$$\|\widehat{\boldsymbol{a}}\|_{\text{mean}} = \frac{1}{K} \sum_{k=1}^{K} |a_k - \widehat{a}_k|. \tag{24}$$

We call $\| \ \|_{\text{mean}}$ the *mean norm* for the restored bit word $\widehat{\boldsymbol{a}}$. Because of Eq.(22), we realize that the sum, $\sum_{k=1}^{K} |a_k - \widehat{a}_k|$, means the total number of $k \in \{1, \cdots, K\}$ satisfying $a_k \neq \widehat{a}_k$. Therefore, we can realize how often bit flips appear in the restored bit word $\widehat{\boldsymbol{a}}$ by

$\|\widehat{\boldsymbol{a}}\|_{\text{mean}}$ representing the statistical probability of the occurrence of the bit flip in the restored bit word. (25)

**Optimization of Parameter** $A$: Regarding $X_t^i$ in Eq.(15) as the velocity of a particle in the air, in terms of the Langevin equation, the air resistance works for the object with the velocity $X_t^i$ in the air as $\dot{X}_t^i = -(1 - A)X_t^i$.

Here, we set $A^i = A$ for all $i$. Thinking this equation to be the leading term of Eq.(15), we may fix the residence coefficient $1 - A > 0$. Before estimating parameters, $b^i = b$ and $b^i_u = b_u$, we show the behavior of the parameter $A$ around 1. Here, we set parameters as $b = b_u = 1$, $\sigma_1 = 0.01$, $\sigma_2 = \sigma_v = 1$.

Referring to our experimental results as in Figures 8, it seems to be better that $A$ is set less than 1 and far from 1. Thus, we set the parameter $A$ as $A = 0.1$ throughout this paper from now on.

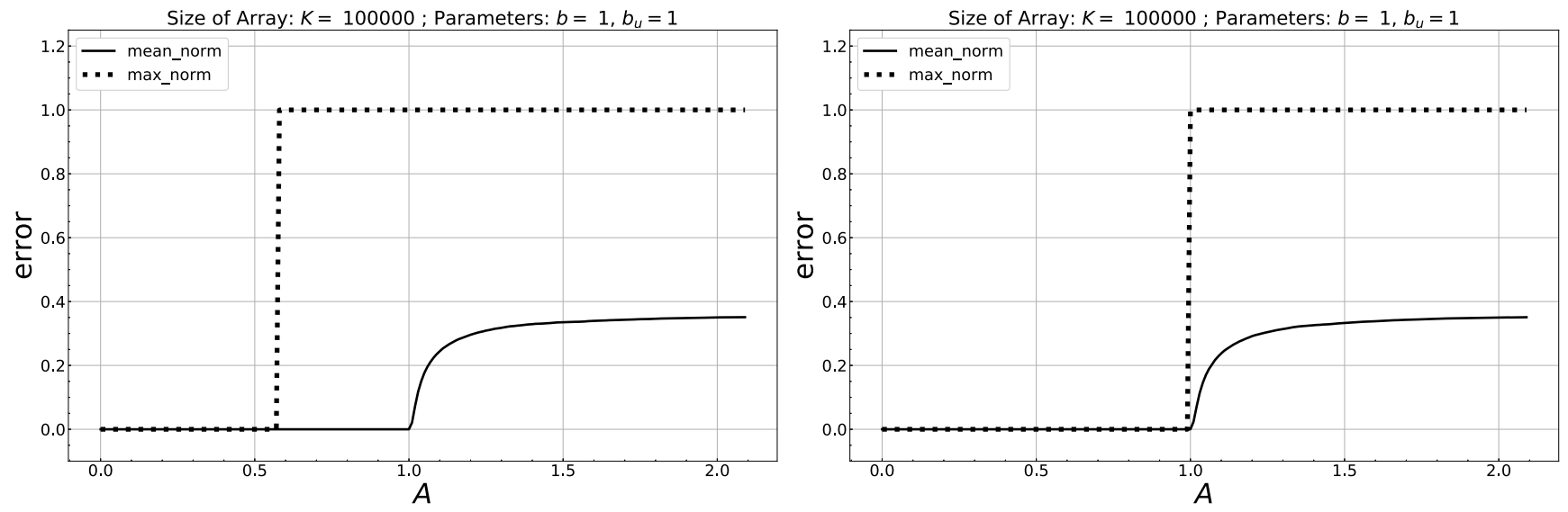

Figure 8: Optimization of Parameter $A$ in Linear CRS. Other parameters are set as $b = b_u = 1$, $\sigma_1 = 0.01$, and $\sigma_2 = \sigma_v = 1$. The size of the bit words is $K = 100,000$. The left graph shows the best case in our several numerical experiments, and the right one the worst case. The continuation in Python is used to obtain the graphs. The dotted line is the maximum norm, and the bold line the mean norm. According to Eqs.(23) and (25), the bit flip does not take place almost surely where both the lines (i.e., norms) indicate 0. There is a possibility of bit flip where the maximum norm is 1, and the probability of the possibility is shown by the corresponding mean norm.

**Optimization of Parameter** $b$: We fix parameters as $A = 0.1$, $\sigma_1 = 0.01$, $\sigma_2 = \sigma_v = 1$. Following our numerical estimates, as in Figures 9 and 10, it is not too much to say that no bit flip takes place in the restored bit word for $0 < b < 100$ if $b_u$ is less than 10, while it rarely happens for $0 < b < 100$ if $b_u$ is more than 11.

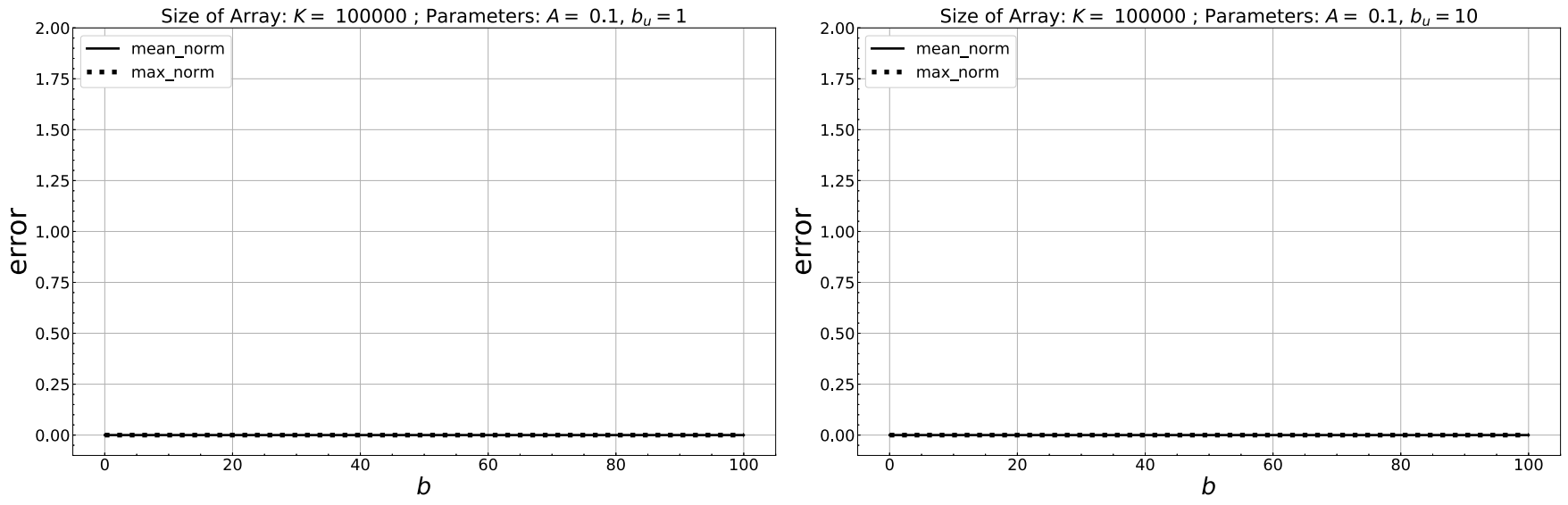

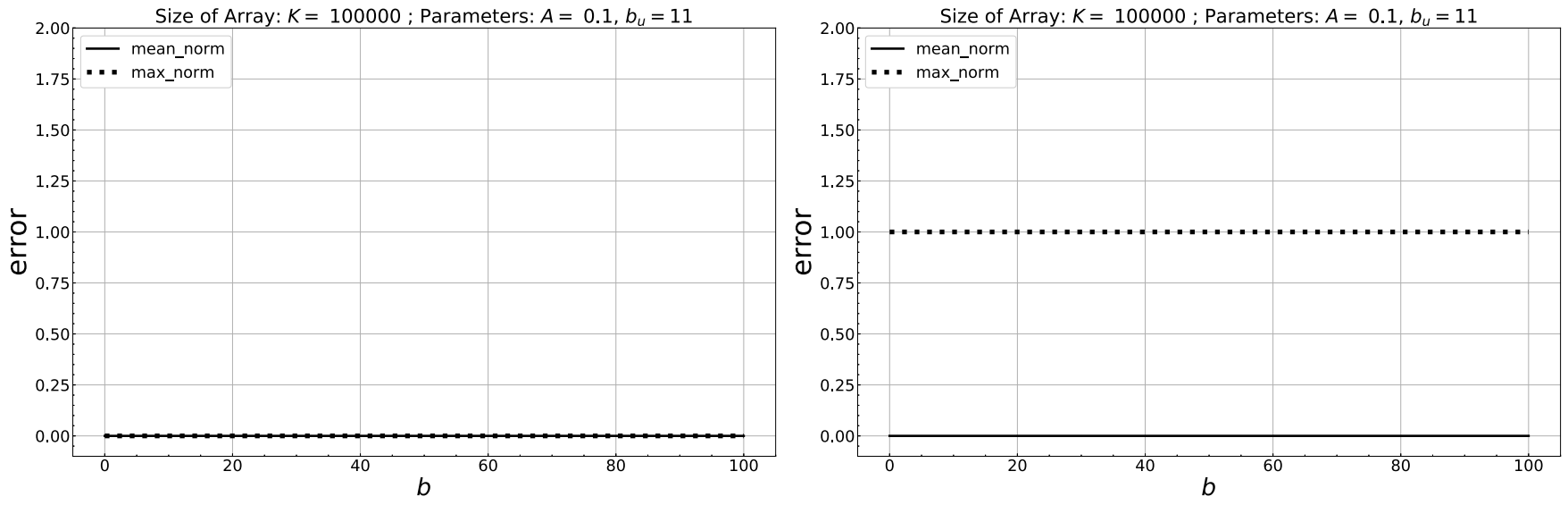

Figure 9: Optimization of Parameter $b$ in Linear CRS. Other parameters are set as $A = 0.1$, $b_u = 1, 10, 11$, $\sigma_1 = 0.01$, and $\sigma_2 = \sigma_v = 1$. The size of the bit words is $K = 100,000$. The continuation in Python is used to obtain the graphs. The dotted line is the maximum norm, and the bold line the mean norm. According to Eqs.(23) and (25), the bit flip does not take place almost surely where both the lines (i.e., norms) indicate 0. There is a possibility of bit flip where the maximum norm is 1, and the probability of the possibility is shown by the corresponding mean norm.

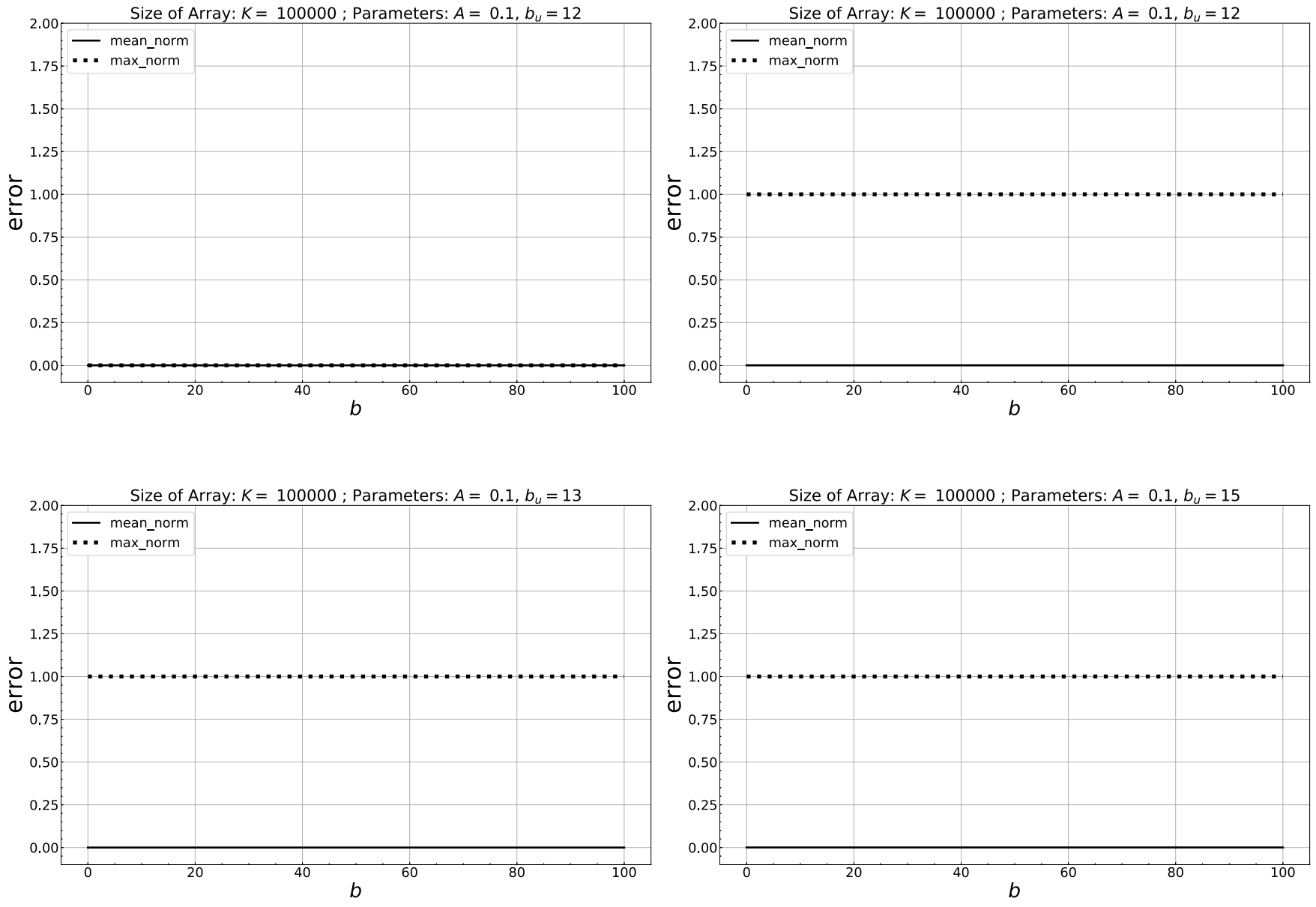

Figure 10: Optimization of Parameter $b$ in Linear CRS. Other parameters are set as $A = 0.1$, $b_u = 12, 13, 15$, $\sigma_1 = 0.01$, and $\sigma_2 = \sigma_v = 1$. The size of the bit words is $K = 100,000$. The continuation in Python is used to obtain the graphs. The dotted line is the maximum norm, and the bold line the mean norm. According to Eqs.(23) and (25), the bit flip does not take place almost surely where both the lines (i.e., norms) indicate 0. There is a possibility of bit flip where the maximum norm is 1, and the probability of the possibility is shown by the corresponding mean norm.

**Optimization of Parameter $b_u$:** We fix parameters as $A = 0.1$, $\sigma_1 = 0.01$, $\sigma_2 = \sigma_v = 1$. As in Figure 11, our numerical estimates say that, if $0 < b < 150$, then no bit flip takes place in the restored bit word for $0 < b_u < 12$, and it is beginning to happen for $12 < b_u$.

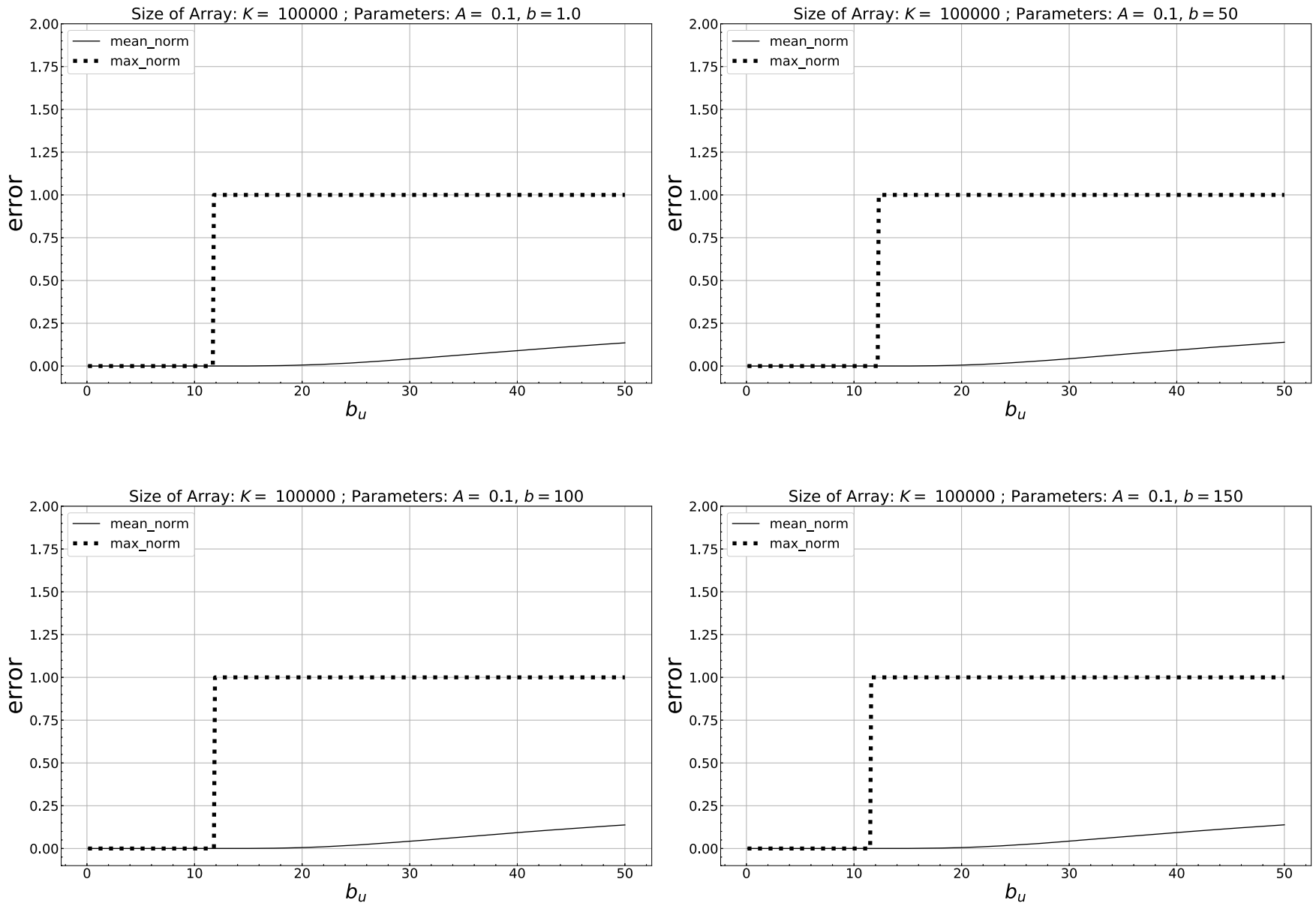

Figure 11: Optimization of Parameter $b$ in Linear CRS. Other parameters are set as $A = 0.1$, $b = 1, 50, 100, 150$, $\sigma_1 = 0.01$, and $\sigma_2 = \sigma_v = 1$. The size of the bit words is $K = 100,000$. The continuation in Python is used to obtain the graphs. The dotted line is the maximum norm, and the bold line the mean norm. According to Eqs.(23) and (25), the bit flip does not take place almost surely where both the lines (i.e., norms) indicate 0. There is a possibility of bit flip where the maximum norm is 1, and the probability of the possibility is shown by the corresponding mean norm.

### *5.2. Statistical Error Estimates for CRS*

We execute many trials of CRS, and make the statistical estimates for CRS. Referring to the results in §5.1, we set our parameters as $A = 0.1$, $b = b_u = 1$, $\sigma_1 = 0.01$, and $\sigma_2 = \sigma_v = 1$ from now on. Each trial is done by a numerical analysis. In the numerical analyses, all the bit words which we conceal are randomly generated.

We introduce two kinds of norm to make the statistical error estimates. For our original bit word, $a_1 \cdots a_K$, we make $T$ trials of the CRS. In $T$ trials, we have $T$ samples of the restored bit word $\widehat{a}_1(\tau)\widehat{a}_2(\tau)\cdots\widehat{a}_K(\tau)$, $\tau = 1, 2, \cdots, T$. Here, $\tau$ is the parameter representing the $\tau$-th trial.

**Maximum Norm**: We define the *maximum norm* $\| \ \|_{\max}$ for the restored bit $\widehat{a}_k$ by

$$\|\widehat{a}_k\|_{\max} = \max_{\tau=1,2,\cdots,T} |a_k - \widehat{a}_k(\tau)| \tag{26}$$

for each $k = 1, \cdots, K$. In the above maximum norm, it makes a function of $k$, i.e., $k \mapsto \|\widehat{a}_k\|_{\max}$, and thus, it is different from the maximum norm defined by Eq.(21). We can realize the bit flip by

$$\|\widehat{a}_k\|_{\max} = \begin{cases} 1 & \text{if there is a trial indicated by } \tau \text{ satisfying } a_k \neq \widehat{a}_k(\tau), \\ 0 & \text{if } a_k = \widehat{a}_k(\tau) \text{ for all } \tau. \end{cases} \tag{27}$$

Thus, we can statistically know whether a bit flip takes place at each $k$.

**Mean Norm**: We define the *mean norm* $\| \ \|_{\mathrm{mean}}$ for the restored bit $\widehat{a}_k$ by

$$\|\widehat{a}_k\|_{\mathrm{mean}} = \frac{1}{T}\sum_{\tau=1}^{T} |a_k - \widehat{a}_k(\tau)| \tag{28}$$

for each $k = 1, \cdots, K$. The above mean norm is different from the mean norm defined by Eq.(24) since it is also a function of $k$, i.e., $k \mapsto \|\widehat{a}_k\|_{\mathrm{mean}}$. We can grasp how often bit flips appear in the $T$ trials at each $k$ by

$\|\widehat{a}_k\|_{\mathrm{mean}}$ representing the statistical probability of the occurrence of the bit flip in the trials at each $k$. (29)

Our all numerical analyses show the same result as in Figure 12. This says that CRS can completely restore the original bit words with the statistical probability 1 under proper parameter conditions.

Since linear RS is based on the linear Kalman filtering theory, it is so sensitive to $\sigma_1$. Our numerical analyses reveal that some bit flips are beginning to take place as $\sigma_1$ exceeds around 0.1, and say that the bit flip seldom takes place at $\sigma_1 = 0.1$. We show an example of the numerical analyses in Figure 13.

We again suppose that Eve succeeds in wiretapping and obtaining the form of SES of the linear CRS but the exact value of the parameter $b$ or $b_u$. We also suppose that Eve uses wrong $b$ to restore the bit word. Our numerical analyses reveal that the statistical probability of the occurrence of the bit flip increases as the gap between the true $b$ and the wrong $b$ gets larger. We show the inclination in Figure 14. Instead of $b$, we suppose that Eve uses wrong $b_u$ to restore the bit word. Our numerical analyses show the results similar to the case for $b$, and reveal the statistical probability of the occurrence of the bit flip increases as the gap between the true $b_u$ and the wrong $b_u$ gets larger. We show the inclination in Figure 15.

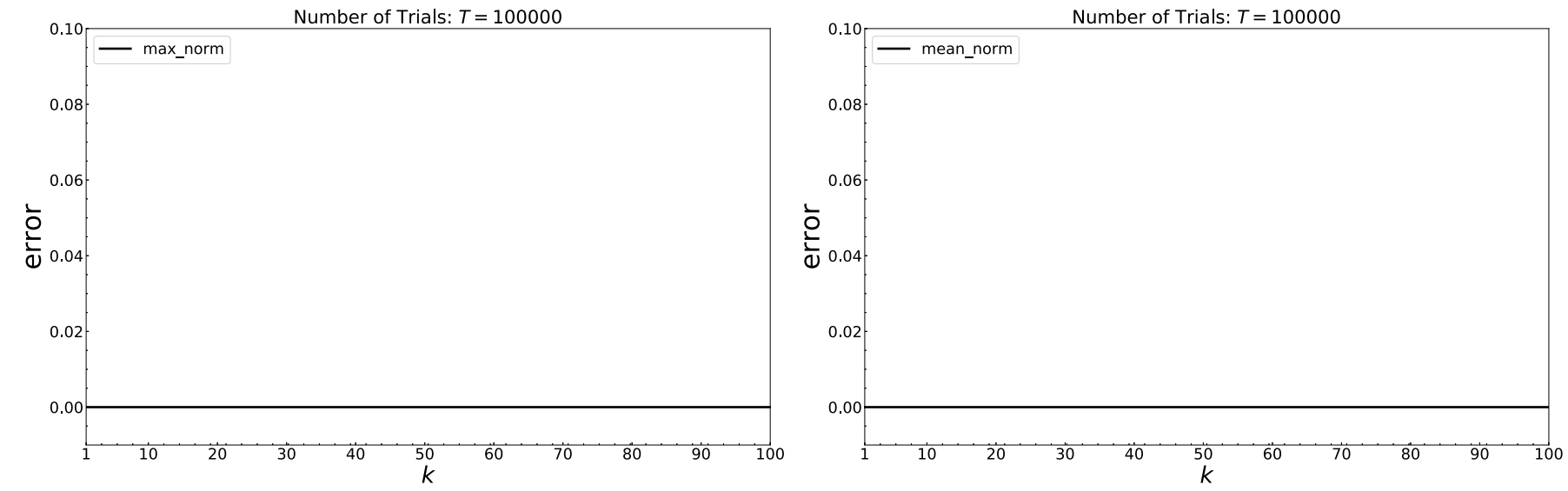

Figure 12: Error Estimates of Restoration. The restoration is obtained from concealed bit words in linear CRS. The number of trials is $100,000$, and we set parameters as $A = 0.1$, $b = b_u = 1$, $\sigma_1 = 0.01$, and $\sigma_2 = \sigma_v = 1$. DA continuation is used to obtain each graph. The two graphs show the maximum and mean norms for linear CRS from the left. According to Eqs.(27) and (29), theses graphs mean there is no possibility of bit flip almost surely.

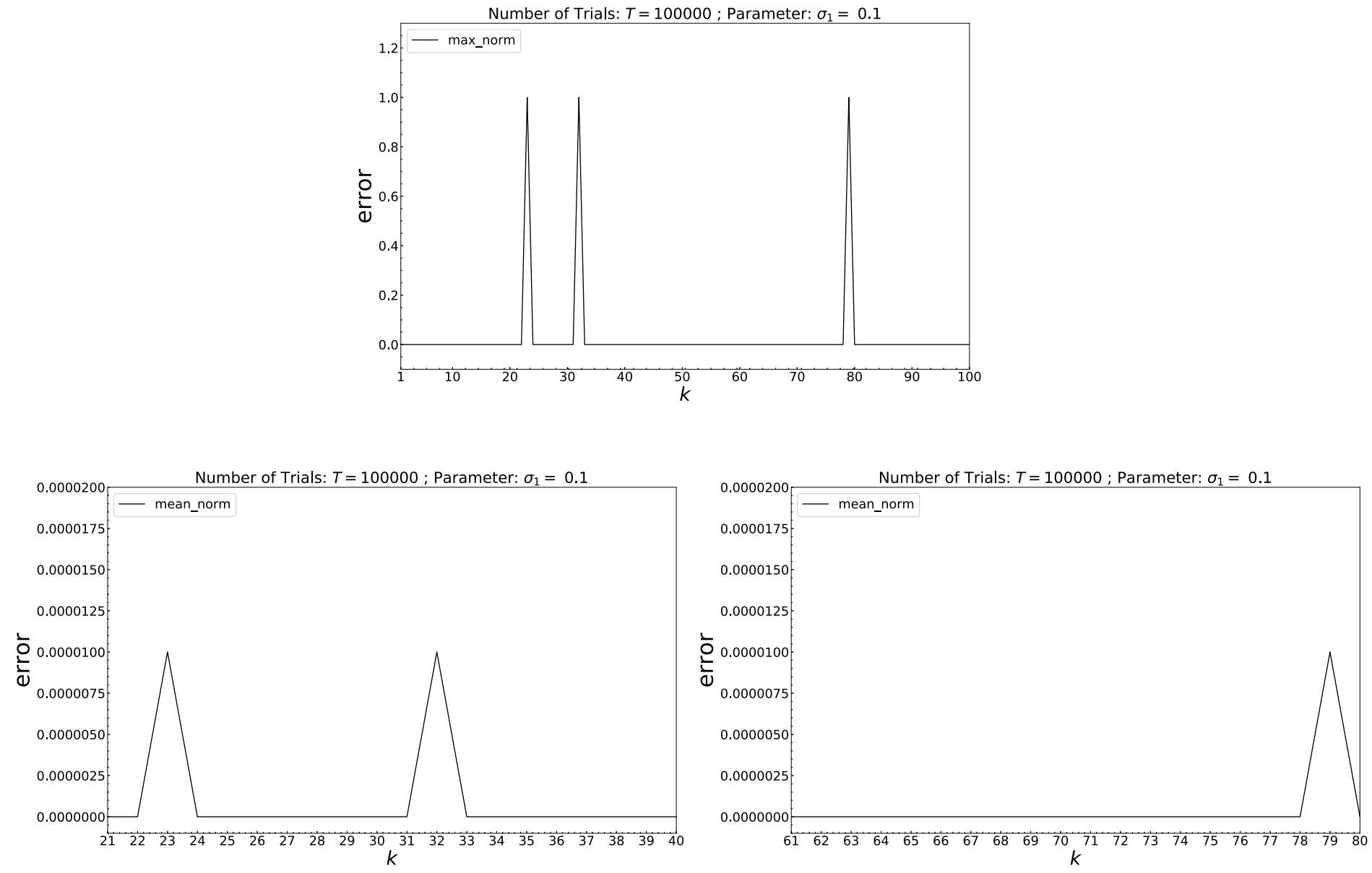

Figure 13: Error Estimates of Restoration in Linear CRS. The restoration is obtained from concealed bit words by linear RS. The number of trials is $100,000$. We set parameters as $A = 0.1$, $b = b_u = 1$, $\sigma_1 = 0.1$, and $\sigma_2 = \sigma_v = 1$. DA continuation is used to obtain each graph. The graph of the maximum norm on the top row tells at which $k$ the bit flip takes place according to Eq.(27). The graphs of the mean norm on the bottom row describe how often the each bit flip takes place according to Eq.(29). The numerical analysis in these graphs says that there is the possibility of bit flip at $k = 23, 32, 79$ with very small statistical probability.

At last of the statistical error estimates, we see the case where Eve succeeds in wiretapping and obtaining the form of SES of the linear CRS but the secret common key $v_k$, $k = 1, \cdots, K$. We suppose that Alice generates $v_k$ by using a white noise with $N(0, (\sigma_v)^2)$. Since Eve does not know the exact $v_k$ between Alice and Bob, she is bound to find the key by guesswork using random noises. We assume that Eve's random noises are white noises with $N(0, (\sigma_{\mathrm{eve}})^2)$. Our numerical analyses in the cases of $\sigma_v = 1$ and $\sigma_{\mathrm{eve}}$ from 0.1 to 10 say that the statistical probability of the occurrence of the bit flip at each $k$ is in the interval between 30% and 45% as in Figures 16 for $\sigma_{\mathrm{eve}} = 1, 2, 3$, and Figure 17 for $\sigma_{\mathrm{eve}} = 0.1, 10$.

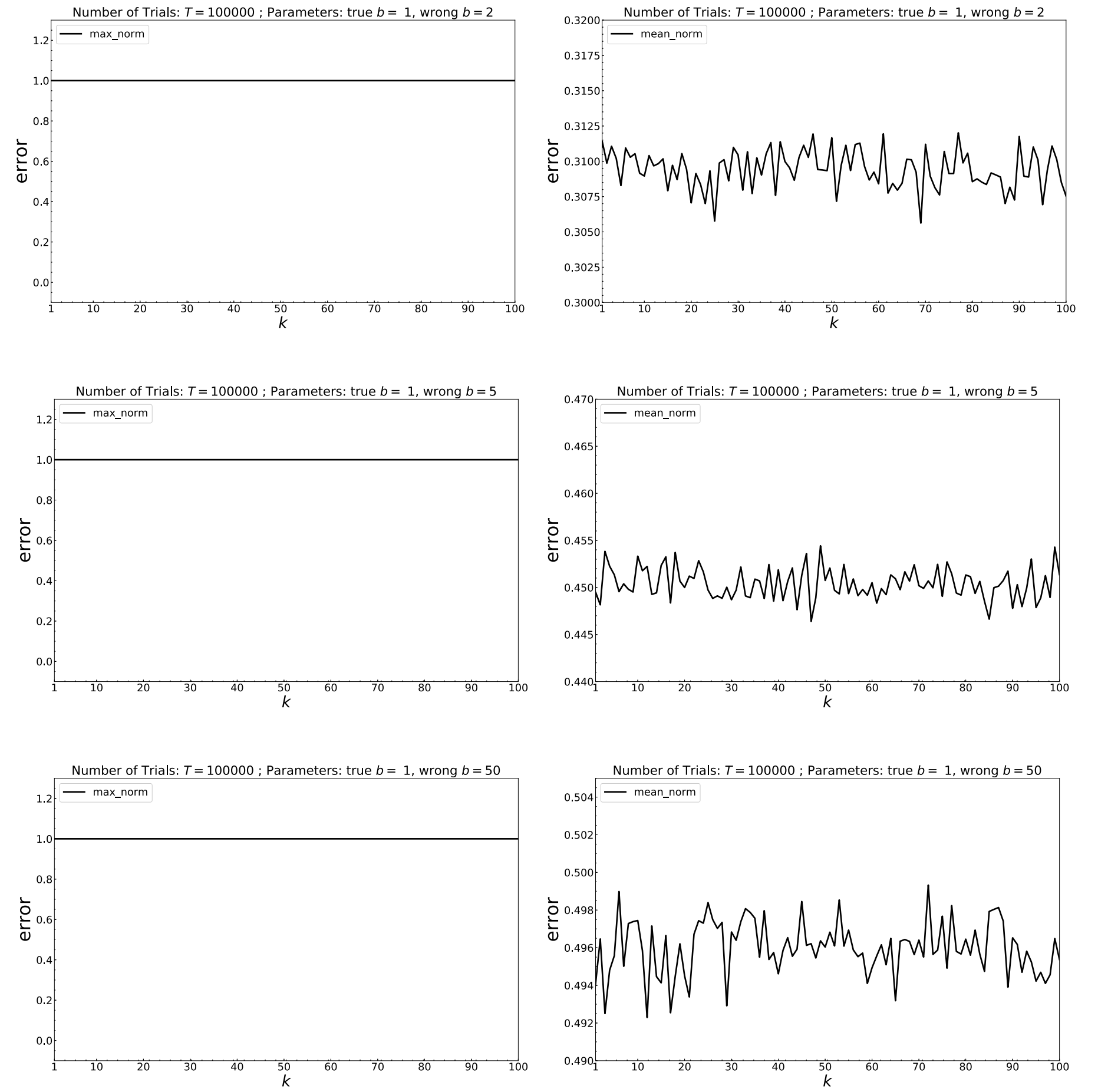

Figure 14: Error Estimates of Restoration in Linear CRS. We set parameters as $A = 0.1$, $b_u = 1$, $\sigma_1 = 0.01$, and $\sigma_2 = \sigma_v = 1$. The number of trials is $100,000$. The restoration is obtained from concealed data by linear RS with wrong $b$. DA continuation is used to obtain the graphs. On each row, the left graph is for the maximum norm, and the right graph for the mean norm. The graph of the maximum norm indicates 1 at all points $k$, which means there is a possibility of bit flip according to Eq.(27). The statistical probability of the possibility is shown in the individually corresponding graph of the mean norm according to Eq.(29).

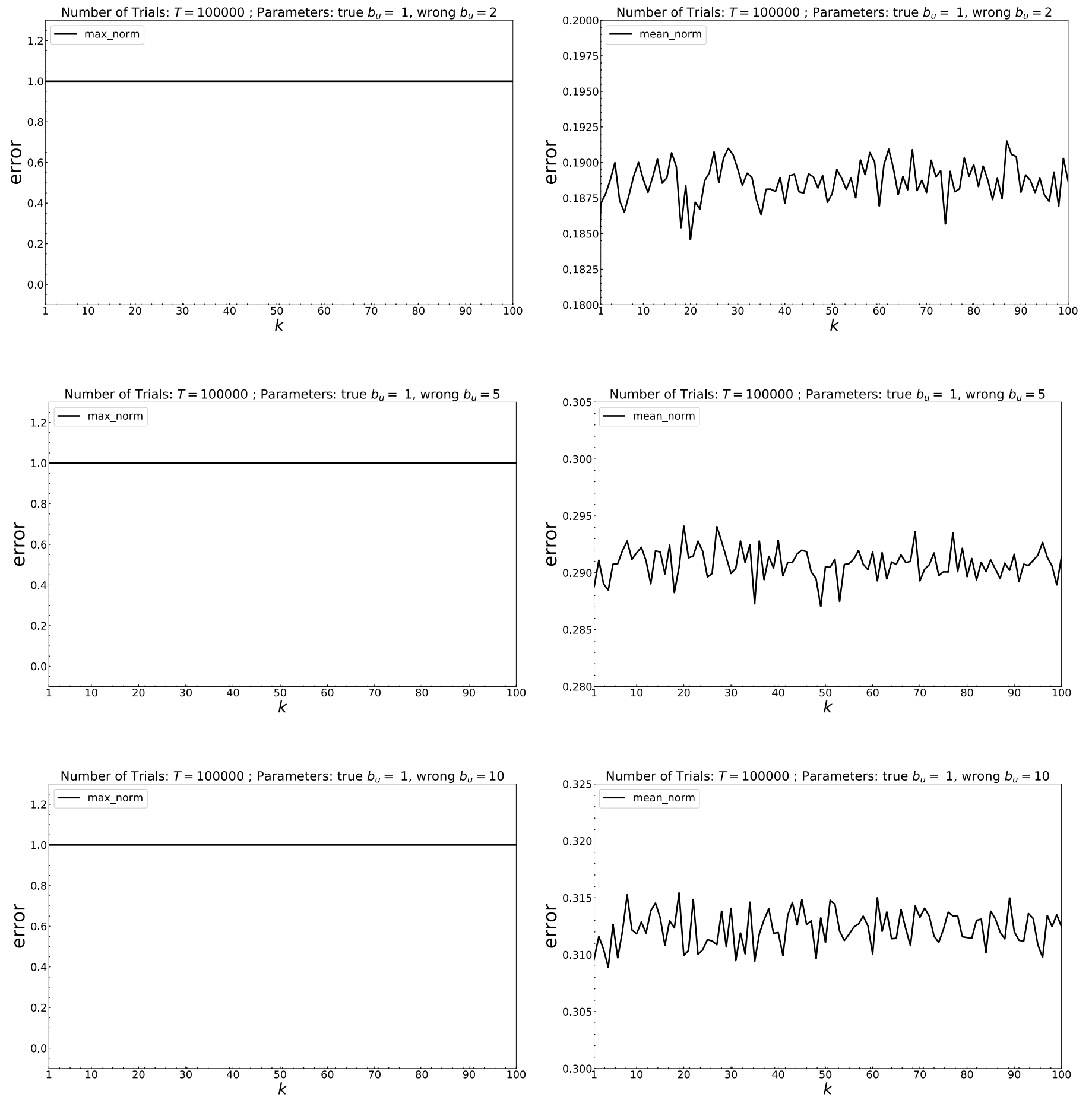

Figure 15: Error Estimates of Restoration in Linear CRS. We set parameters as $A = 0.1$, $b = 1$, $\sigma_1 = 0.01$, and $\sigma_2 = \sigma_v = 1$. The number of trials is $100,000$. The restoration is obtained from concealed data by the linear RS with wrong $b_u$. The number of trials is $100,000$. DA continuation is used to obtain the graphs. On each row, the left graph is for the maximum norm, and the right graph for the mean norm. The graph of the maximum norm indicates 1 at all points $k$, which means there is a possibility of bit flip according to Eq.(27). The statistical probability of the possibility is shown in the individually corresponding graph of the mean norm according to Eq.(29).

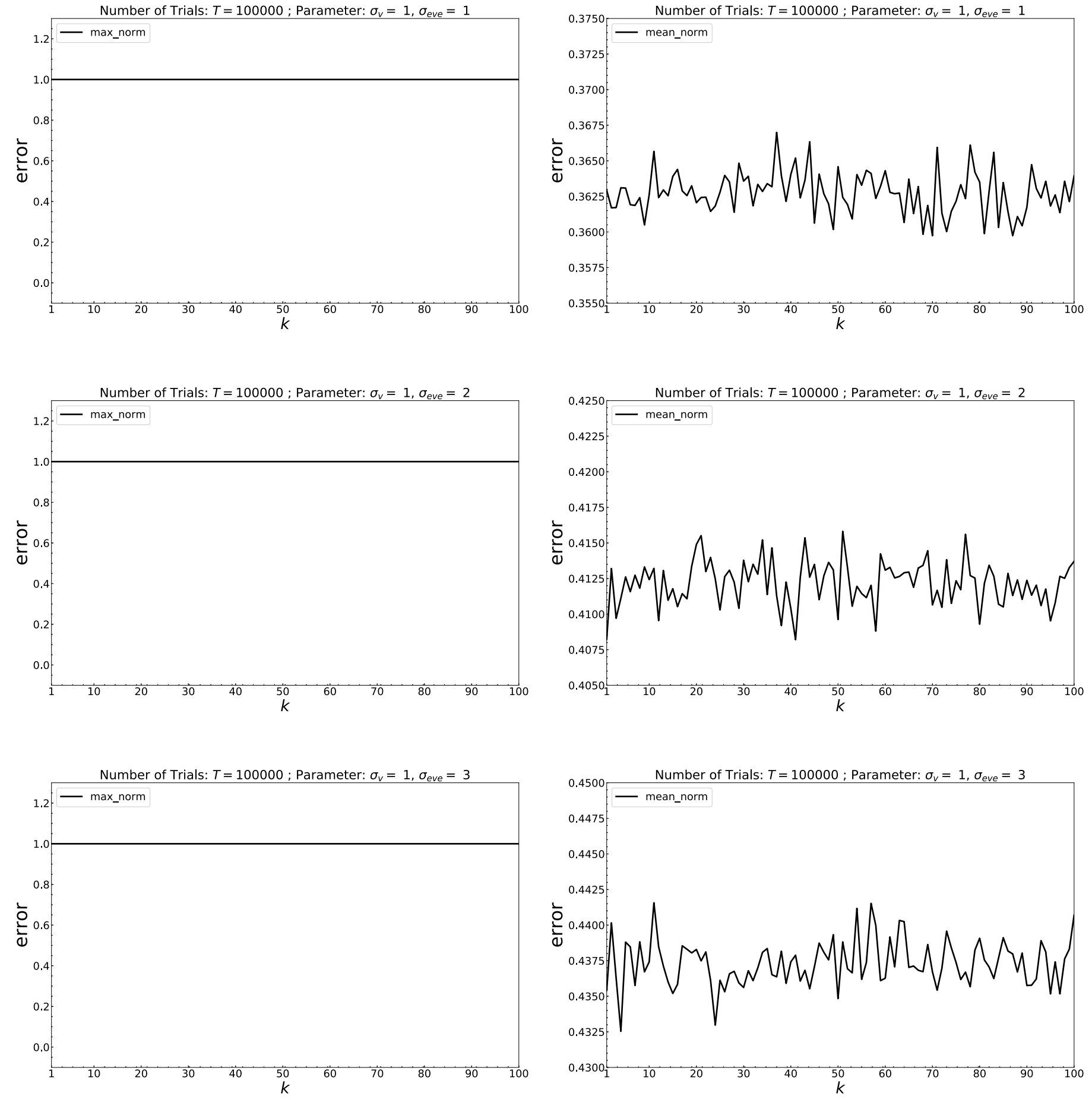

Figure 16: Error Estimates of Restoration in Linear CRS. We set parameters as $A = 0.1$, $b = b_u = 1$, $\sigma_1 = 0.01$, $\sigma_2 = \sigma_v = 1$, and $\sigma_{\rm eve} = 1, 2, 3$. The restoration is obtained from concealed data by linear RS with wrong $v_k$. The number of trials is $100,000$. DA continuation is used to obtain the graphs. On each row, the left graph is for the maximum norm, and the right graph for the mean norm. The graph of the maximum norm indicates 1 at all points $k$, which means there is a possibility of bit flip according to Eq.(27). The statistical probability of the possibility is shown in the individually corresponding graph of the mean norm according to Eq.(29).

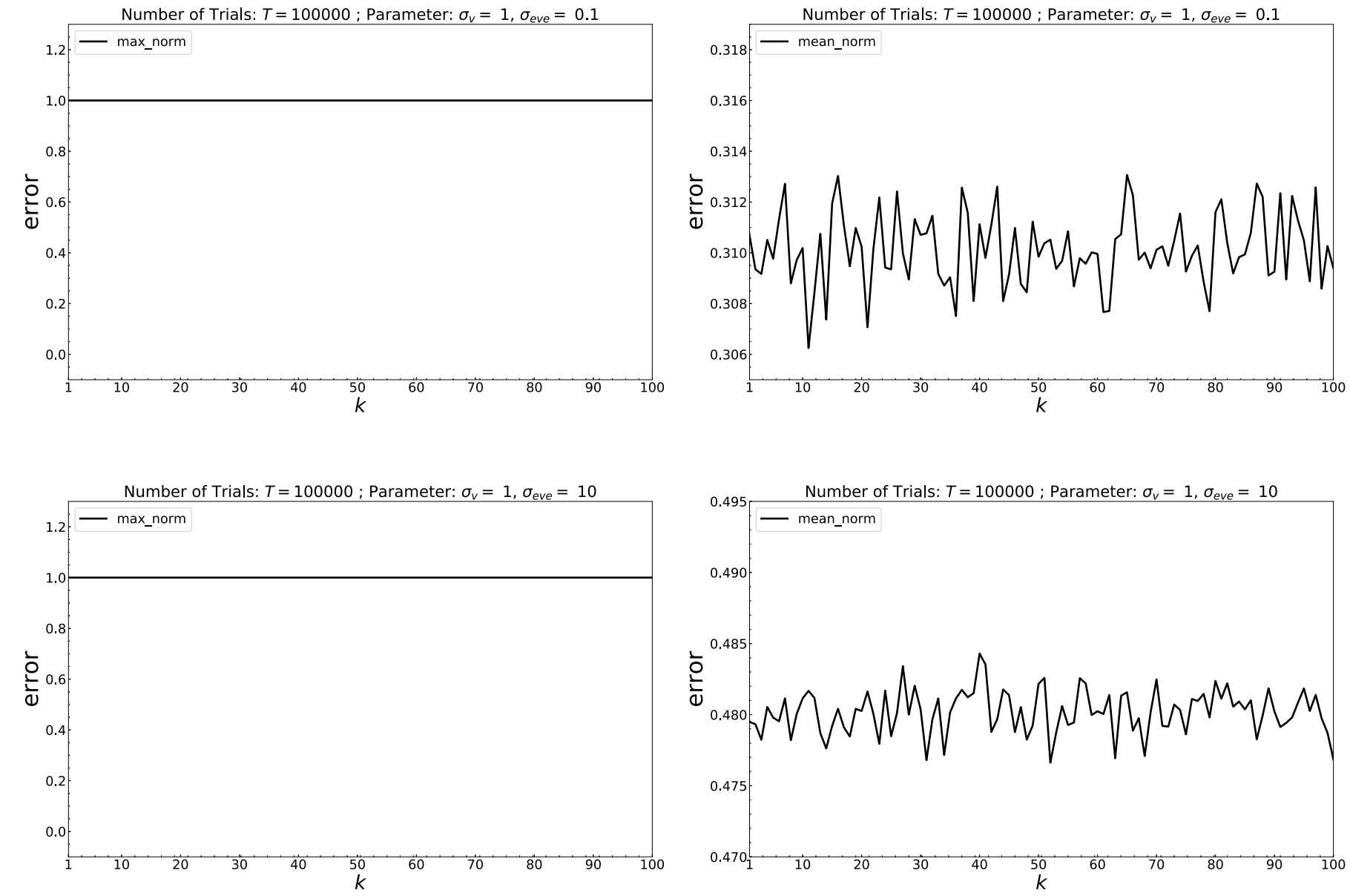

Figure 17: Error Estimates of Restoration in Linear RS. We set parameters as $A = 0.1$, $b = b_u = 1$, $\sigma_1 = 0.01$, $\sigma_2 = \sigma_v = 1$, and $\sigma_{\text{eve}} = 0.1, 10$. The number of trials is $100,000$. The restoration is obtained from concealed data by linear RS with wrong $v_k$. DA continuation is used to obtain the graphs. On each row, the left graph is for the maximum norm, and the right graph for the mean norm. The graph of the maximum norm indicates 1 at all points $k$, which means there is a possibility of bit flip according to Eq.(27). The statistical probability of the possibility is shown in the individually corresponding graph of the mean norm according to Eq.(29).

### *5.3. Nonlinearity Estimation for CRS*

We investigate the case where Eve succeeds in wiretapping and obtaining the knowhow of the linear CRS but the nonlinearity. by using the numerical analyses. Here, all the bit words are randomly generated. Our numerical analyses with the nonlinear function, $g_{\mathrm{c}}$ defined by Eq.(12), reveal that the bit flip takes place with the statistical probability near $1/2$ at each $k$ if she restores the bit word from the data concealed by nonlinear CS without knowing the nonlinearity. The maximum and mean norms of an example of numerical analyses are in Figure 18. According to the example, the statistical probability of the occurrence of the bit flip is almost 0.48 at each $k$. Therefore, we can say that the statistical probability that the bit word with the length $K$ is completely restored without knowing the nonlinearity is almost $(0.52)^K$.

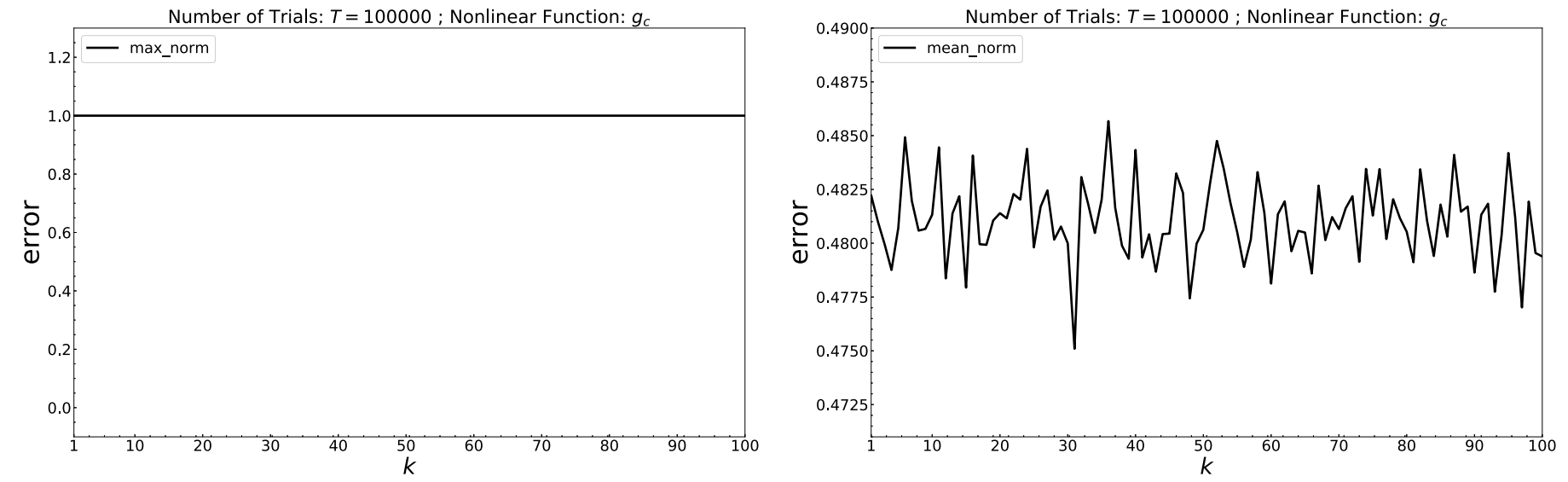

Figure 18: Nonlinearity in Restoration. The nonlinearity is introduced by $g_{\mathrm{c}}$ defined by Eq.(12). The error estimates are made with $A = 0.1$, $b = b_u = 1$, $\sigma_1 = 0.01$, and $\sigma_2 = \sigma_v = 1$ in nonlinear CS. The restoration is made by linear RS without knowing the nonlinearity. The number of trials is $100,000$. DA continuation is used to obtain the graphs. The left graph is for the maximum norm, and the right graph for the mean norm. The graph of the maximum norm indicates 1 at all points $k$, which means there is a possibility of bit flip according to Eq.(27). The statistical probability of the possibility is shown in the graph of the mean norm according to Eq.(29).

In order to investigate how the sort of the nonlinearity works in CRS, we also consider the nonlinear function $g_{\mathrm{s}}$ defined by Eq.(11). In addition to $g_{\mathrm{c}}$ and $g_{\mathrm{s}}$, we define another nonlinear function $g_{\mathrm{ss}}$ by

$$g_{\mathrm{ss}}(x) = \begin{cases} x + 0.5 & \text{if } 0 \leq x < 0.5, \\ x - 0.5 & \text{if } 0.5 \leq x \leq 1. \end{cases} \tag{30}$$

This nonlinear function is simpler than $g_{\mathrm{c}}$ and $g_{\mathrm{s}}$. We make numerical analyses for the maximum norm defined by Eq.(26) and the mean norm defined by Eq.(28) for $g_{\mathrm{s}}$ and $g_{\mathrm{ss}}$, respectively.

In the case of $g_{\mathrm{s}}$, the numerical analyses reveals that the statistical probability of the occurrence of the bit flip is almost 0.54 at each $k$ as in Figure 19. Therefore, we expect that the statistical probability that the bit word with the length $K$ is completely restored without knowing the nonlinearity is almost $(0.46)^K$.

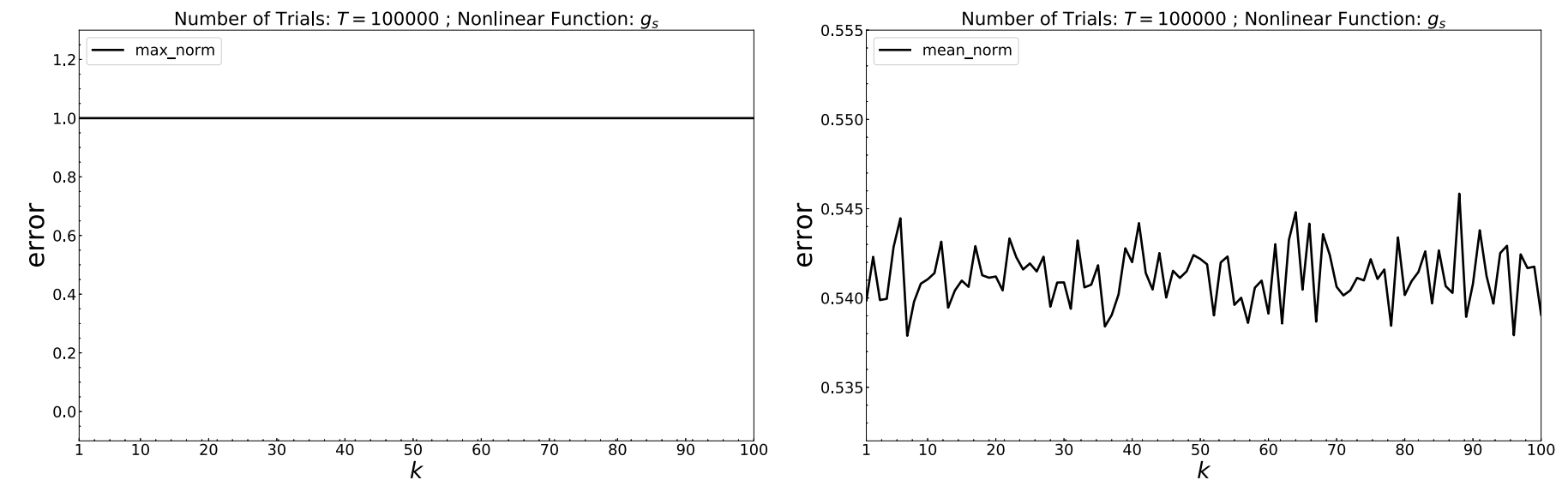

Figure 19: Nonlinearity in Restoration. The nonlinearity is introduced by $g_{\mathrm{s}}$ defined by Eq.(11). The error estimates are made with $A^i = 0.1$, $b^i = b_u^i = 1$, $\sigma_1 = 0.01$, and $\sigma_2 = \sigma_v = 1$ in nonlinear CS. The restoration is made by linear RS without knowing the nonlinearity. The number of trials is $100,000$. DA continuation is used to obtain the graphs. The left graph is for the maximum norm, and the right graph for the mean norm. The graph of the maximum norm indicates 1 at all points $k$, which means there is a possibility of bit flip according to Eq.(27). The statistical probability of the possibility is shown in the graph of the mean norm according to Eq.(29).

In the case of $g_{\mathrm{ss}}$, based on the numerical analyses, it is thought that each bit in the bit word is flipped with the statistical probability near 0.65 as in Figure 20. Therefore, we can see that the statistical probability that the bit word with the length $K$ is completely restored without knowing the nonlinearity is almost $(0.35)^K$.

In order to deceive Eve, the probability of the occurrence of the bit flip is desirably around 0.5. The nonlinear function $g_{\mathrm{ss}}$ makes the possibility gets higher than the nonlinear functions, $g_{\mathrm{c}}$ and $g_{\mathrm{s}}$, make. We, however, note that the high probability of the occurrence of the bit flip at each $k$ means the high probability of the availability of the NOT operation. Due to use of the NOT operation, therefore, the concealing effect of $g_{\mathrm{ss}}$ is essentially equivalent to that of the nonlinear function with the probability 0.35 at each $k$.

## *5.4. Noise Margin Estimation for CRS & Other External Noises*

We estimate the noise margin for ADE of CRS. We define the *maximum norm* for the restoration in the following to estimate LLR and LHR. In $T$ trials, we have $T$ samples of the restoration $\widehat{X}_t(\tau)$, $\tau = 1, 2, \cdots, T$, where $\tau$ is the parameter representing the $\tau$-th trial. Here, all the bit words in the individual numerical analyses are randomly generated. The maximum norm of $\widehat{X}_k$ is given by

$$\|\widehat{X}_k\|_{\max} = \max_{\tau=1,2,\cdots,T} |X_k - \widehat{X}_k(\tau)| \tag{31}$$

for each $k = 1, \cdots, K$. Due to this definition, we should take $L_{\text{NM}}$, the length of the noise margin in Figure 2, so that

$$L_{\text{NM}} > \|\widehat{X}_k\|_{\max}, \quad k = 1, \cdots, K, \quad \text{for } \textit{arbitrary trial } \tau. \tag{32}$$

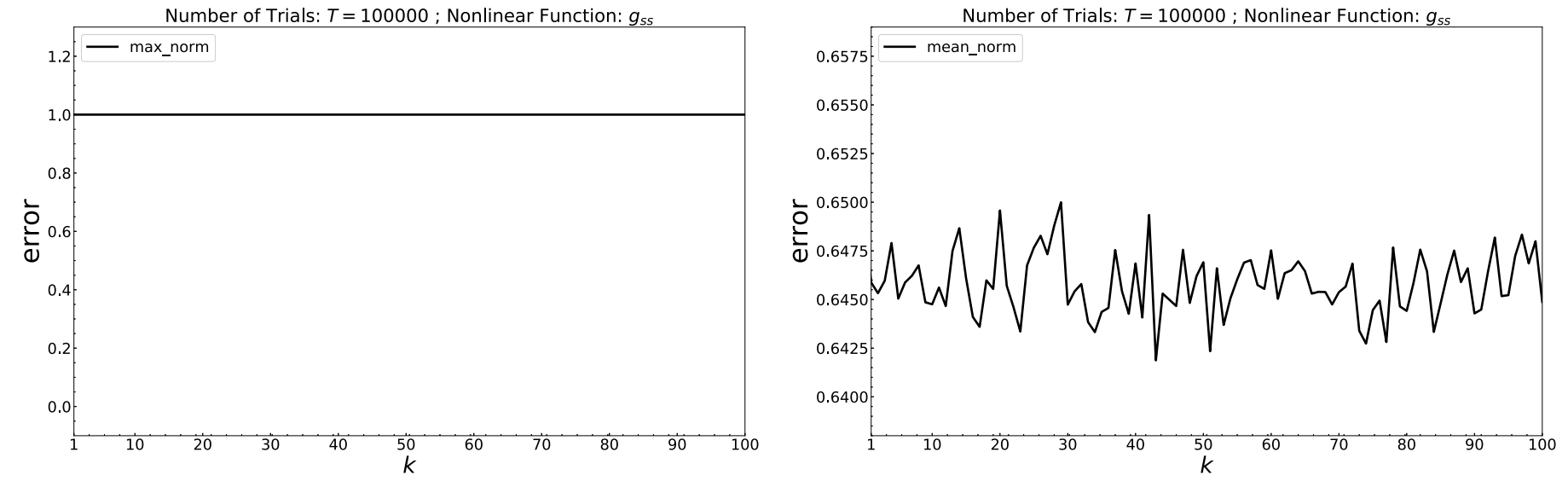

Figure 20: Nonlinearity in Restoration. The nonlinearity is introduced by $g_{\text{ss}}$ defined by Eq.(30). The error estimates are made with $A^i = 0.1$, $b^i = b^i_u = 1$, $\sigma_1 = 0.01$, and $\sigma_2 = \sigma_v = 1$ in nonlinear CS. The restoration is made by linear RS without knowing the nonlinearity. The number of trials is $100,000$. DA continuation is used to obtain the graphs. The left graph is for the maximum norm, and the right graph for the mean norm. The graph of the maximum norm indicates 1 at all points $k$, which means there is a possibility of bit flip according to Eq.(27). The statistical probability of the possibility is shown in the graph of the mean norm according to Eq.(29).

Based on our numerical analyses, as in Figure 21, it seems to be enough to set $L_{\text{NM}}$ as $L_{\text{NM}} = 0.1$, and thus, to define LLR and LHR by $V_{\text{LLR}} = 0.1$ and $V_{\text{LHR}} = 0.9$, respectively, for the deviation coming from $W^{j,i}_t$.

Numerically to check this conjecture, *only for Figures 22 and 23*, we add an extra value to the restored bit $\widehat{a}_k$, $k = 1, \cdots, K$, in the following. We set the extra value as $\widehat{a}_k = 0.5$ if $\widehat{X}_k$ is in FZ (i.e., $V_{\text{LLR}} < \widehat{X}_k < V_{\text{LHR}}$). We numerically analyze the maximum norm defined by Eq.(26) and the mean norm defined by Eq.(28) for the revised version of the restored bit $\widehat{a}_k$. Here, we note that the value of $|a_k - \widehat{a}_k(\tau)|$ is 0, 0.5, or 1 for the revised version of the restored bit $\widehat{a}_k$. Thus, double the value of the mean norm, $2\|\widehat{a}_k\|_{\text{mean}}$, means the statistical probability that $\widehat{X}_k$ is in FZ when the maximum norm $\|\widehat{a}_k\|_{\max}$ is 0.5. Because there is no trial $\tau_1$ satisfying $|a_k - \widehat{a}_k(\tau_1)| = 1$ if $\|\widehat{a}_k\|_{\max} = 0.5$, and thus, $2\sum_{\tau=1}^{T} |a_k - \widehat{a}_k(\tau_1)|$ indicates the total number of the occurrence of $|a_k - \widehat{a}_k| = 0.5$.

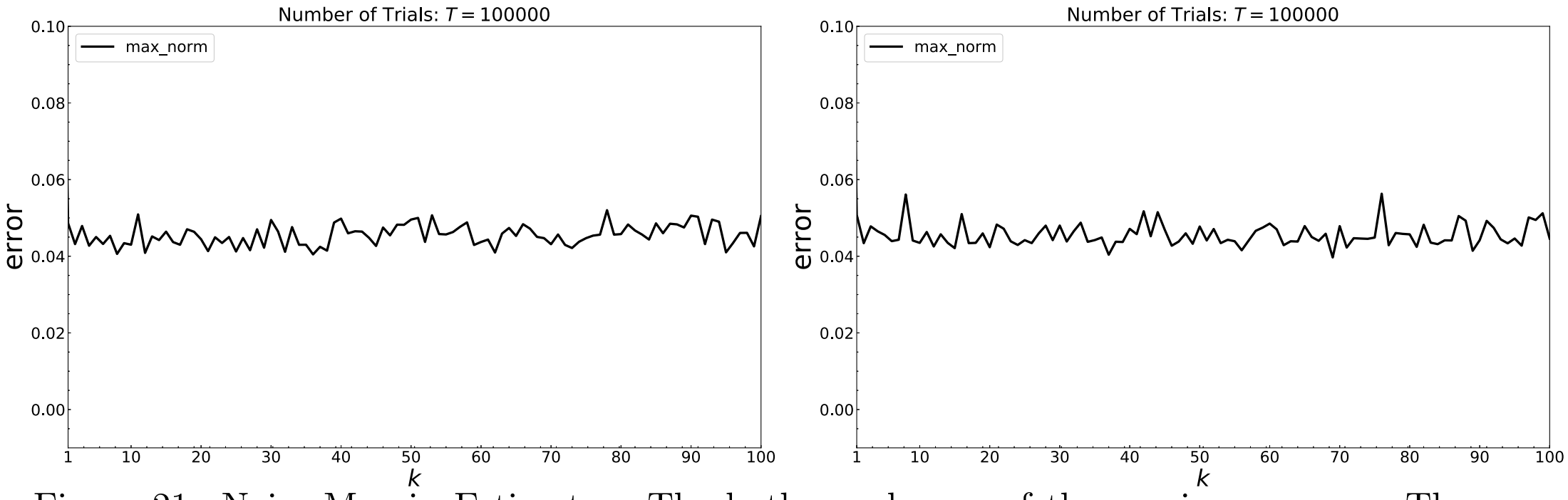

Figure 21: Noise Margin Estimates. The both graphs are of the maximum norm. The left shows the error estimate for the restoration $\widehat{X}_t$ in linear CRS, and the right shows the error estimate for the restoration $\widehat{X}_t$ in nonlinear CRS. The parameters are set as $A = 0.1$, $b = b_u = 1$, $\sigma_1 = 0.01$, and $\sigma_2 = \sigma_v = 1$. DA continuation is used to obtain the graphs. The number of trials is $100,000$. This example of numerical analyses supports the setting $L_{\mathrm{NM}} = 0.1$ for Eq.(32).

Our numerical analyses support the conjecture as in Figure 22. Meanwhile, if we set the LHR and LLR so narrow as in Figure 23 for instance, $\widehat{X}_k$ in FZ appears for each $k$.

If, for instance, we use CRS just for data saving, it is seldom necessary to mind other noises $W_t^{\mathrm{ext}}$ separately from $W_t^{j,i}$. However, when we use CRS, for instance, in wireless (tele)communication, we may meet the situation that the concealed data have some external noises $W_t^{\mathrm{ext}}$ from the outside of CRS, and we are bound to mind the deviation coming from $W_t^{\mathrm{ext}}$. In this case, CRS allows $W_t^{\mathrm{ext}}$ until it is broken that $\widehat{X}_k$ for 'high' is more than $V_{\mathrm{LHR}}$ and $\widehat{X}_k$ for 'low' is less than $V_{\mathrm{LHR}}$ for each $k = 1, \cdots, K$. We show some examples of the results of numerical analyses below.

We suppose that Eve disturbs the communication between Alice and Bob, and tampers with the concealed data $U_t^i$ by injecting an external noise $W_t^{\mathrm{ext}}$ into the concealed data $U_t^i$ that Alice sends to Bob. In this case, Bob receives the data $\widetilde{U}_t^i$ polluted with noise $W_t^{\mathrm{ext}}$ *unknown to Alice and Bob*,

$$\widetilde{U}_t^i = U_t^i + W_t^{\mathrm{ext}},$$

and restores the bit words with the current RS based on the linear Kalman filtering theory. Since the state equation is given by Eq.(14) (or Eq.(15)), the Langevin equation, the external noise $W_t^{\mathrm{ext}}$ is processed as the part of its random force. We estimate the restoring ability of CRS, which depends on the ability of the noise elimination. We assume that Eve uses the external noise $W_t^{\text{ext}}$ with the distribution $N(0, (\sigma_{\text{ext}})^2)$, but Alice and Bob *use CRS without any information about* $W_t^{\text{ext}}$.

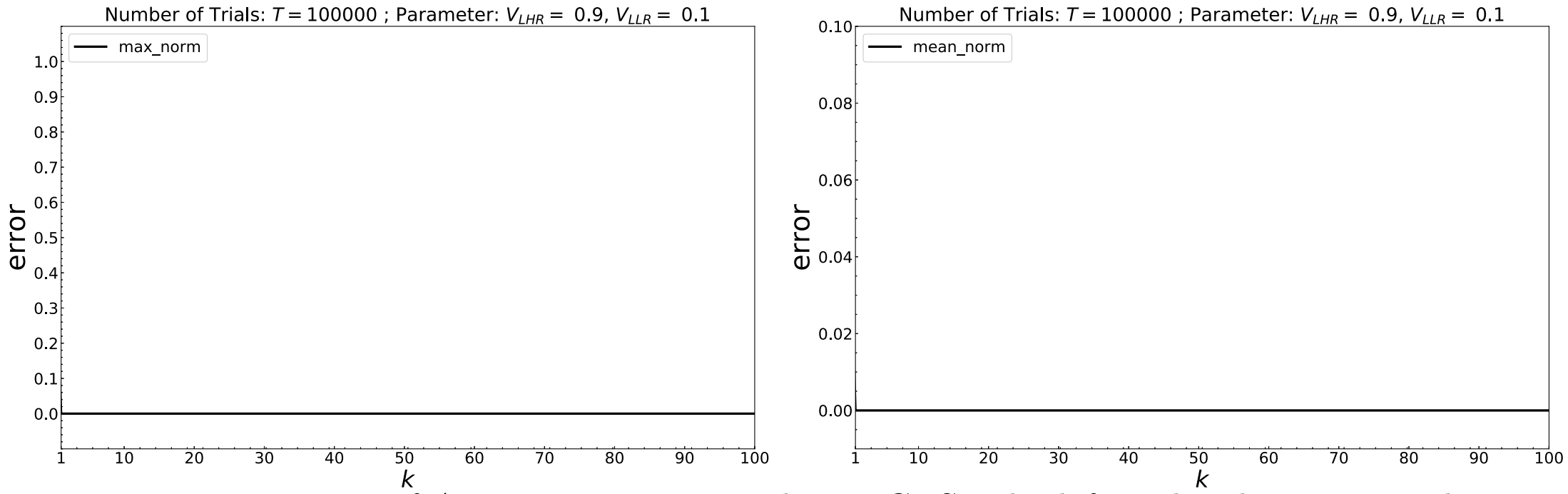

Figure 22: Estimates of Appearance in FZ in linear CRS. The left and right respectively show the maximum norm and the mean norm of error estimate for the restoration $\widehat{X}_t$. Here, $V_{\mathrm{LHR}} = 0.9$ and $V_{\mathrm{LLR}} = 0.1$. The other parameters are set as $A = 0.1$, $b = b_u = 1$, $\sigma_1 = 0.01$, and $\sigma_2 = \sigma_v = 1$. DA continuation is used to obtain the graphs. The number of trials is $100,000$. The left graph is for the maximum norm, and the right graph for the mean norm. According to Eqs.(27) and (29), the both graphs indicates 0 at all points $k$, which means there is no possibility of bit flip almost surely.

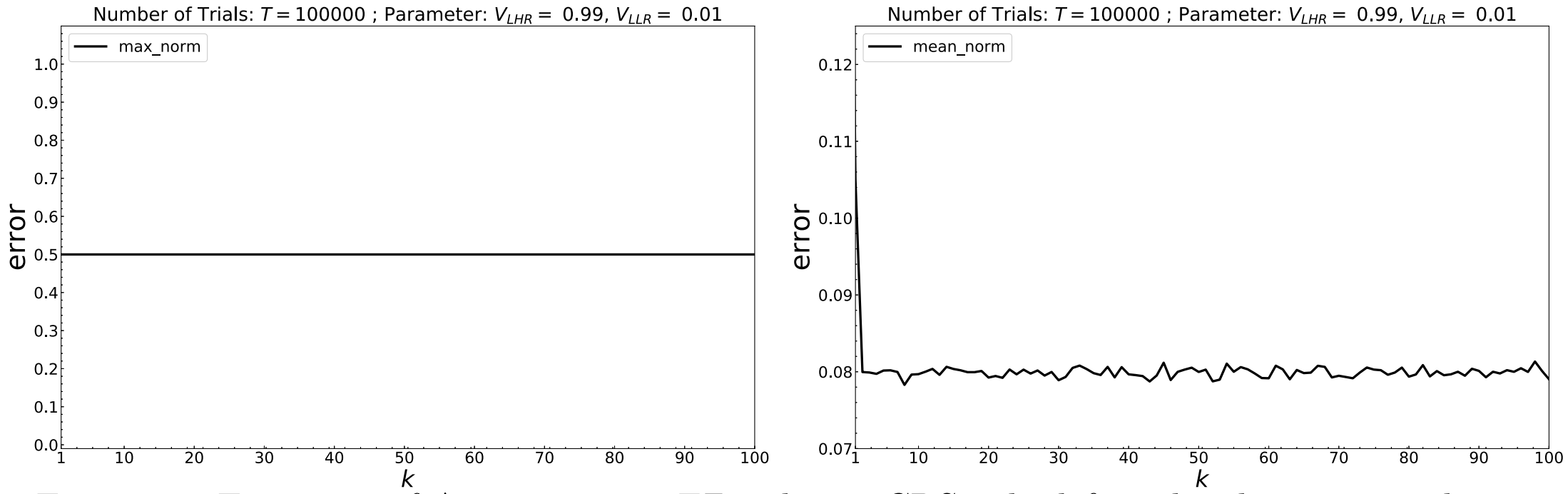

Figure 23: Estimates of Appearance in FZ in linear CRS. The left and right respectively show the maximum norm and the mean norm of error estimate for the restoration $\widehat{X}_t$. Here, $V_{\mathrm{LHR}} = 0.99$ and $V_{\mathrm{LLR}} = 0.01$. The other parameters are set as $A = 0.1$, $b = b_u = 1$, $\sigma_1 = 0.01$, and $\sigma_2 = \sigma_v = 1$. DA continuation is used to obtain the graphs. The number of trials is $100,000$. The left graph is for the maximum norm, and the right graph for the mean norm. Revising Eqs.(27) and (29) to meet the revised version of the restored bits, the graph of the maximum norm indicates 0.5 at all points $k$, which means there is a possibility that $\widehat{X}_t$ falls in FZ. The statistical probability of the possibility is given by double the value of the graph of the mean norm.

**Linear RS**: We here use linear CRS. Our numerical analyses reveal that the restoring ability of linear CRS is decreasing as the standard deviation $\sigma_{\text{ext}}$ of the unknown external noise $W_t^{\text{ext}}$ grows larger. We show some examples of the numerical analyses in Figures 24 and 25. The bit flips gradually appear at some $k$ as $\sigma_{\text{ext}}$ grows, while linear RS functions well against $W_t^{\text{ext}}$ with small $\sigma_{\text{ext}}$. According to our numerical analyses, the bit flips rarely appears at $\sigma_{\text{ext}} \approx 0.1$ as in Figure 25.

**Nonlinear RS**: It is easy to conceive that Nonlinearization Schemes A & B do not function well against the unknown external noise $W_t^{\text{ext}}$ because the nonlinearity introduced by $g_\nu$ enhances the random noise disturbance coming from $W_t^{\text{ext}}$ through the process, $g_\nu^{-1}(\widetilde{U}_t^i) = g_\nu^{-1}(U_t^i + W_t^{\text{ext}})$, which is beyond the control of the 'linear' Kalman filtering theory. Actually, we can see such examples in Figure 26. Even if the standard deviation $\sigma_{\text{ext}}$ is small, the bit flips rarely take place at each $k$. If we have enough computing power for CRS, we can handle the nonlinearity as well as non-Gaussian distribution by employing the so-called particle filtering [47] instead of the linear Kalman filtering under some proper conditions. We will show some results on this noise elimination in the near future.

We make a remark at the end of this section. In our previous paper [1] we propose a *tampering detection method* in CRS. The method uses the message digests, the values returned by a hash function, and is already implemented.

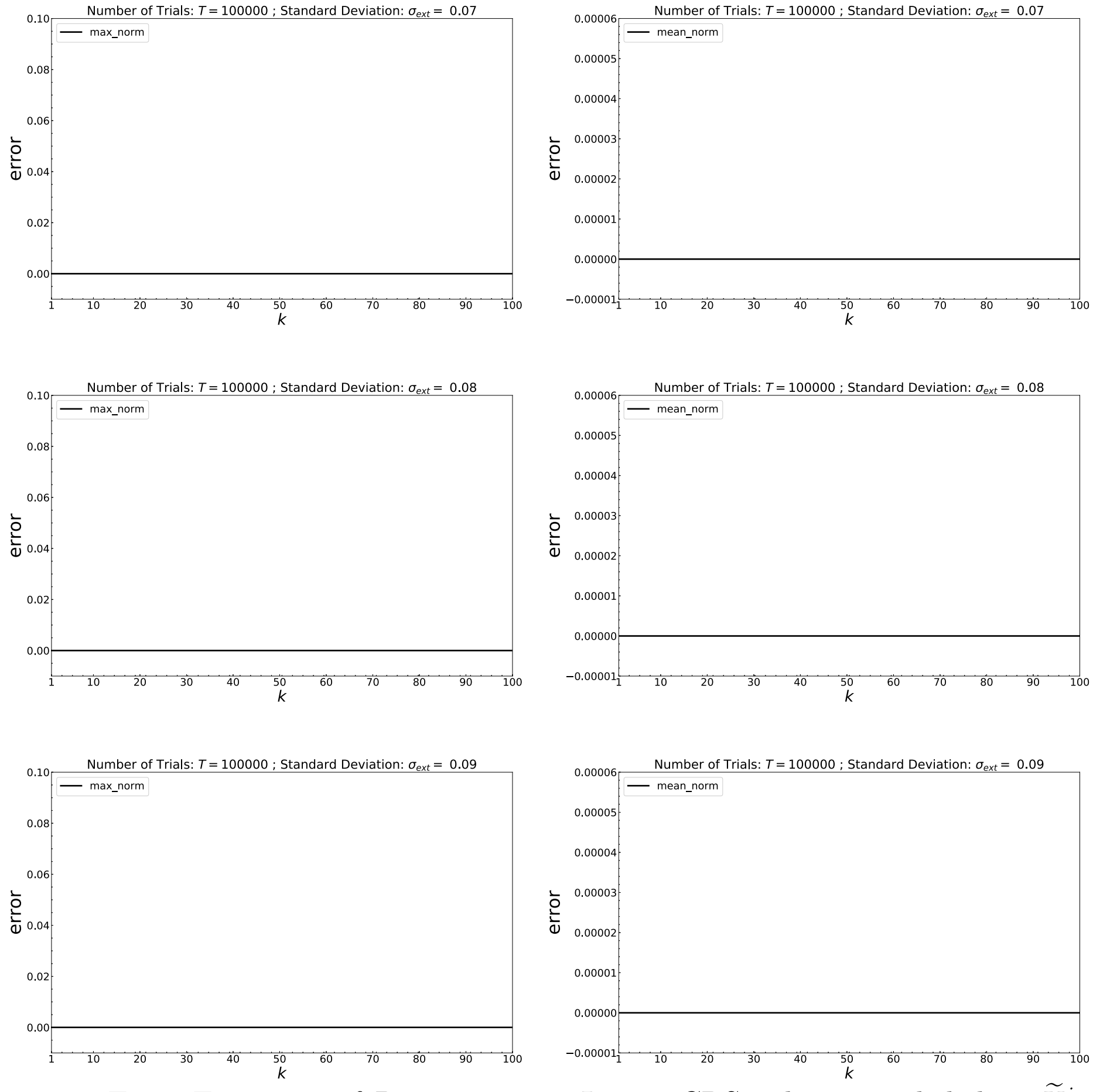

Figure 24: Error Estimates of Restoration in Linear CRS. The concealed data $\widetilde{U}_t^i$ are tampered with by the unknown external noise $W_t^{\text{ext}}$ with $\sigma_{\text{ext}} = 0.07, 0.08, 0.09$. The parameters are $A = 0.1$, $b = 1$, $\sigma_1 = 0.01$, $\sigma_2 = \sigma_v = 1$ in the linear CRS, and the number of trials is $100,000$. The noise margin is determined by $V_{\text{LHR}} = V_{\text{LLR}} = V_{\text{thd}} = 0.5$. DA continuation is used to obtain the graphs. On each row, the left graph is for the maximum norm, and the right graph for the mean norm. According to Eqs.(27) and (29), the both graphs indicates 0 at all points $k$, which means there is no possibility of bit flip almost surely.

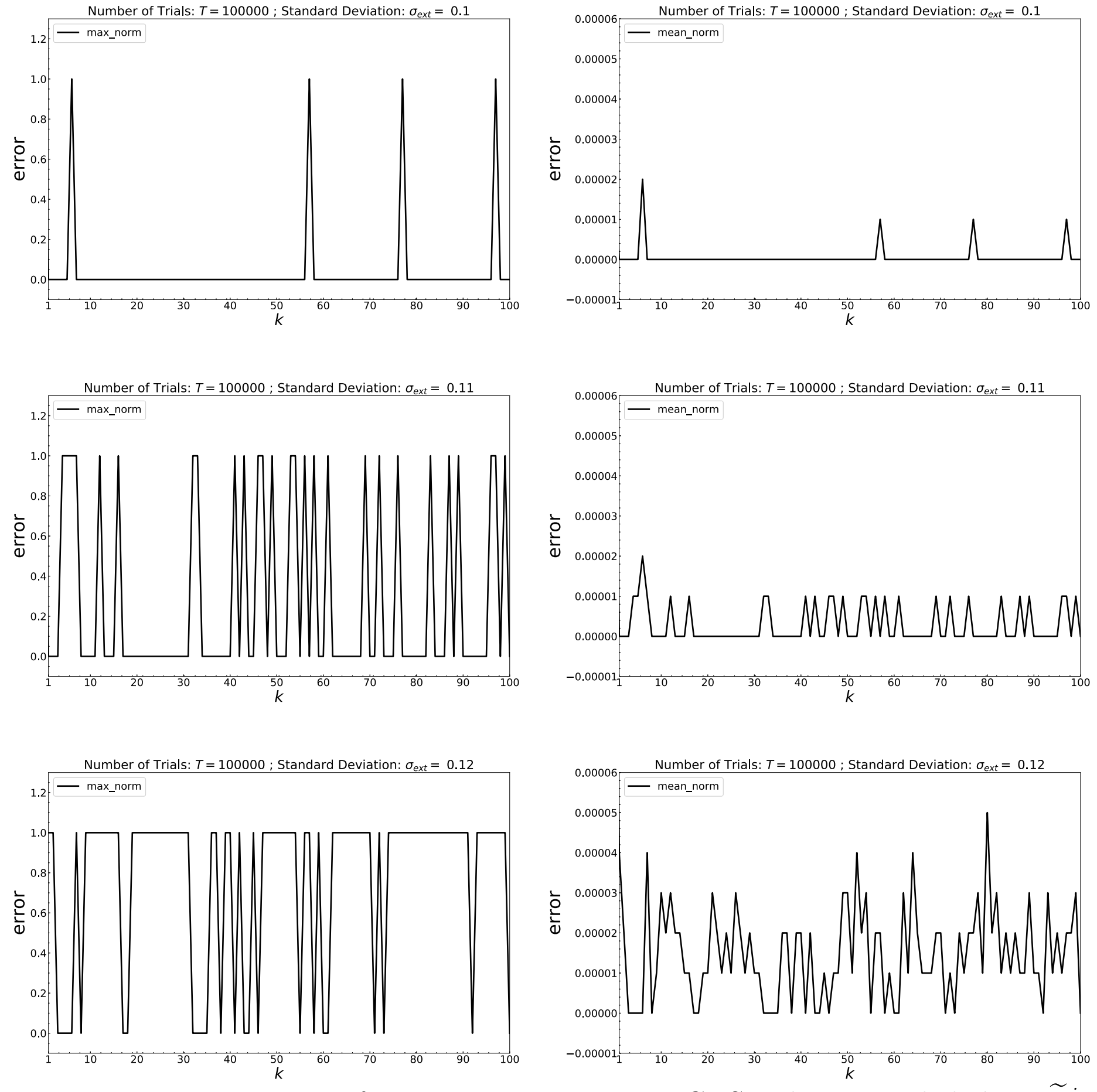

Figure 25: Error Estimates of Restoration in Linear CRS. The concealed data $\widetilde{U}_t^i$ are tampered with by the unknown external noise $W_t^{\rm over}$ with $\sigma_{\rm ext} = 0.10, 0.11, 0.12$. The parameters are $A = 0.1$, $b = 1$, $\sigma_1 = 0.01$, $\sigma_2 = \sigma_v = 1$ in the linear CRS, and the number of trials is $100,000$. The noise margin is determined by $V_{\rm LHR} = V_{\rm LLR} = V_{\rm thd} = 0.5$. DA continuation is used to obtain the graphs. On each row, the left graph is for the maximum norm, and the right graph for the mean norm. The graph of the maximum norm indicates 1 at some points $k$, which means there is a possibility of bit flip at those points according to Eq.(27). The statistical probability of the possibility is shown by the individual corresponding graph of the mean norm according to Eq.(29).

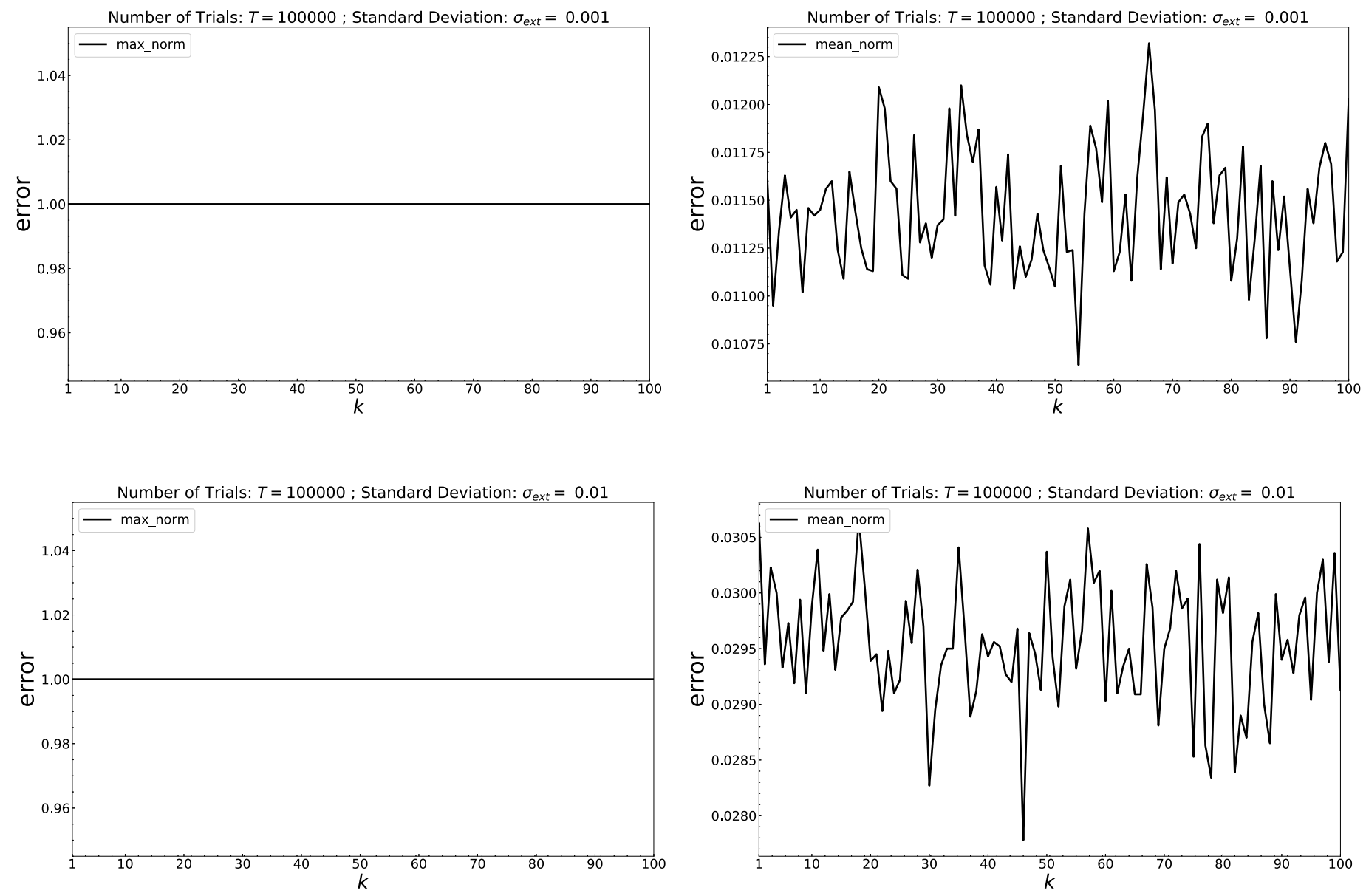

Figure 26: Error Estimates of Restoration in Nonlinear CRS. The concealed data $\widetilde{U}_t^i$ are tampered with by the unknown external noise $W_t^{\mathrm{over}}$ with $\sigma_{\mathrm{ext}} = 0.001, 0.01$. The parameters are $A = 0.1$, $b = 1$, $\sigma_1 = 0.01$, $\sigma_2 = \sigma_v = 1$ in the linear CRS, and the number of trials is $100,000$. The noise margin is determined by $V_{\mathrm{LHR}} = V_{\mathrm{LLR}} = V_{\mathrm{thd}} = 0.5$. DA continuation is used to obtain the graphs. On each row, the left graph is for the maximum norm, and the right graph for the mean norm. The graph of the maximum norm indicates 1 at some points $k$, which means there is a possibility of bit flip at those points according to Eq.(27). The statistical probability of the possibility is shown by the individual corresponding graph of the mean norm according to Eq.(29).

## 6. Application to Digital Pictorial Image

We apply CRS to a digital pictorial image. We use binary data of a digital pictorial image in the ORL Database of Faces, the archive of AT&T Laboratories Cambridge [52]. The data have the greyscale value of 256 gradations (8bit/pixel). We set our parameters as $A^i = A = 0.1$, $b^i = b = 1$, $b_u^i = b_u = 1$, $c^i = c = 1$, $\sigma_1^i = \sigma_1 = 0.1$, $\sigma_2^i = \sigma_2 = 1$, $\sigma_v^i = \sigma_v = 1$, and $V_{\mathrm{LHR}} = V_{\mathrm{LHR}} = V_{\mathrm{thd}} = 0.5$. Here, we dare to use $\sigma_1 = 0.1$ because we know that the bit flip seldom take place with this value of $\sigma_1$ in §5.2. The original pictorial image and the first part of its binary pulse $X_t$ are obtained as in Figure 27. The upper bound of $t$ is given by the number of its total pixels, $92 \times 112 = 10304$, and thus, $t$ runs over $[0, 10304 \times 8]$ actually because we need 8 bits per pixel. In other words, we need the bit word with the length $10304 \times 8 + 1$. We obtain the data $U_t^i$, $i = 1, 2, 3$, concealed by linear CS. In

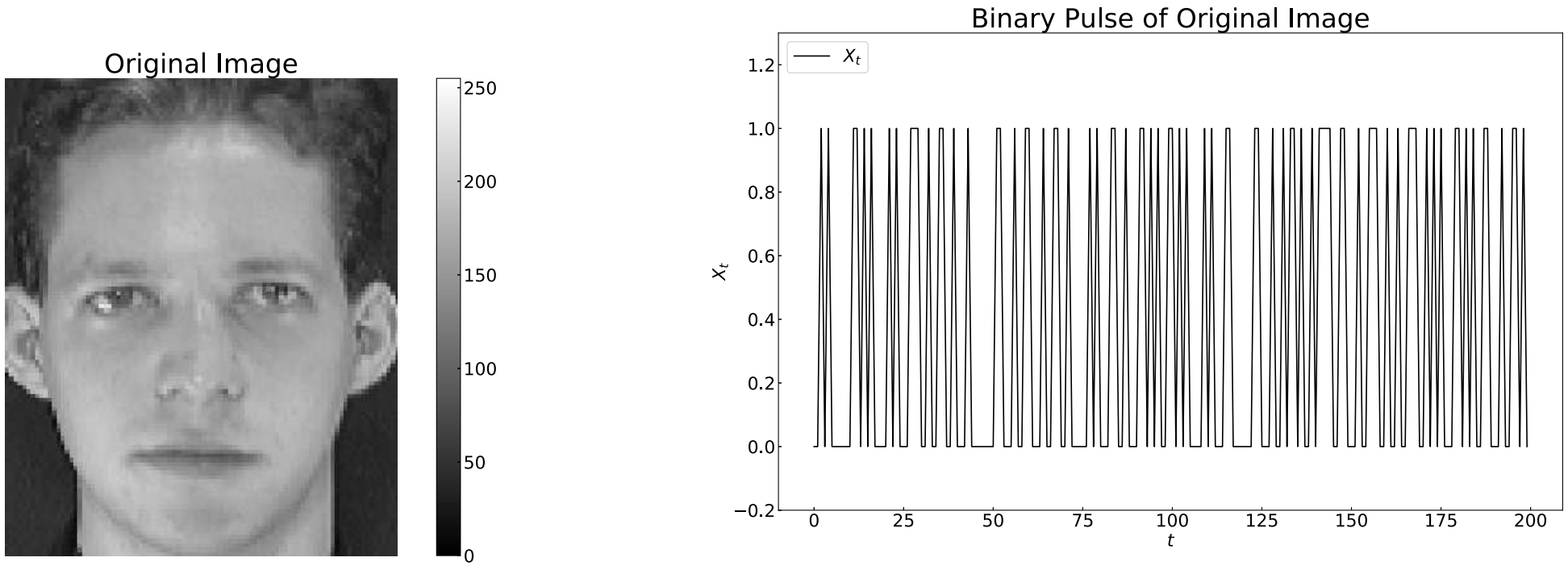

Figure 27: Pictorial Image & Binary Pulse. The original pictorial image with the digital data is on the left. Its binary pulse $X_t$ is on the right, and shown only for $t \in [0, 200]$. Actually, $t$ runs over $t \in [0, 10304 \times 8]$.

Figure 28, we show the first part of each $U_t^i$. We are interested in how much we can recover the pictorial image from the concealed data. We assume that Eve succeeds in wiretapping and obtaining one of $U_t^1$, $U_t^2$, $U_t^3$, and tries to get a pictorial image from it without the noise elimination. Then, she must basically know relevant ADE since the concealed data, $U_t^1$, $U_t^2$, and $U_t^3$, are analog signals as in Figure 28.

Meanwhile, in the OSI reference model, ADE is done for the data on the physical layer, and is handled from the data link layer. Our transformation from the digital data to a pictorial image should be done on an upper layer of the OSI reference model. We denote this transformation by TDP.

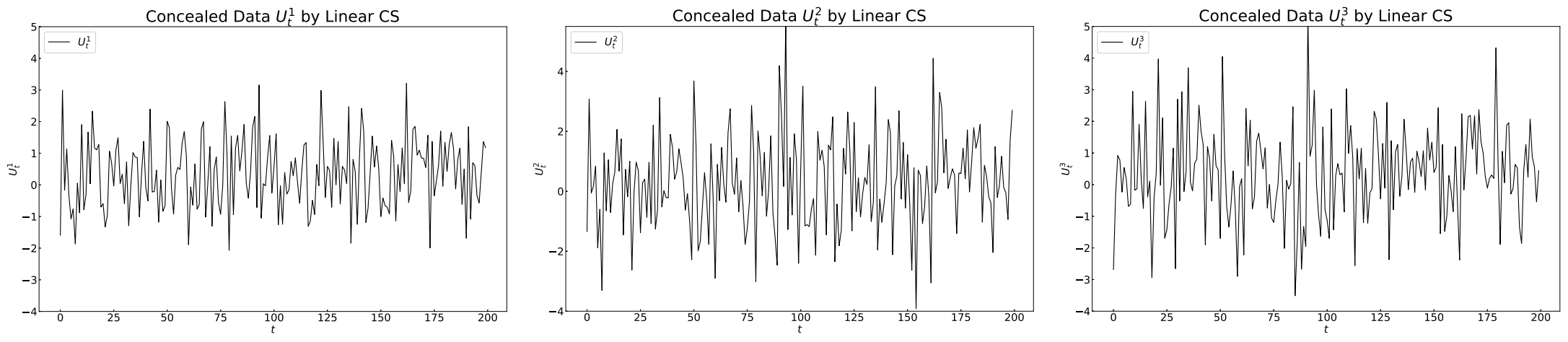

Figure 28: Concealed Data by Linear CS. The graphs of $U_t^1$, $U_t^2$, $U_t^3$ obtained by linear CS for the binary pulse $X_t$ in Figure 27 are respectively given from the left. Here, we show each graph for $t \in [0, 200]$ only; though $t \in [0, 10304 \times 8]$ actually. DA continuation is used to obtain each graph of $U_t^i$.

We moreover assume that Eve succeeds in obtaining both ADE and TDP. Then, she gets the individual pictorial images from the concealed data, $U_t^1$, $U_t^2$, and $U_t^3$, as in Figure 29.

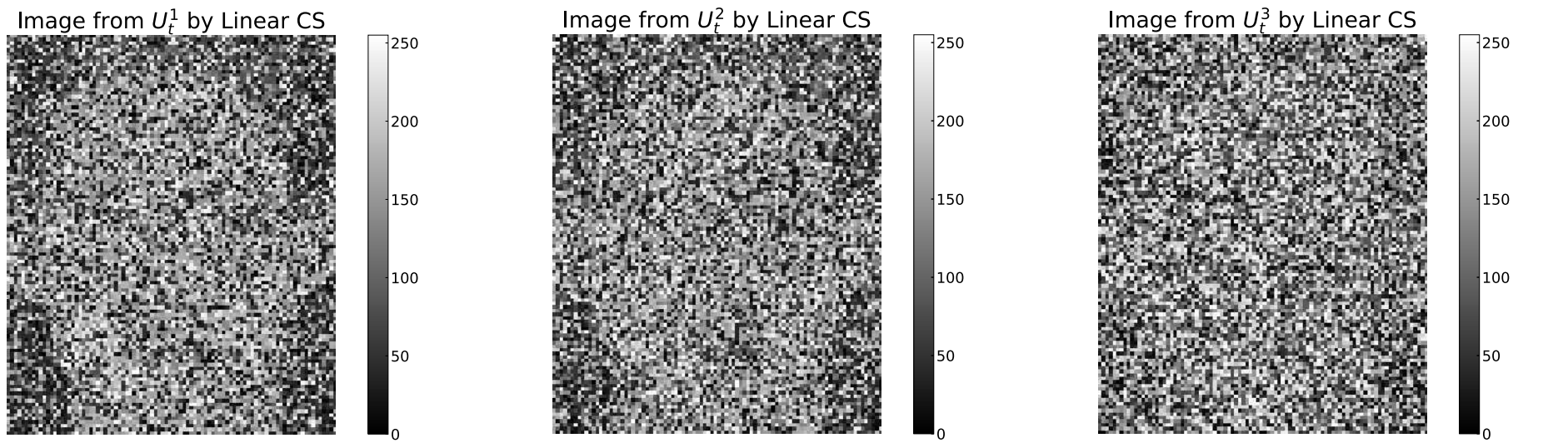

Figure 29: Pictorial Images Restored from Concealed Data by Linear CS. The individual pictorial images are respectively recovered from the concealed data, $U_t^1$, $U_t^2$, $U_t^3$, by linear CS from the left. Here, we assume that Eve employs our ADE as TDP.

Eve somewhat finds the faint contour of the face by the light (white) and shade (black) in the pictorial images in Figure 29, though the faint contour gradually fades out of the pictorial image as repeating the noise disturbances from $U_t^1$ to $U_t^3$. We will improve this appearance of the faint contour by introducing nonlinearity later on. On the other hand, Bob succeeds in restoring the binary pulse and recovering the pictorial image. The restoration $\widehat{X}_t$ by linear RS and the recovered pictorial image from it are in Figures 30 and 31, respectively. In Figure 30, we also show the first part of the comparison of the original binary pulse $X_t$ and the restoration $\widehat{X}_t$ by linear RS.

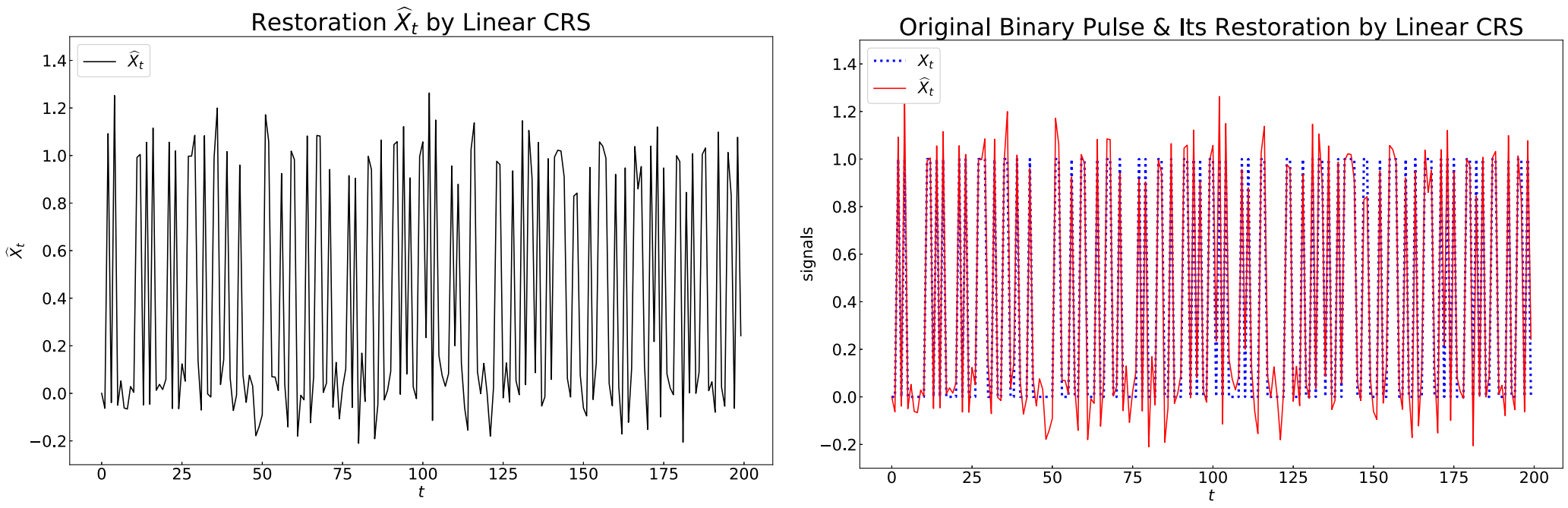

Figure 30: Bob's Restoration. The left is the restoration $\widehat{X}_t$ by linear RS, and the right is the comparison of the original binary pulse $X_t$ and its restoration $\widehat{X}_t$. Here, we show the graph only for $t \in [0, 200]$; $t \in [0, 10304 \times 8]$ actually.

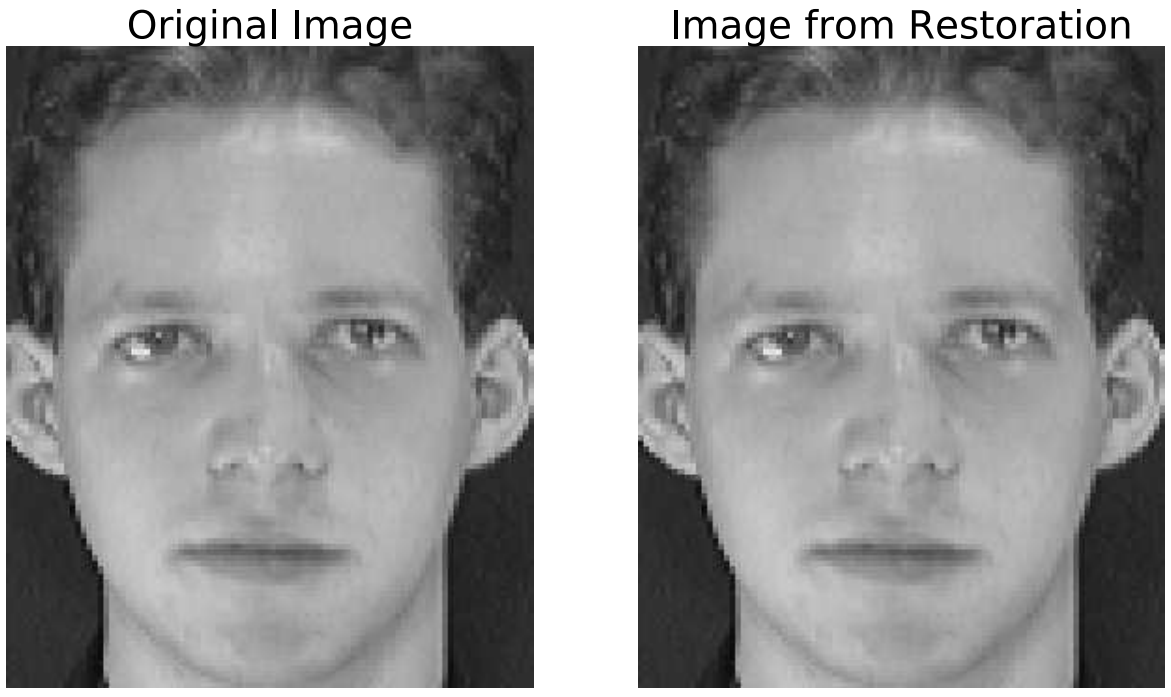

Figure 31: Bob's Success in Recovery of Pictorial Image. The left is the original pictorial image in Figure 27, and the right is the pictorial image obtained from the restoration $\widehat{X}_t$ obtained by linear RS.

We conceal the original binary pulse $X_t$ by nonlinear CS with $g_{\mathrm{c}}$ defined by Eq.(12). We have the concealed data, $U_t^i$, $i = 1, 2, 3$, as in Figure 32.

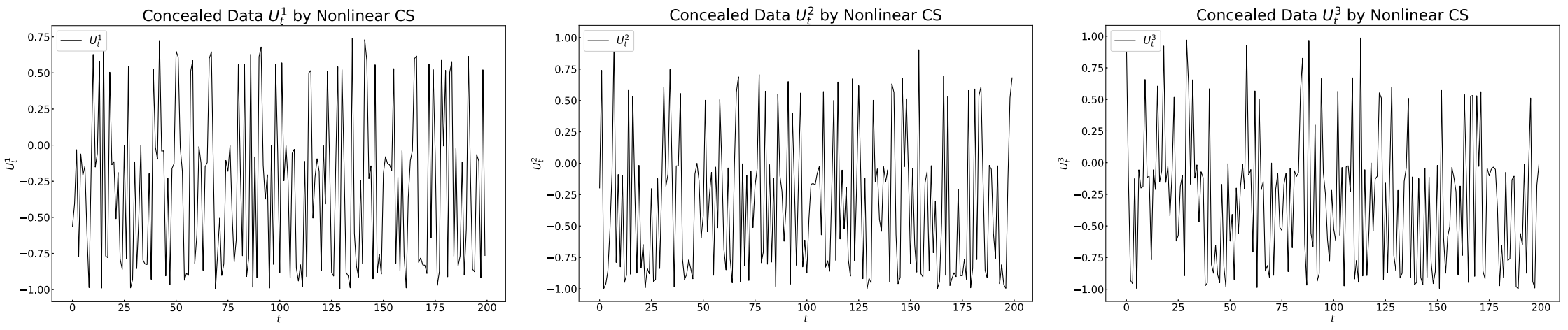

Figure 32: Concealed Data by Nonlinear CS. The graphs of $U_t^1$, $U_t^2$, $U_t^3$ obtained by nonlinear CS for the binary pulse $X_t$ in Figure 27 are respectively given from the left. Here, we denote $U_t^{\mathrm{nl},i}$ just by $U_t^i$ following the abbreviation in Nonlinearization Strategy in Section 4, and $t \in [0, 200]$ only; though $t \in [0, 10304 \times 8]$ actually. DA continuation is used to obtain each graph of $U_t^i$.

In the same way as the linear case, we assume that Eve tries to obtain a pictorial image from $U_t^i$. Then, she reach the individual pictorial images in Figure 33.

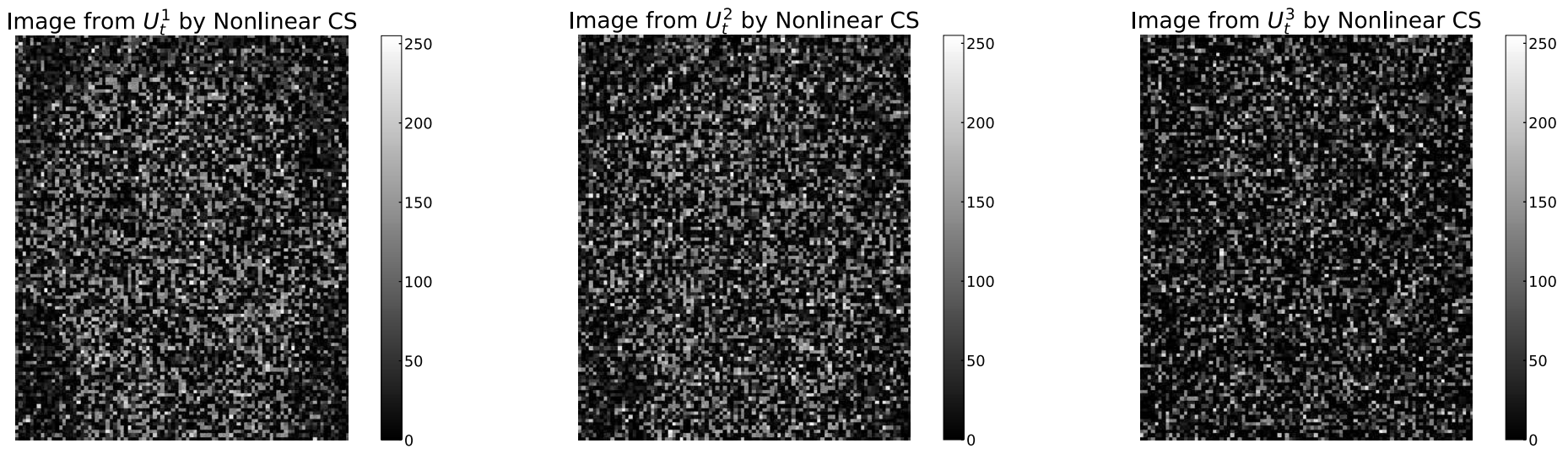

Figure 33: Pictorial Images Restored from Concealed Data by Nonlinear CS. The individual pictorial images recovered from the concealed data, $U_t^1$, $U_t^2$, $U_t^3$, by nonlinear CS from the left, where we denote $U_t^{\mathrm{nl},i}$ just by $U_t^i$ following the abbreviation in Nonlinearization Strategy in Section 4. Here, we assume that Eve employs our ADE as TDP.

Comparing Figure 29 and Figure 33, we realize the faint contour of the face in Figure 29, but it fades away in Figure 33. This fade-away is an effect of nonlinearity. This effect becomes remarkable for the pictorial images with big pixel. Compare Figure 39 and Figure 40 below.

We now assume that Eve knows Alice and Bob's linear CRS and others, but she does not know how Alice and Bob introduce the nonlinearity into their linear CRS. Then, Eve uses linear RS and has the restoration $\widehat{X}_t$ as in Figure 34 from all the data $U_t^i$, $i = 1, 2, 3$, concealed by nonlinear CS. She obtains its pictorial image as in Figure 35.

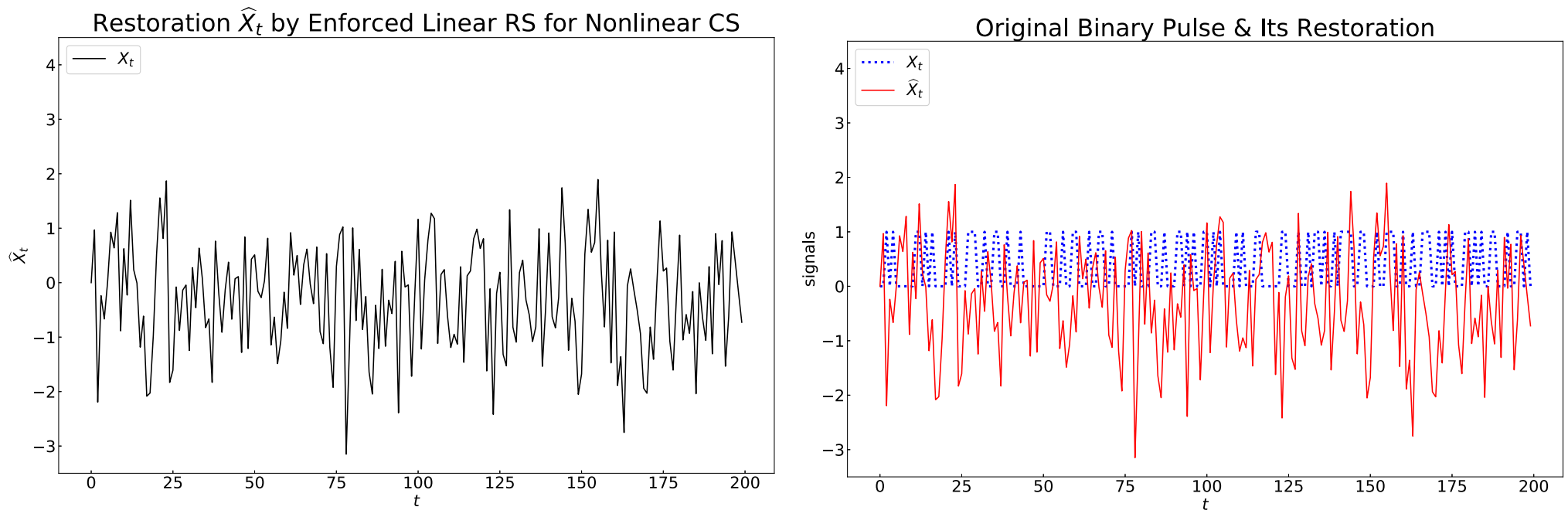

Figure 34: Eve's Restoration. The left is the restoration $\widehat{X}_t$ by linear RS from the concealed data by nonlinear CS, and the right is the comparison of the original pulse $X_t$ and that restoration $\widehat{X}_t$. Here, $t \in [0, 200]$ only; though $t \in [0, 10304 \times 8]$ actually.

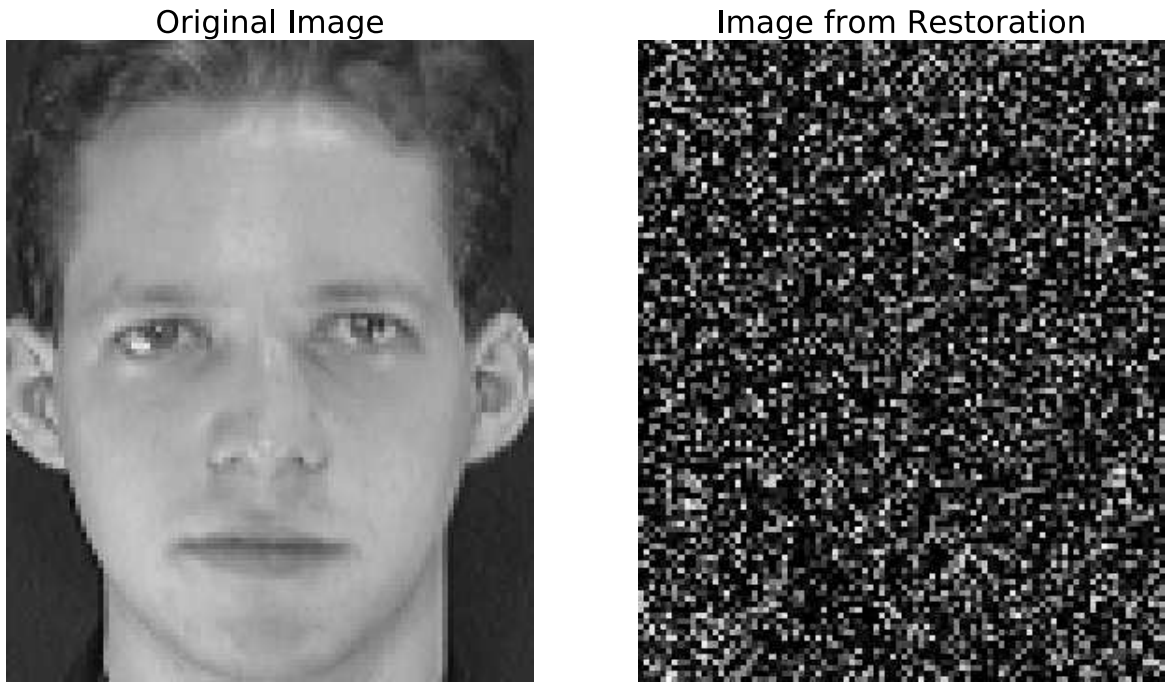

Figure 35: Eve's Failure in Recovery of Pictorial Image. The left is the original pictorial image in Figure 27, and the right is the pictorial image obtained from Eve's restoration $\widehat{X}_t$ as in the left graph of Figure 34.

Since Bob knows the nonlinearity that Alice uses as the common secret key, he can make nonlinear RS with the nonlinearity, which is actually based on the 'linear' Kalman filtering theory due to Nonlinearization Strategy in Section 4. Thus, he obtains the restoration $\widehat{X}_t$ as its first part is in Figure 36, and its pictorial image is in Figure 37.

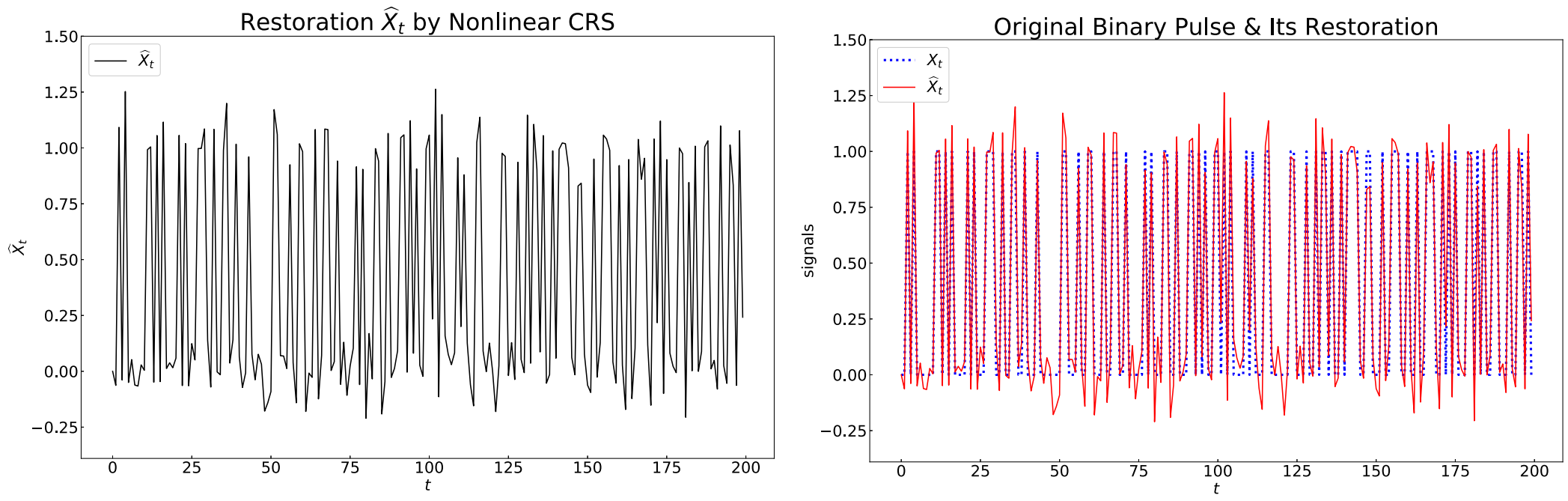

Figure 36: Bob's Restoration. The left is the restoration $\widehat{X}_t$ by nonlinear RS, and the right is the comparison of the original pulse $X_t$ and its restoration $\widehat{X}_t$. Here, we show the graphs only for $t \in [0, 200]$ though we actually need $t \in [0, 10304 \times 8]$ for the whole graphs.

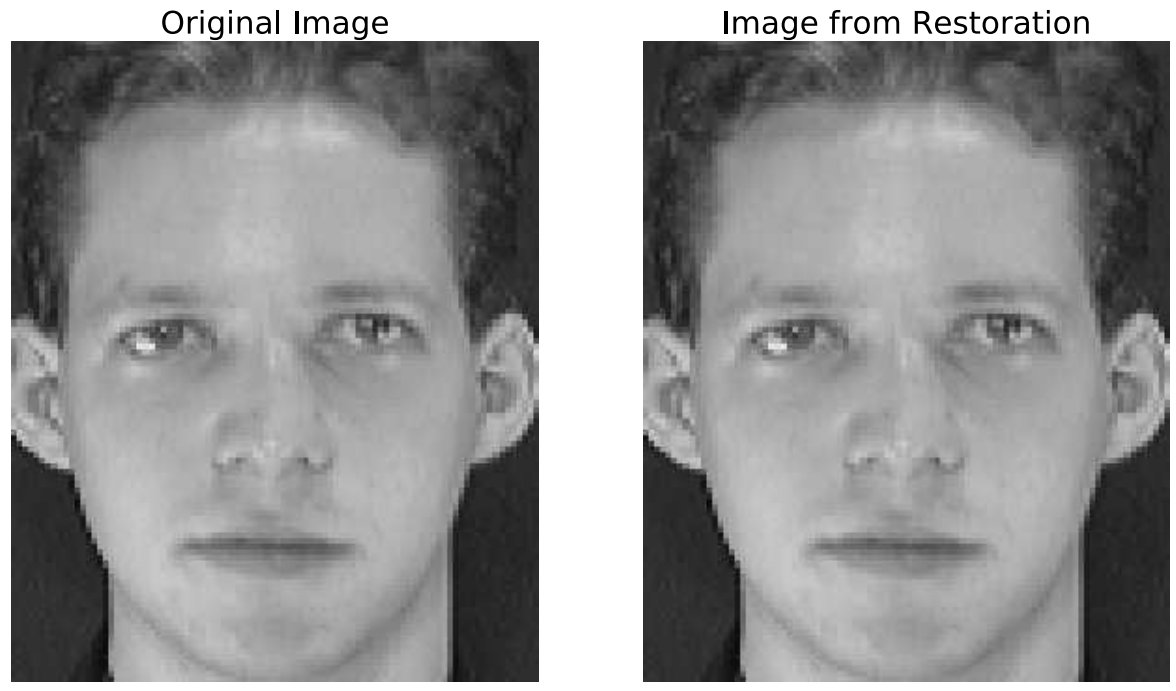

Figure 37: Bob's Success in Recovery of Pictorial Image. The left is the original pictorial image in Fig.27, and the right is the pictorial image obtained from Bob's restoration $\widehat{X}_t$ as in the left graph of Fig.36.

In order to clarify the difference between linear CRS and nonlinear CRS, we apply the both CRSs to the pictorial image with bigger pixel than the pictorial image in Figure 27 with $10,304$ pixels. We use the pictorial image of a baboon in the Standard Image Data-Base. The number of its total pixels is $512 \times 512 = 262,144$. The pictorial image and the first part of its binary pulse is in Figure 38.

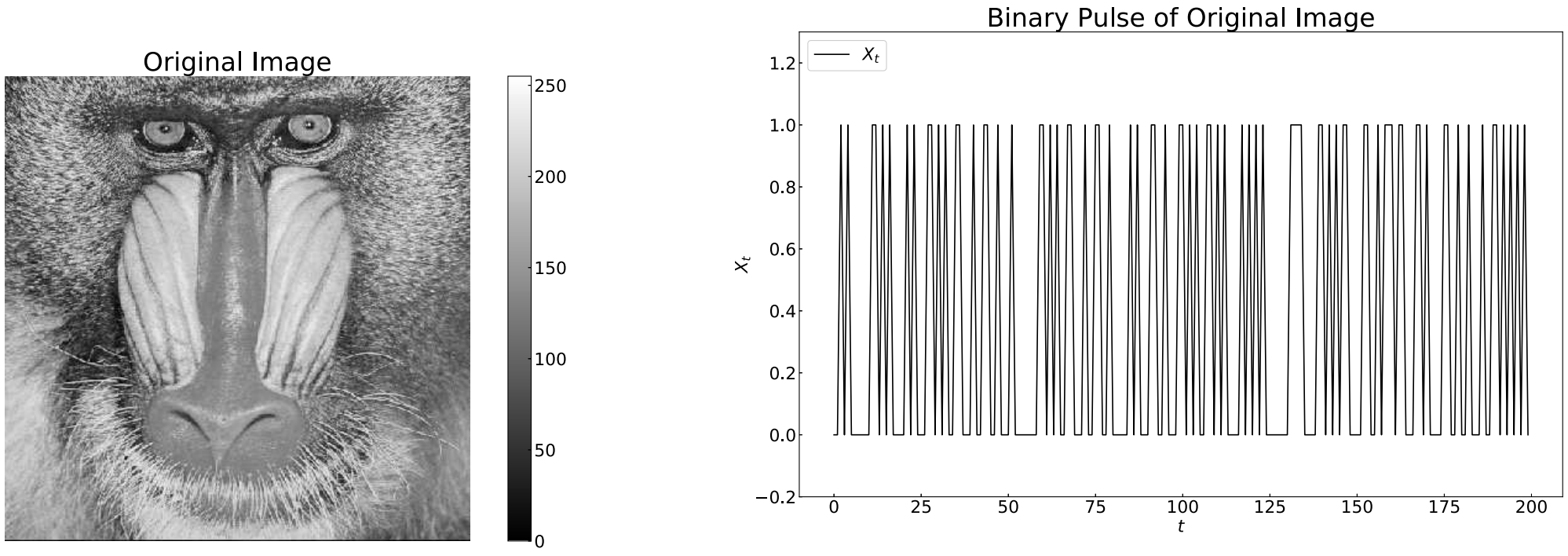

Figure 38: Pictorial Image & Binary Pulse. The left is the original pictorial image with the digital data, and the right is its binary pulse $X_t$ only for $t \in [0, 200]$. The time $t$ actually runs over $[0, 262144 \times 8]$.

The pictorial images in Figure 39 are respectively recovered from the concealed data $U_t^1$, $U_t^2$, and $U_t^3$ from the left, where each $U_t^i$ is obtained by linear CS . We can realize the faint contour of baboon's face in them.

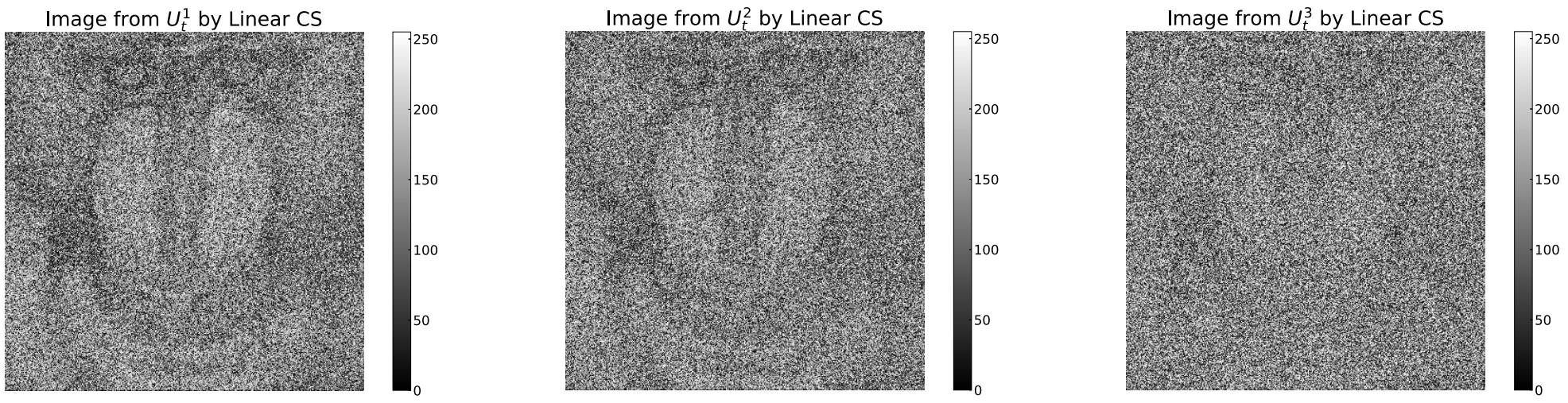

Figure 39: The individual pictorial images recovered from the concealed data, $U_t^1$, $U_t^2$, and $U_t^3$ by linear CS from the left.

The pictorial images in Figure 40 are recovered from the data $U_t^i$, $i = 1, 2, 3$, concealed by nonlinear CS. We can realize almost no contour of baboon's face in them.

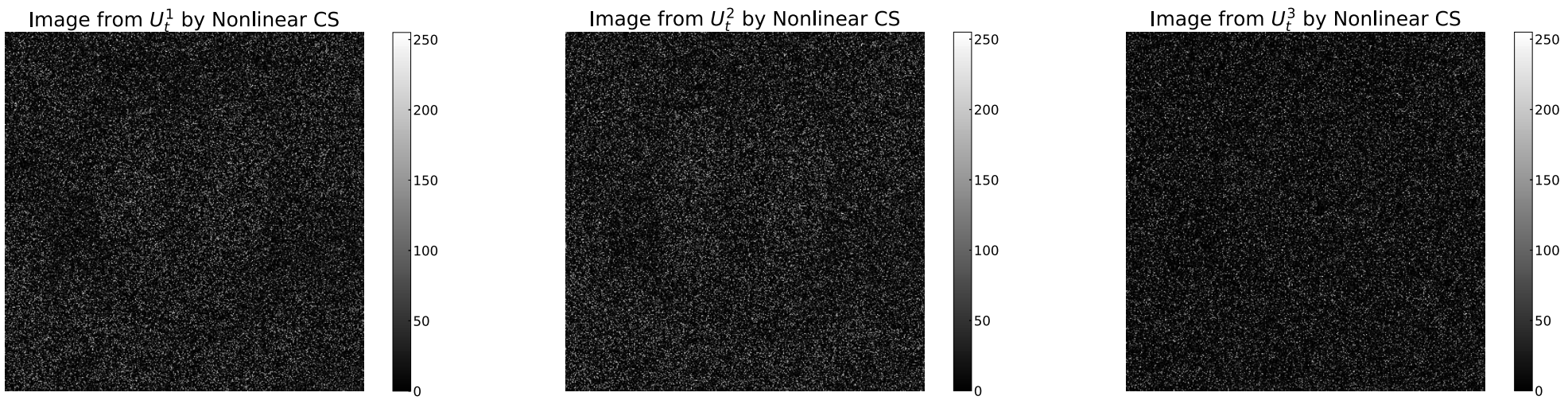

Figure 40: The individual pictorial images recovered from the concealed data, $U_t^1$, $U_t^2$, and $U_t^3$ by nonlinear CS from the left.

We recover the baboon's pictorial image after we restore its binary pulse with nonlinear RS, and then, we obtain the restoration as its first part is in Figure 41 and the recovered pictorial image from it as in Figure 42.

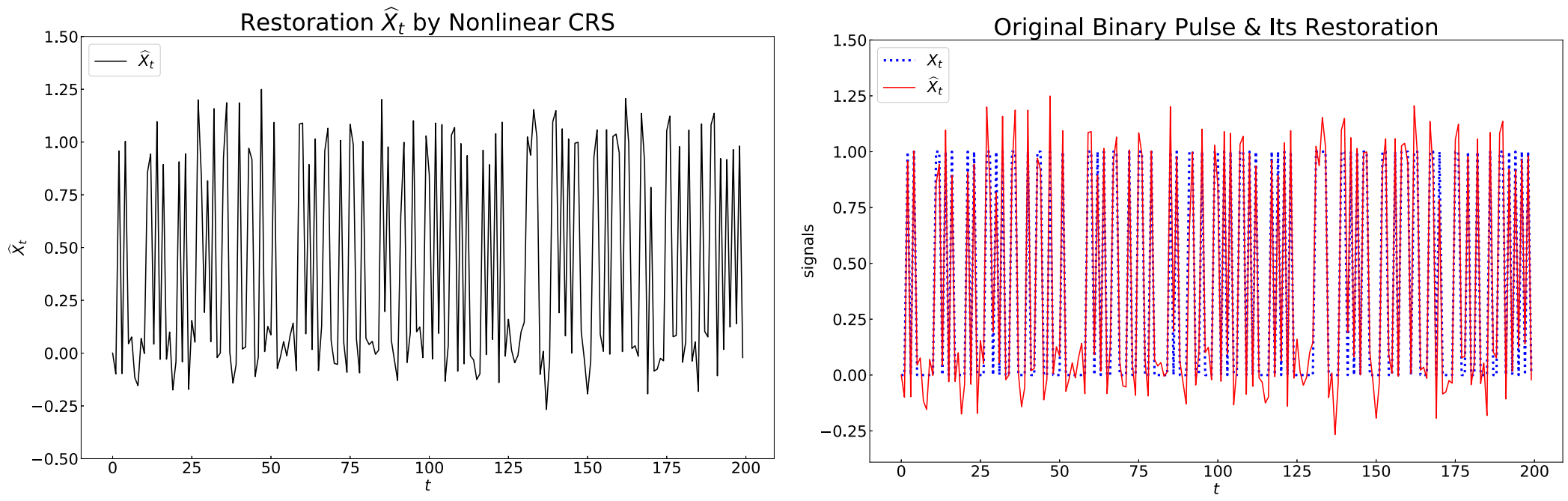

Figure 41: Restoration. The left is the restoration $\widehat{X}_t$ by nonlinear RS, and the right is the comparison of the original pulse $X_t$ and its restoration $\widehat{X}_t$ by nonlinear CRS. Here, we show the first part of the graph only for $t \in [0, 200]$ though whole graph needs $t$ in $[0, 262144 \times 8]$.

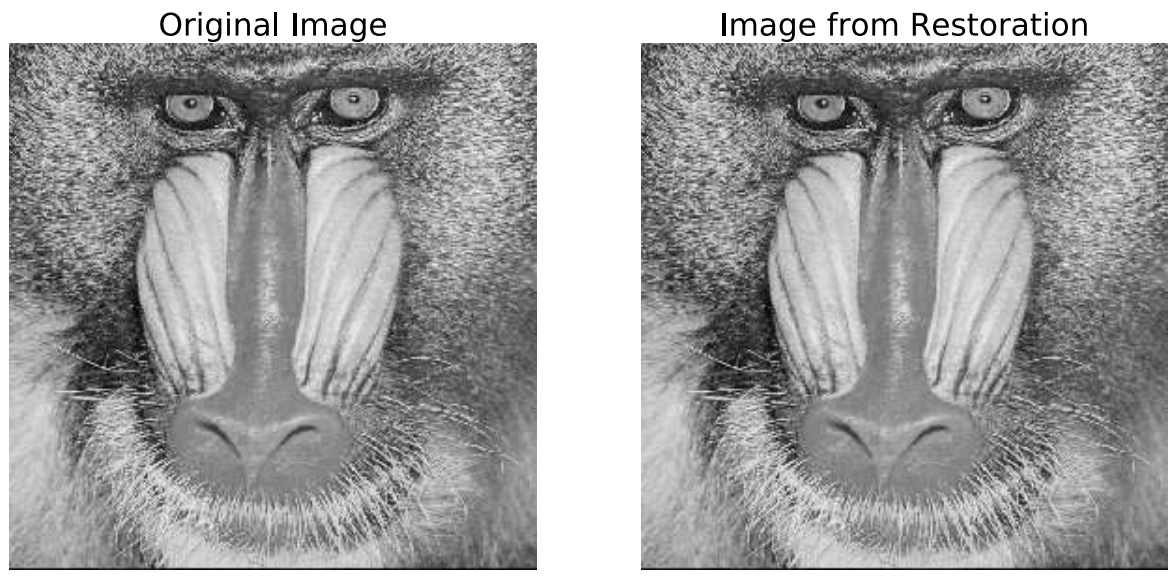

Figure 42: Recovery of Pictorial Image. The left is the original pictorial image in Figure 38, and the right is the pictorial image obtained from the restoration $\widehat{X}_t$ in the left graph of Figure 41.

## 7. Conclusion

We have proposed a nonlinearization of CRS for the data on the physical layer of the OSI reference model. CRS conceals the data by using random noise disturbance and the nonlinearity, and restores the data from the concealed ones to the original ones by using the noise elimination based on a proper stochastic filtering. Making our own SES and finding our own stochastic filtering theory, we can obtain our own CRS for the data on the physical layer of the OSI reference model. We are planning to install CRS on Arduino or FPGA boards [53, 54].

### Acknowledgements

For useful comments and discussion, the authors thank the following: Kirill Morozov (University of North Texas), Shuichi Ohno (Hiroshima University), Teruo Tanimoto (Kyushu University), Kouichi Sakurai (Kyushu University), and Tatsuya Tomaru (HITACHI Ltd.). A part of this paper consists of the thesis for master's degree submitted to Hiroshima University by T. Fujii, and the conception and theoretical parts of this work are based on the research by M. Hirokawa during his tenure at Hiroshima University.